\documentclass{aa} 
\usepackage{natbib}
\usepackage{graphicx}
\usepackage{txfonts}
\usepackage{enumitem}

\begin{document} 
\title{The deepest \textit{Herschel}-PACS far-infrared survey: number counts and infrared luminosity functions from combined PEP/GOODS-H observations\thanks{ Based on observations carried out by the \textit{Herschel} space observatory. \textit{Herschel} is an ESA space observatory with science instruments provided by European-led Principal Investigator consortia and with important participation from NASA.}}

\author{B.~Magnelli\inst{1,2}
	\and
	P.~Popesso\inst{1}
	\and
	S.~Berta\inst{1}
	\and
	F.~Pozzi\inst{3}
	\and
	D.~Elbaz\inst{4}
	\and
	D.~Lutz\inst{1}
	\and
	M.~Dickinson\inst{5}
	\and 
	B.~Altieri\inst{6}
	\and
	P.~Andreani\inst{7,8}
	\and
	H.~Aussel\inst{4}
	\and
	M.~B\'ethermin\inst{4}
	\and
	A.~Bongiovanni\inst{9,10}
	\and
	J.~Cepa\inst{9,10}
	\and
	V.~Charmandaris\inst{11,12,13}
	\and
	R.-R.~Chary\inst{14}
	\and
	A.~Cimatti\inst{3}
	\and
	E.~Daddi\inst{4}
	\and
	N.~M.~F{\"o}rster Schreiber\inst{1}
	\and
	R.~Genzel\inst{1}
	\and
	C.~Gruppioni\inst{15}
	\and
	M.~Harwit\inst{16,17}
	\and
	H.~S.~Hwang\inst{18}
	\and
	R.~J.~Ivison\inst{19}
	\and
	G.~Magdis\inst{20}
	\and
	R.~Maiolino\inst{21}
	\and
	E.~Murphy\inst{22}
	\and
	R.~Nordon\inst{23}
	\and
	M.~Pannella\inst{4}
	\and
	A.~P{\'e}rez Garc{\'\i}a\inst{9,10}
	\and
	A.~Poglitsch\inst{1}
	\and
	D.~Rosario\inst{1}
	\and
	M.~Sanchez-Portal\inst{6}
	\and
	P.~Santini\inst{24}
	\and
	D.~Scott\inst{25}
	\and
	E.~Sturm\inst{1}
	\and
	L.~J.~Tacconi\inst{1}
	\and
	I.~Valtchanov\inst{6}
               }
\institute{
Max-Planck-Institut f\"{u}r Extraterrestrische Physik (MPE), Postfach 1312, 85741 Garching, Germany
\and
Argelander-Institut f\"{u}r Astronomie, Universit\"{a}t Bonn, Auf dem H\"ugel 71, D-53121 Bonn, Germany\\  \email{magnelli@astro.uni-bonn.de}
\and
Dipartimento di Astronomia, Universit\`a di Bologna, via Ranzani 1, 40127 Bologna, Italy
\and
Laboratoire AIM, CEA/DSM-CNRS-Universit{\'e} Paris Diderot, IRFU/Service
d'Astrophysique,
B\^at.709, CEA-Saclay, 91191 Gif-sur-Yvette Cedex, France.
\and
National Optical Astronomy Observatory, Tucson, AZ 85719, USA
\and
Herschel Science Centre, ESAC, Villanueva de la Ca\~nada, 28691 Madrid, Spain 
\and
ESO, Karl-Schwarzschild-Str. 2, 85748 Garching, Germany 
\and
INAF - Osservatorio Astronomico di Trieste, via Tiepolo 11, 34143 Trieste, Italy 
\and
Instituto de Astrof\'isica de Canarias (IAC), C/v{\'\i}a L{\'a}ctea S/N, 38200 La Laguna, Spain 
\and
Departamento de Astrof\'isica, Universidad de La Laguna, Spain 
\and
Department of Physics and Institute of Theoretical and Computational Physics, University of Crete, GR-71002, Heraklion, Greece
\and
Chercheur Associ\'e, Observatoire de Paris, F-75014 Paris, France
\and
IESL/Foundation for Research and Technology-Hellas, P.O. Box 1527, 71110 Heraklion, Greece
\and
Infrared Processing and Analysis Center, California Institute of Technology, Pasadena, CA 91125, USA
\and
INAF - Osservatorio Astronomico di Bologna, via Ranzani 1, I-40127 Bologna, Italy
\and
Center for Radiophysics and Space Research, Cornell University, Ithaca, NY, 14853, USA
\and
511 H street, SW, Washington, DC 20024-2725, USA
\and
Smithsonian Astrophysical Observatory, 60 Garden Street, Cambridge, MA 02138, USA
\and
UK Astronomy Technology Centre, Royal Observatory, Blackford Hill, Edinburgh EH9 3HJ, UK
\and
Department of Physics, University of Oxford, Keble Road, Oxford OX1 3RH, UK
\and
Cavendish Laboratory, University of Cambridge, 19 J.J. Thomson Ave., Cambridge CB3 0HE, UK
\and
Observatories of the Carnegie Institution for Science, 813 Santa Barbara Street, Pasadena, CA 91101, USA
\and
School of Physics and Astronomy, The Raymond and Beverly Sackler Faculty of Exact Sciences, Tel Aviv University, Tel Aviv 69978, Israel
\and
INAF - Osservatorio Astronomico di Roma, via di Frascati 33, 00040 Monte Porzio Catone, Italy
\and
Department of Physics \& Astronomy, University of British Columbia, 6224 Agricultural Road, Vancouver, BC V6T 1Z1, Canada
}

\date{Received ??; accepted ??}
\abstract{
We present results from the deepest \textit{Herschel}-PACS (Photodetector Array Camera and Spectrometer) far-infrared blank field extragalactic survey, obtained by combining observations of the GOODS (Great Observatories Origins Deep Survey) fields from the PACS Evolutionary Probe (PEP) and GOODS-\textit{Herschel} key programmes. 
We describe data reduction and the construction of images and catalogues.
In the deepest parts of the GOODS-S field, the catalogues reach  3$\sigma$ depths of 0.9, 0.6 and 1.3 mJy at 70, 100 and 160$\,\mu$m, respectively,  and resolve $\thicksim\,$75\% of the cosmic infrared background at 100$\,\mu$m and 160$\,\mu$m into individually detected sources. 
We use these data to estimate the PACS confusion noise, to derive the PACS number counts down to unprecedented depths, and to determine the infrared luminosity function of galaxies down to $L_{\rm IR}=10^{11}\,$L$_{\odot}$ at $z$$\,\thicksim\,$$1$ and $L_{\rm IR}=10^{12}\,$L$_{\odot}$ at $z$$\,\thicksim\,$$2$, respectively.
For the infrared luminosity function of galaxies, our deep \textit{Herschel} far-infrared observations are fundamental because they provide more accurate infrared luminosity estimates than those previously obtained from mid-infrared observations.
Maps and source catalogues ($>\,$$3\sigma$) are now publicly released.
Combined with the large wealth of multi-wavelength data available for the GOODS fields, these data provide a powerful new tool for studying galaxy evolution over a broad range of redshifts.
}
\keywords{Galaxies: evolution $-$ Infrared: galaxies $-$ Galaxies: starburst $-$ Galaxies: statistics}
\authorrunning{Magnelli et al. }
\titlerunning{PEP/GOODS-H far infrared survey}
\maketitle
\section{introduction\label{sec:intro}}
The detection of a cosmic infrared background, as energetic as the optical/near-infrared background \citep{puget_1996,hauser_1998}, has revealed the importance of the energy absorbed by the dust in galaxies and re-emitted at mid- to far-infrared wavelengths.
From this finding it became clear that a complete census on the formation and evolution of galaxies could not be obtained without accounting for this dust emission.
Since then, many studies have confirmed the importance of dust emission using individual detections of galaxies from infrared facilities such as the \textit{Infrared Space Observatory}, ISO, or the \textit{Spitzer Space Telescope}.
However, because of their relatively small onboard optics, the far-infrared capabilities of these observatories were strongly limited by source confusion.
At the time, far-infrared studies were restricted to the analysis of local galaxies or to the analysis of rare high-redshift very luminous galaxies.
As a consequence, only a small fraction of the cosmic far-infrared (i.e., $\lambda>40\,\mu$m) background was resolved into individual objects, so our knowledge of the high-redshift Universe at far-infrared wavelengths was very incomplete.

Thanks to the advent of the PACS \citep{Poglitsch_2010} and SPIRE \citep{griffin_2010} instruments onboard the \textit{Herschel Space Observatory} \citep{pilbratt_2010}, this limitation has been largely overcome.
Indeed, using the relatively high spatial resolution (provided by a $3.5\,$m mirror) and sensitivity of \textit{Herschel}, deep extragalactic surveys can be pursued, thereby resolving a large fraction of the cosmic far-infrared \citep[i.e., $\thicksim\,$58\% and $\thicksim\,$74\% with PACS at 100 and 160$\,\mu$m, respectively;][]{berta_2010,berta_2011} and submillimetre \citep[i.e., $\thicksim\,$15\% with SPIRE at 250$\,\mu$m;][]{oliver_2010} background into individually detected galaxies.
From these observations, one can study the origin and nature of the cosmic infrared background through, e.g., the determination of the infrared luminosity functions of galaxies \citep[e.g.,][]{gruppioni_2010,gruppioni_2013,casey_2012} and the examination of their spectral energy distributions \citep[e.g.,][]{hwang_2010,elbaz_2010,elbaz_2011, nordon_2010,nordon_2012,magnelli_2010,magnelli_2012,symeonidis_2013,berta_2013}.
All these studies point towards the diversity and redshift evolution, both in term of numbers and properties, of the infrared luminous galaxy population.
Relatively rare in the local Universe, the infrared luminous galaxies dominate the cosmic star-formation history at $z>1$ \citep[e.g.,][]{gruppioni_2010,gruppioni_2013}, and their physical properties differ significantly from those of their local counterparts \citep[e.g.,][]{elbaz_2011,wuyts_2011b,nordon_2012}.

In this context, we study here the infrared luminous galaxy population further by deriving the PACS numbers counts and infrared luminosity functions down to unprecedented depths using the combination of the two main extragalactic surveys designed to take advantage of the full PACS capabilities: the \textit{PACS Evolutionary Probe} \citep[PEP\footnote{\texttt{http://www.mpe.mpg.de/ir/Research/PEP}};][]{lutz_2011} guaranteed time key programme; and the \textit{GOODS-Herschel} \citep[GOODS-H\footnote{\texttt{http://hedam.oamp.fr/GOODS-Herschel}};][]{elbaz_2011} open time key programme.
The PEP survey is structured as a ``wedding cake'' (i.e., with large area shallow images and smaller deep images) and includes many widely studied blank and lensed extragalactic fields, such as the Great Observatories Origins Deep Surveys North (GOODS-N) and South (GOODS-S) fields and the cosmological evolution survey (COSMOS).
The GOODS-H survey only focuses on the GOODS fields, but using very deep observations, close to the \textit{Herschel} confusion limit.
The extensive observations of the GOODS fields made by the PEP and GOODS-H surveys is explained by the availability of a deep multi-wavelength database including X-ray \citep{alexander_2003,xue_2011}, optical \citep{Giavalisco_2004}, near-infrared \citep[CANDELS;][]{grogin_2011}, mid-infrared \citep[GOODS-\textit{Spitzer}; PI: M. Dickinson; see][]{magnelli_2011a}, (sub)mm \citep[e.g.,][]{oliver_2012,borys_2003,weiss_2009b} and radio \citep{morrison_2010,miller_2008} observations.

We present here the combination of the PEP and GOODS-H observations of the GOODS fields.
This combination provides the deepest observations of the GOODS fields obtained by PACS.
In particular, in the GOODS-South field the combination of these observations is not limited by the exposure time but by confusion.
In this field, we thus obtain the deepest blank field observations achievable with PACS onboard the Herschel space observatory.
In this paper we present in detail the data analysis method used to produce the publicly available PEP/GOODS-H maps and catalogues\footnote{\texttt{http://www.mpe.mpg.de/ir/Research/PEP/public\_data\_releases.php}}.
Then we use these deep observations to constrain the PACS-100$\,\mu$m and -160$\,\mu$m confusion noises and number counts, and to study the evolution of the infrared luminosity function and of the star-formation rate history of the Universe up to $z$$\,\thicksim\,$$2$.

The paper is structured as follows.
Observations are described in Section \ref{sec:observations}.
Section \ref{sec: PACS images} presents the method used to produce the PACS maps.
In Section \ref{sec:source extraction}, we present our source extraction methods and contents of the released package.
In Section \ref{sec:confusion}, we estimate the PACS-100 and -160$\,\mu$m confusion noise.
Number counts are presented in Section \ref{sec:counts} while in Section \ref{sec:lf} we derive the infrared luminosity functions of galaxies as well as the star-formation rate history of the Universe up to $z$$\,\thicksim\,$$2$.
Finally, we summarise our results in Section \ref{sec:conclusion}.
Throughout the paper we use a cosmology with $H_{0}$$\,=\,$$71\ \rm{km\ s^{-1}\ Mpc^{-1}},\Omega_{\Lambda}$$\,=\,$$0.73$ and $\Omega_{\rm M}$$\,=\,$$0.27$.
\section{Observations\label{sec:observations}}
The PACS maps and catalogues used and released in this paper are obtained from the combination of the PEP and GOODS-H observations of the GOODS-N and GOODS-S fields (see Table \ref{tab:observations}).
Both programmes have observed the GOODS fields using the scan mode of the PACS photometer on board \textit{Herschel}.
This mode consists of slewing the spacecraft back and forth along parallel lines at a constant speed of 20\arcsec$\,$sec$^{-1}$.
Using this scan mode, astronomical observing requests (AORs) were designed to observe the GOODS fields in both nominal and orthogonal directions.
To reach the desired depth many AORs per field were required.
The central position of each AOR was dithered by $\thicksim8$\arcsec\ in order to improve the spatial redundancy of the data.
\begin{table*}
\center
\footnotesize
\caption{ \label{tab:observations}
Properties of the \textit{Herschel} observations combined for the PEP/GOODS-H data release.}
\begin{tabular}{ccccccc}
\hline\hline
Field & Survey & RA & Dec & Wavelengths & Size & Time \\
\rule{0pt}{2ex} &  & \multicolumn{2}{c}{{\small [Degree, J2000]}} & {\small [$\mu$m]} & {\small[arcmin]} & {\small [h]} \\
 \hline
GOODS-N & PEP & $189.22862$ & $62.23867$ & $100, 160$ & $11\arcmin$$\,\times\,$$17\arcmin$ & $25.8$ \\
GOODS-N & GOODS-H & $189.22862$ & $62.23867$ & $100, 160$ & $11\arcmin$$\,\times\,$$17\arcmin$ & $124.0$ \\
\rule{0pt}{3ex}GOODS-S & PEP & $53.12654$ & $-27.80467$ & $70, 100, 160$ & $11\arcmin$$\,\times\,$$17\arcmin$ & $226$ \\
ECDFS$^{\rm a}$ (GOODS-S) & PEP & $53.10417$ & $-27.81389$ & $100, 160$ & $30\arcmin$$\,\times\,$$30\arcmin$ & $32.8$ \\
GOODS-S & GOODS-H & $53.12654$ & $-27.80467$ & $100, 160$ & $10\arcmin$$\,\times\,$$10\arcmin$ & $206.3$ \\
GOODS-S$^{\rm b}$ & PV & $53.12654$ & $-27.80467$ & $100, 160$ & $11\arcmin$$\,\times\,$$17\arcmin$ & $7.9$ \\
\hline
\end{tabular}
\begin{list}{}{}
\item[\textbf{Notes.} ]
\item[$^{\mathrm{a}}$] Prior to be combined, observations of the ECDFS were trimmed to match the GOODS-S layout of the PEP observations.
\item[$^{\mathrm{b}}$] These observations were taken during the \textit{Herschel} performance verification (PV) phase with preliminary instrument settings and thus were appropriately underweighted for the map creation (Section \ref{sec: PACS images}).
\end{list}
\end{table*}

Using this strategy, the PEP and GOODS-H surveys covered the entire $11\arcmin$$\,\times\,$$17\arcmin$ GOODS-N field with PACS at 100 and 160$\,\mu$m\footnote{GOODS-H has also covered the entire GOODS-N field with SPIRE at 250, 350 and 500$\,\mu$m. SPIRE observations and catalogues are described in \citet{elbaz_2011} and are publicly available at \texttt{http://hedam.oamp.fr/GOODS-Herschel}}.
The total observing time (i.e., including overheads) in GOODS-N was $25.8$ and $124$ hours for PEP and GOODS-H, respectively (Table \ref{tab:observations}). 

The entire $11\arcmin$$\,\times\,$$17\arcmin$ GOODS-S field was observed by PEP with PACS at 70, 100 and 160$\,\mu$m\footnote{SPIRE-250/350/500$\,\mu$m observations of the GOODS-S field have been performed by the HerMES survey \citep{oliver_2012}. These observations are publicly available through the \textit{Herschel} Database in Marseille (HeDaM) at \texttt{http://hedam.oamp.fr/HerMES}}.
The total observing times were $113$, $113$ and $226$ hours at 70, 100 and 160$\,\mu$m, respectively, including 10 hours guaranteed time contributed by \textit{Herschel} mission scientist Martin Harwit.
The Extended Chandra Deep Field South (ECDFS), containing in its centre the GOODS-S field, has also been observed by PEP with a total observing time of 32.8 hours over a $30\arcmin$$\,\times\,$$30\arcmin$ region.
Observations of ECDFS are included in our combined maps, trimmed to match the GOODS-S layout of the PEP observations.
GOODS-H has observed a $10\arcmin$$\,\times\,$$10\arcmin$ region centred on the GOODS-S field with PACS at 100 and 160$\,\mu$m.
The choice of covering only a fraction of the GOODS-S field was made in order to obtain, in a reasonable amount of time, a PACS-100$\,\mu$m map close to the confusion limit of \textit{Herschel}.
The GOODS-H observations of the GOODS-S field lasted for a total of $206.3$ hours.
Finally in our combined maps we also include 7.9 hours of observations of the GOODS-S field taken with preliminary instrument settings during the \textit{Herschel} performance verification phase (Table \ref{tab:observations}).
\begin{figure}
	\includegraphics[width=8.5cm]{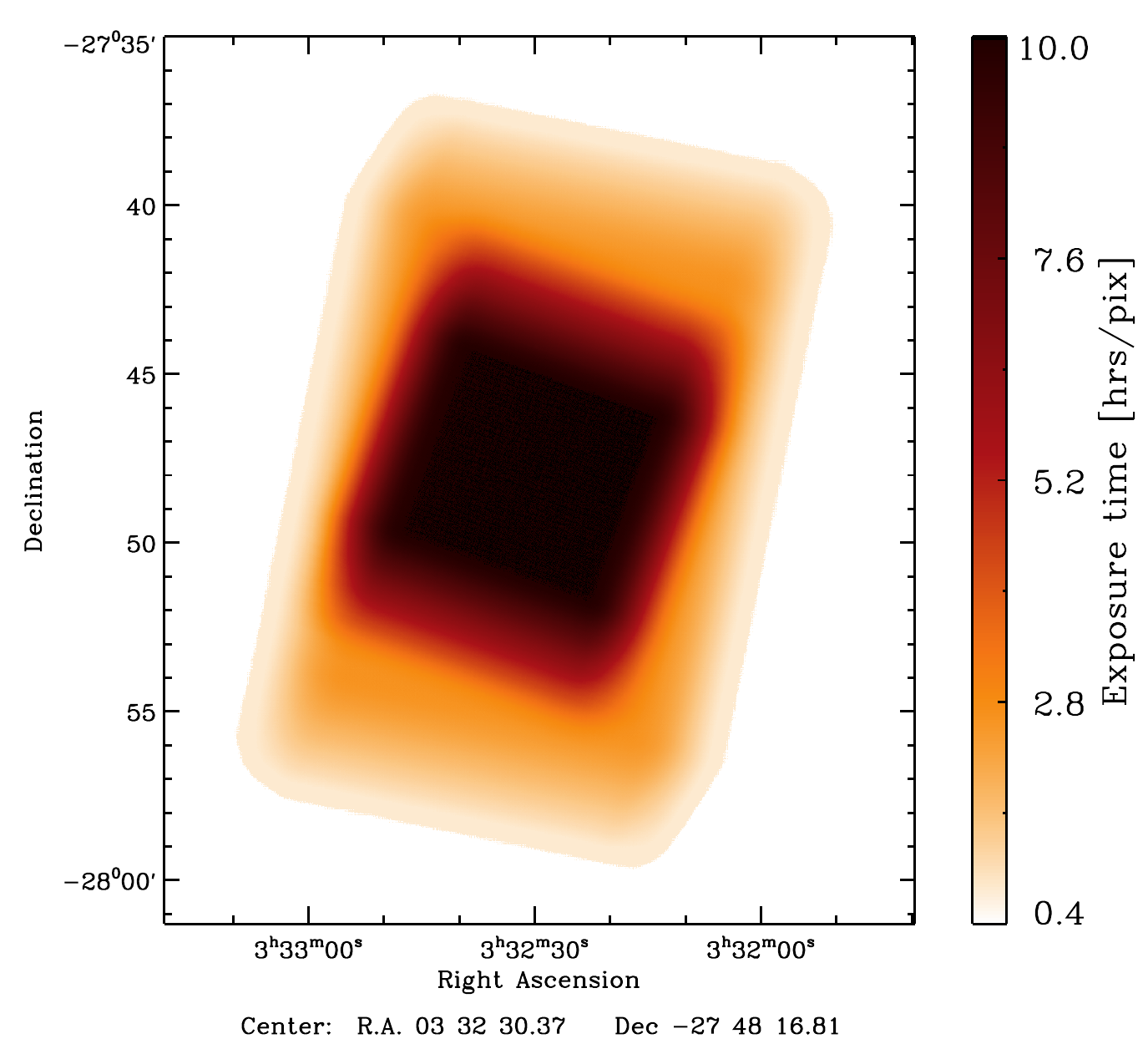}
	\caption{\label{fig:coverage}
	PACS-$100\,\mu$m coverage map of the GOODS-S field.
	Due to the different layout of the PEP and GOODS-H observations, the $10\arcmin$$\,\times\,$$10\arcmin$ region centred on the GOODS-S field has much deeper PACS-100$\,\mu$m  observations than the outskirts.
	We observe the same pattern in the PACS-$160\,\mu$m coverage map of the GOODS-S field.
		}
\end{figure}

Because PEP and GOODS-H observations were executed using the same observing mode, a combination of these data sets can easily be performed.
Of course, due to the different layout of PEP and GOODS-H observations in GOODS-S, this field is not homogeneously covered at 100 and 160$\,\mu$m.
As illustrated by Fig.~\ref{fig:coverage}, a $10\arcmin$$\,\times\,$$10\arcmin$ region centred on the GOODS-S field has higher PACS-100 and 160$\,\mu$m coverage than the outskirts.
In contrast, the GOODS-S 70$\,\mu$m coverage is uniform across the field, with the exception of fall-off at the edges. 
In the rest of the paper, the centred region of the GOODS-S field with ultradeep observations is referred as ``GOODS-S-ultradeep'', while the outskirts are referred to as ``GOODS-S-deep''.
This combined data set provides the deepest \textit{Herschel} far-infrared observations of the GOODS-N and -S fields.
In the rest of the paper this combined data set is referred to as the ``PEP/GOODS-H'' observations.

Our combined PACS-100$\,\mu$m maps reach a total observing time per sky position of $\thicksim\,$2.6 hours, $\thicksim\,$2.6 hours and $\thicksim\,$10.0 hours in GOODS-N, GOODS-S-deep and GOODS-S-ultradeep, respectively. 
The PACS-160$\,\mu$m maps reach a total observing time per sky position of $\thicksim\,$2.6 hours, $\thicksim\,$4.7 hours and $\thicksim\,$12.1 hours in GOODS-N, GOODS-S-deep and GOODS-S-ultradeep, respectively.
In the GOODS-S field, the PACS-70$\,\mu$m map reaches a total observing time per sky position of $\thicksim\,$2.1 hours.

We note that a detailed description of the observational strategies adopted by PEP and GOODS-H is provided in \citet{lutz_2011} and \citet{elbaz_2011}, respectively.

\begin{figure*}
	\includegraphics[height=9.cm]{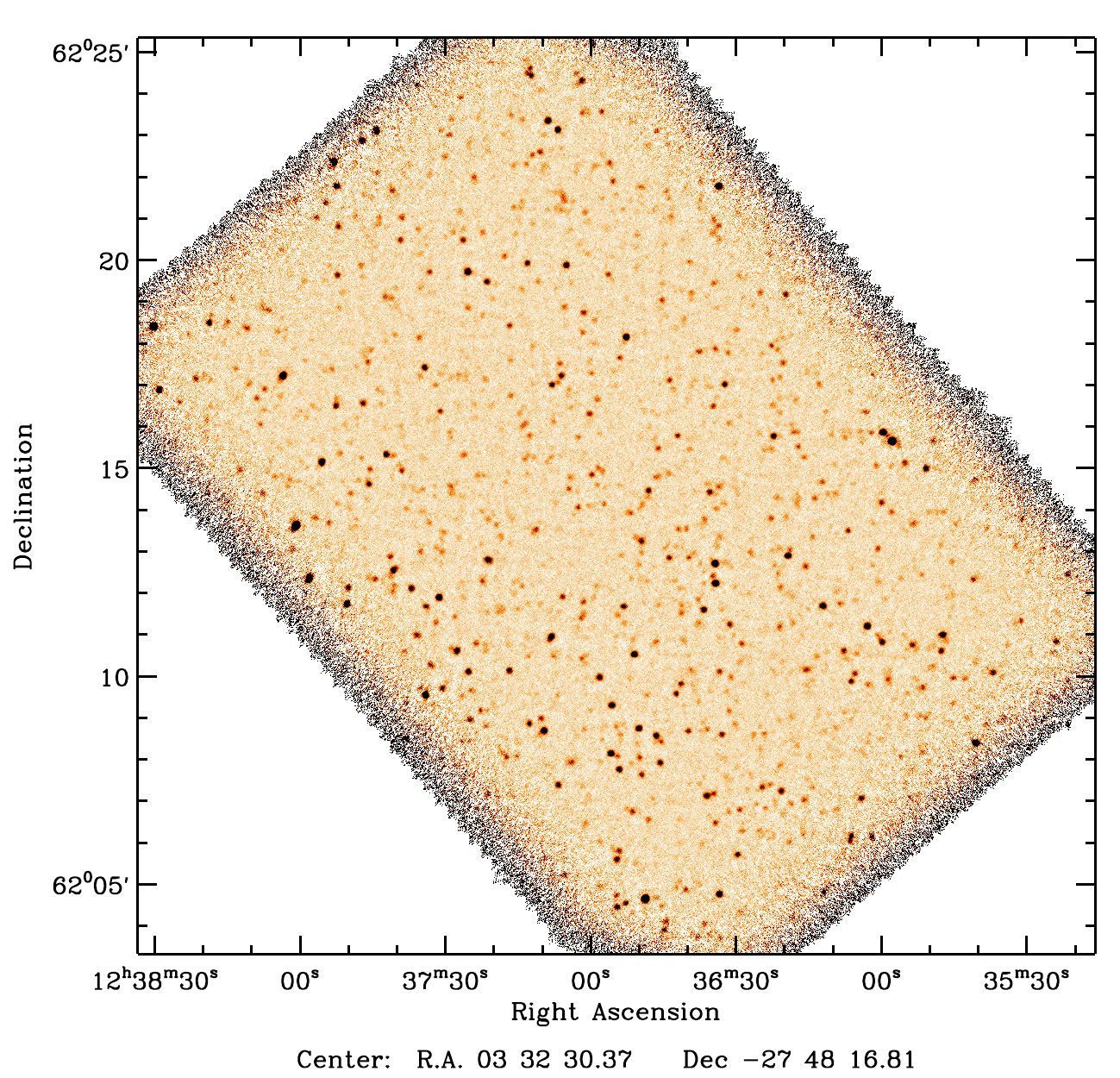}
		\includegraphics[height=9.cm]{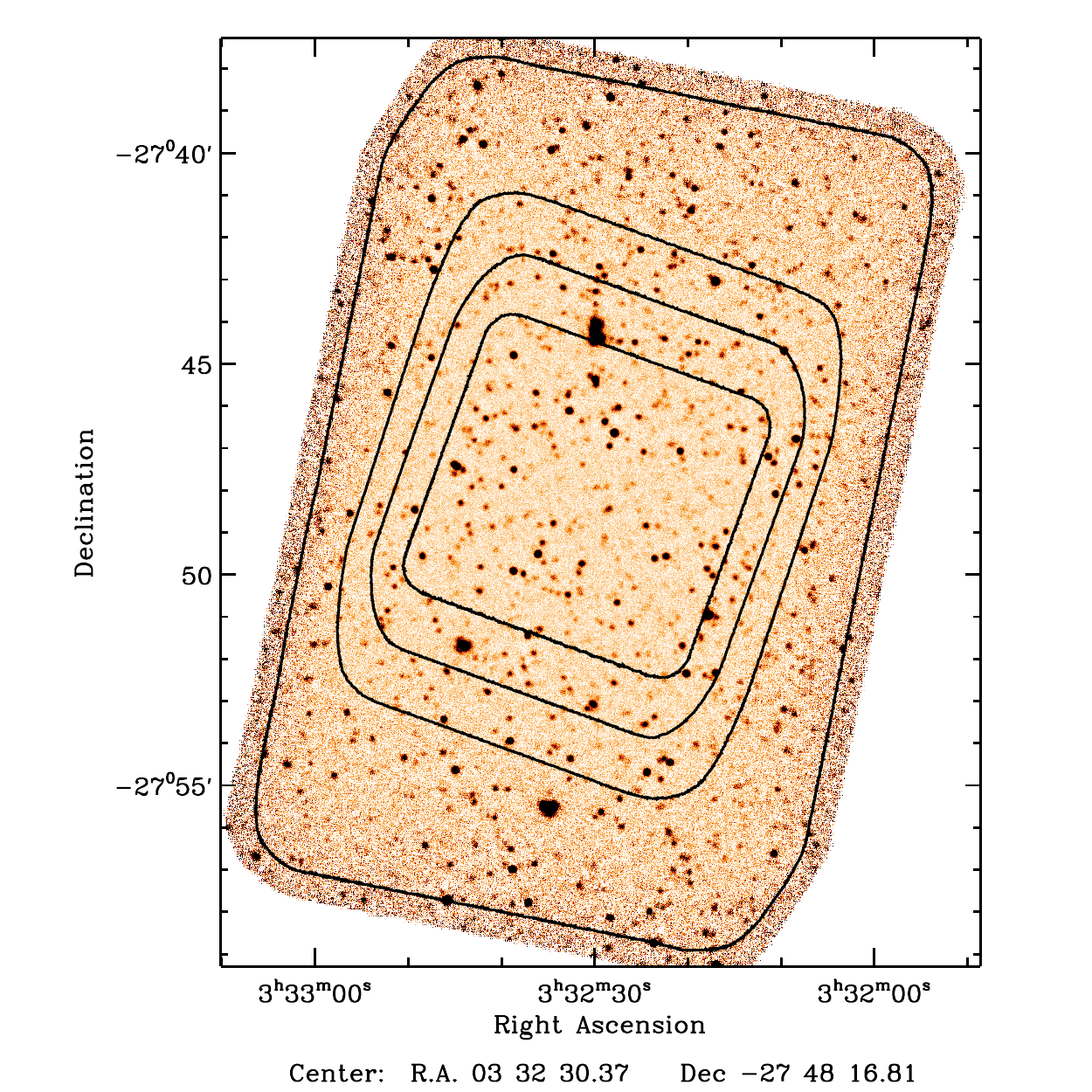}
	\includegraphics[height=9.cm]{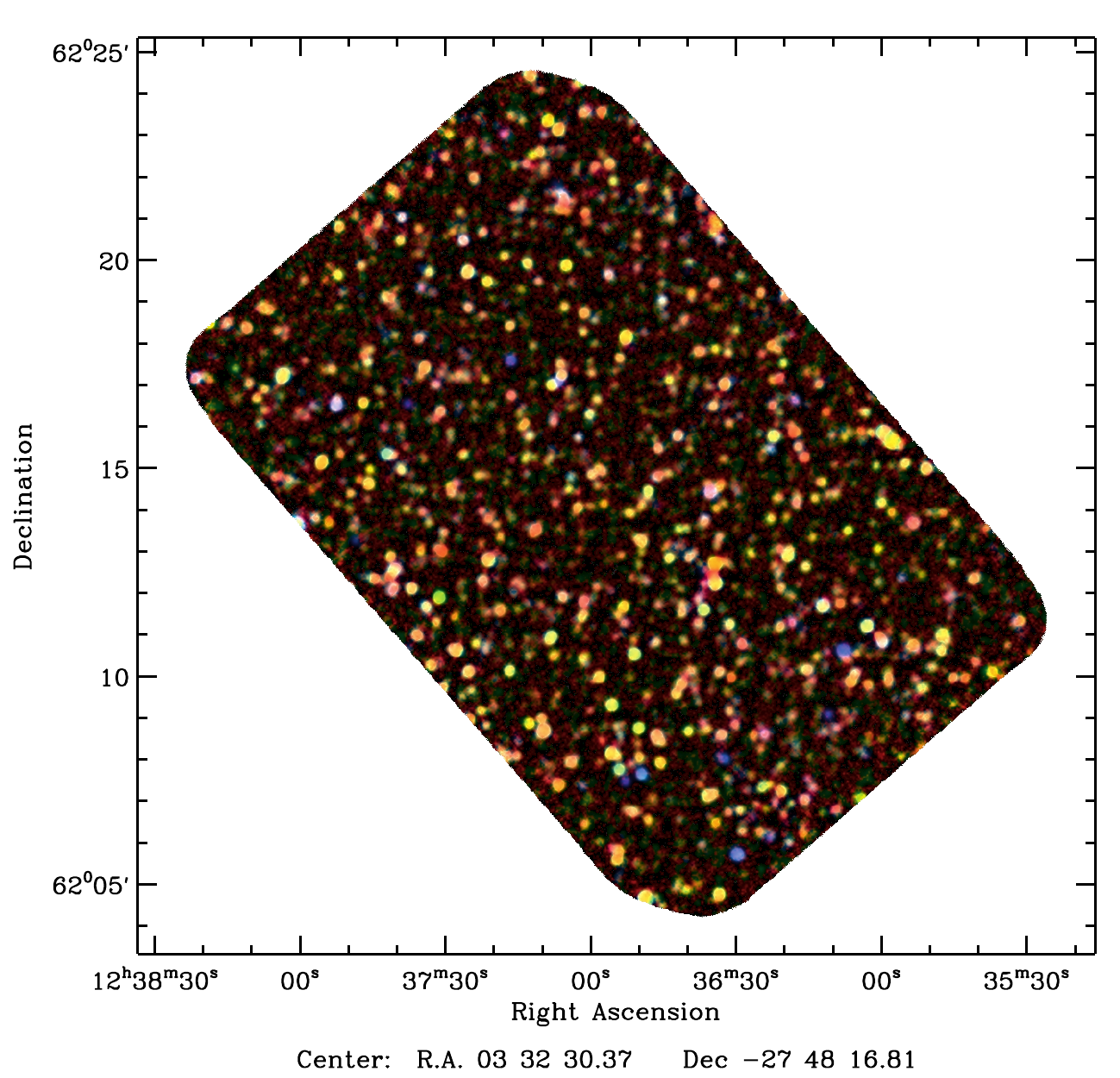}
		\includegraphics[height=9.cm]{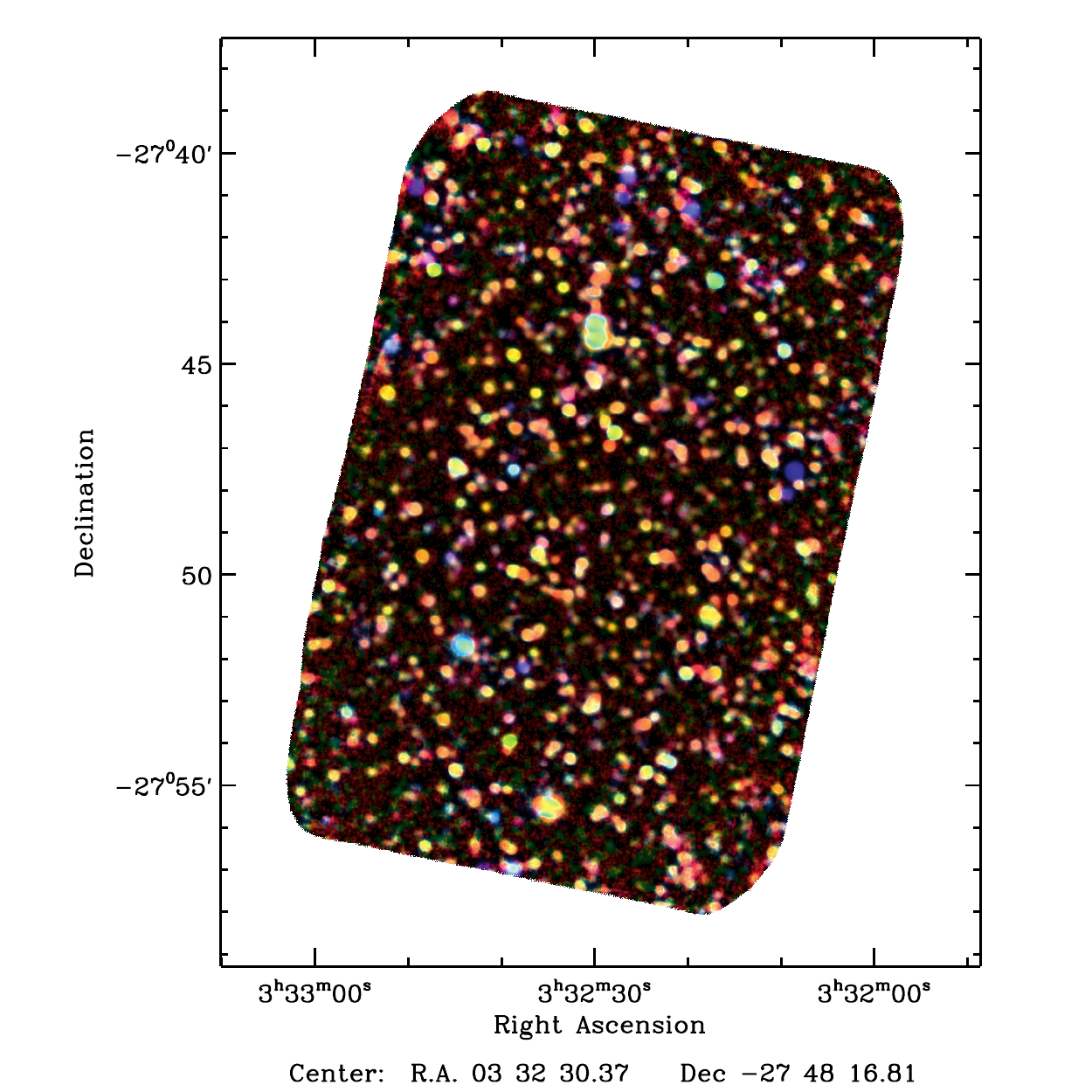}
	\caption{\label{fig:3 colors}
	\textit{(Top left)} PEP/GOODS-H map of the GOODS-north field ($11\arcmin$$\,\times\,$$17\arcmin$) at 100$\,\mu$m. 
	\textit{(Top right)} PEP/GOODS-H map of the GOODS-south field ($11\arcmin$$\,\times\,$$17\arcmin$) at 100$\,\mu$m.
	Contours correspond to exposure times greater than 0.5 (at the edge), 3, 6 and 9 (centre) hrs/pix.
	\textit{(Bottom left)} Colour composite image of the GOODS-north field at 24$\,\mu$m (blue), 100$\,\mu$m (green) and 160$\,\mu$m (red).
	\textit{(Bottom right)} Colour composite image of the GOODS-south field at 24$\,\mu$m (blue), 100$\,\mu$m (green) and 160$\,\mu$m (red).
	The 24$\,\mu$m images (PI: M. Dickinson) were obtained by the \textit{Spitzer Space Telescope}, while the 100 and 160$\,\mu$m images were obtained by the \textit{Herschel Space Observatory}.
	In the colour composite images, sources with contribution from the 24+100$\,\mu$m, 24+160$\,\mu$m, 100+160$\,\mu$m and 24+100+160$\,\mu$m bands would correspond to a cyan, magenta, yellow and white colours, respectively. 
	The relatively noisy edges of the colour composite images have been trimmed.
	}
\end{figure*}
\section{Map creation\label{sec: PACS images}}
Observations were reduced using the standard PACS photometer pipeline \citep{wieprecht_2009} and some custom procedures, all implemented within the \texttt{HIPE} environment in the \textit{Herschel} common science system (HCSS).
This data reduction process is described in more detail in \citet{lutz_2011} and \citet{popesso_2012}.
Here we summarise the main steps.

The data reduction process starts at the AOR level and is based on the scanmap script of the PACS photometer pipeline.
First, the pipeline flags bad or saturated pixels, converts detector signals from digital units to volts, finds pixels affected by short glitches and replaces their values using a standard interpolation method, and finally it applies a recentring correction to the pointing product of \textit{Herschel} using reference positions of 24$\,\mu$m sources with accurate astrometry \citep[see][]{lutz_2011} from deep observations with the Multiband Imaging Photometer \citep[MIPS;][]{rieke_2004} onboard the \textit{Spitzer Space Telescope}.
Thanks to these recentring corrections, positions of PACS sources measured in a ``blind'' catalogue (see Section \ref{subsec:blind}) are offset by less than $0.2\arcsec$ and have a RMS difference of $\thicksim\,$$1\arcsec$ with respect to the position of their MIPS-24$\,\mu$m counterparts \citep[see also][]{lutz_2011}.
We note that the MIPS-24$\,\mu$m astrometry, and therefore our PACS maps, match the GOODS ACS version 2 coordinate system\footnote{see the GOODS ACS data release at MAST:\\ \texttt{http://archive.stsci.edu/prepds/goods/}}.
Second, the pipeline removes from the timeline of each AOR the ``1/\textit{f}'' noise which is the main source of instrumental noise in PACS data.
For deep cosmological surveys such as ours, the PACS pipeline does so by using a high-pass filtering method which subtracts, from each timeline, a version of the timeline filtered by a running box median of a given radius (expressed in readouts, i.e., in numbers of points of the timeline).
The presence of sources in the timeline affects the high-pass filtering method by artificially boosting this running box median, i.e., leading to the subtraction of part of the source flux \citep{popesso_2012}.
Consequently, sources have to be ``masked'' from the timelines.
Using results from \citet{popesso_2012}, we choose a masking strategy based on circular patches at prior positions.
This method reduces the amount of flux loss due to the high-pass filter and more importantly leads to flux losses which are independent of the PACS flux densities \citep{popesso_2012}.
We note that this improved masking strategy differs from that adopted in \citet{lutz_2011}.
Timelines are masked at the position of 24 $\mu$m sources with $S_{24}$$\,>\,$$60\,\mu$Jy using circular patches with radius of 4$\arcsec$, 4$\arcsec$ and 6$\arcsec$ at 70, 100 and 160 $\mu$m, respectively.
This strategy allows us to mask almost all PACS detections and corresponds to the masking of about $\thicksim\,$12\% of the timeline.
A small number of resolved sources (mainly in the GOODS-S field) are masked using larger patches which are adjusted visually.
After being masked, timelines are high-pass filtered using a running box median with radius of 12 readouts (24\arcsec), 12 readouts (24\arcsec) and 20 readouts (40\arcsec) at 70, 100 and 160 $\mu$m, respectively.
These radii are chosen based on results from \citet{popesso_2012} .
Although we use an optimized high-pass filtering strategy, PACS flux densities still have to be corrected for flux losses.
The corrections are provided in \citet{popesso_2012} for our high-pass filter setting, i.e., 13\%, 12\% and 11\% at 70, 100 and 160$\,\mu$m, respectively (see Section \ref{sec:source extraction}).

After the removal of the ``1/$f$'' noise, AORs are flat-fielded and flux calibrated.
Then, the map of each AOR is created using a ``drizzle'' method \citep{fruchter_2002} implemented in the HCSS.
Because all AOR maps are projected on the same world coordinate system, they are coadded into a final map using, as weight, the effective exposure time of each pixel, appropriately underweighting the performance verification phase data which were taken with preliminary instrument settings.
Taking advantage of the large number of AORs, we create uncertainty maps from the standard deviation of this weighted mean in each pixel.
Although, a ``drizzle'' method is applied for the creation of each AOR map, the final map still contains some correlated noise.
Thanks to the high redundancy of our data at each sky position, we are able to calculate the mean correlated noise across the map and within our PSFs \citep[see][]{lutz_2011}.
These mean correlation corrections were taken into account while measuring uncertainties from the uncertainty maps (i.e., uncertainties have to be corrected upward by a factor $\thicksim$$1.5$; Section \ref{subsec:blind}).

Figure \ref{fig:3 colors} shows the combined PEP/GOODS-H PACS-$100\,\mu$m maps of the GOODS-N (top left panel) and GOODS-S (top right panel) fields.
Three-colour composite images of the GOODS-N (bottom left panel) and GOODS-S (bottom right panel) fields at 24-100-160$\,\mu$m are also shown in this figure.
\begin{figure*}
		\begin{center}
	\includegraphics[width=8.cm]{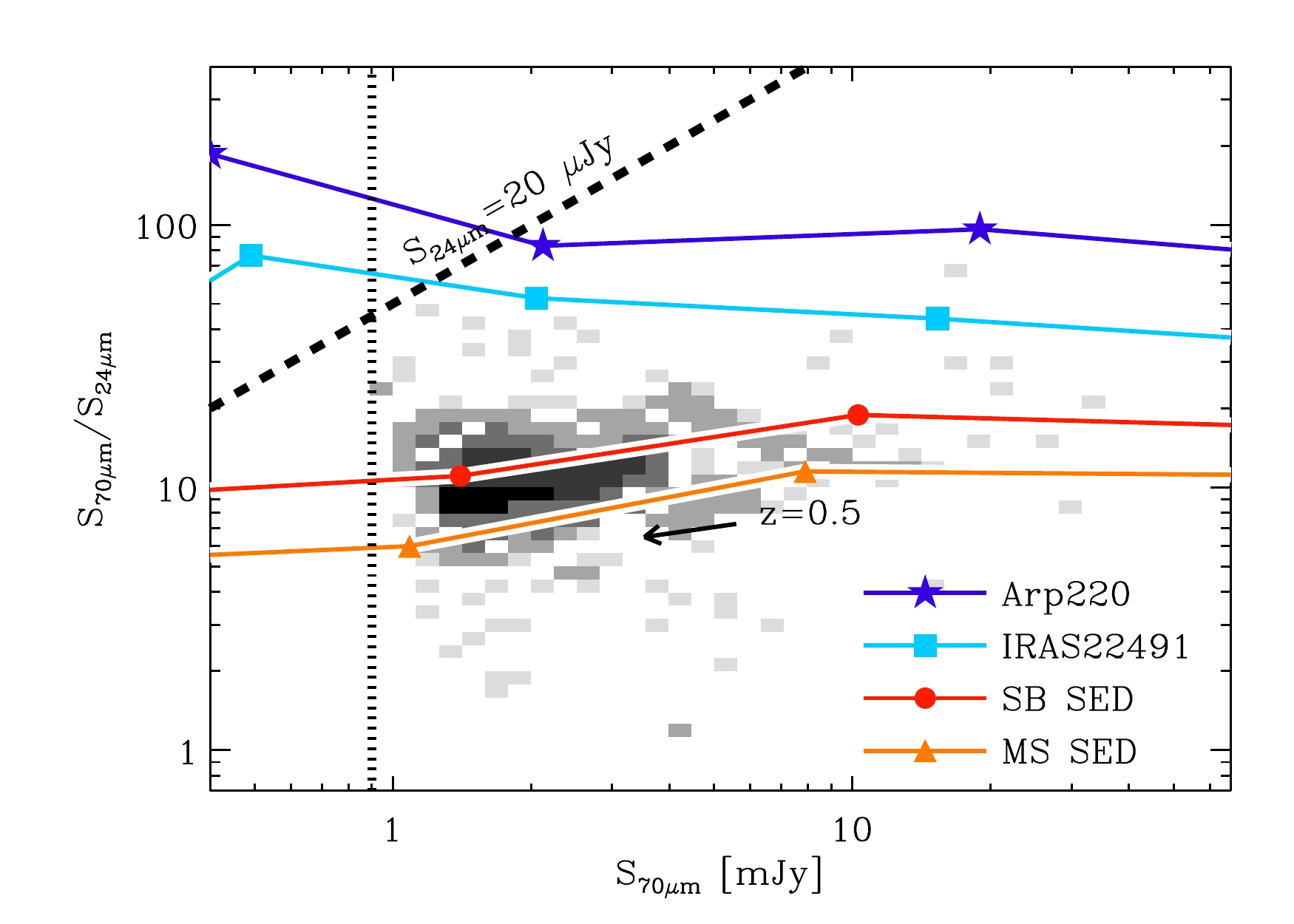}
		\includegraphics[width=8.cm]{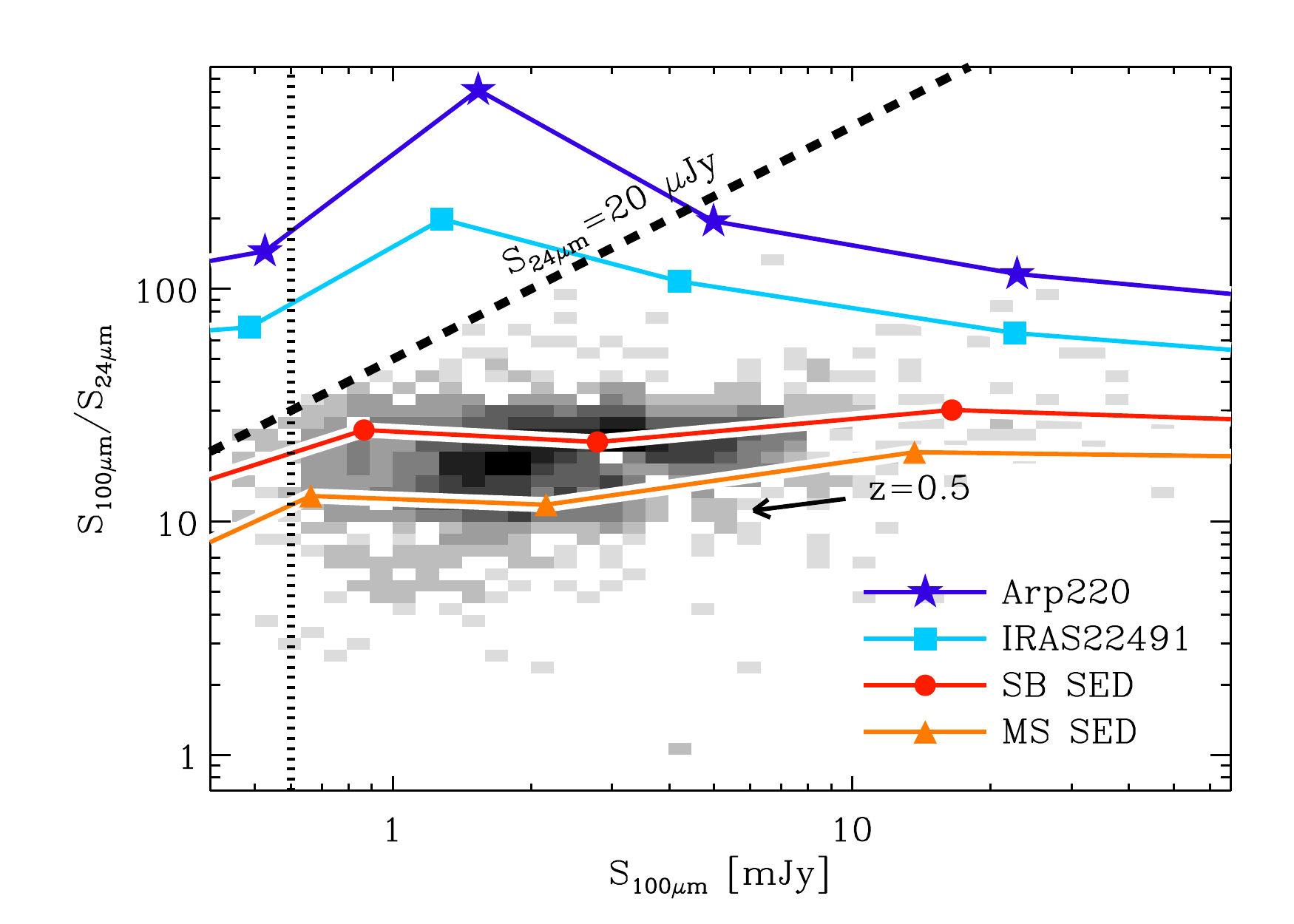}
	\includegraphics[width=8.cm]{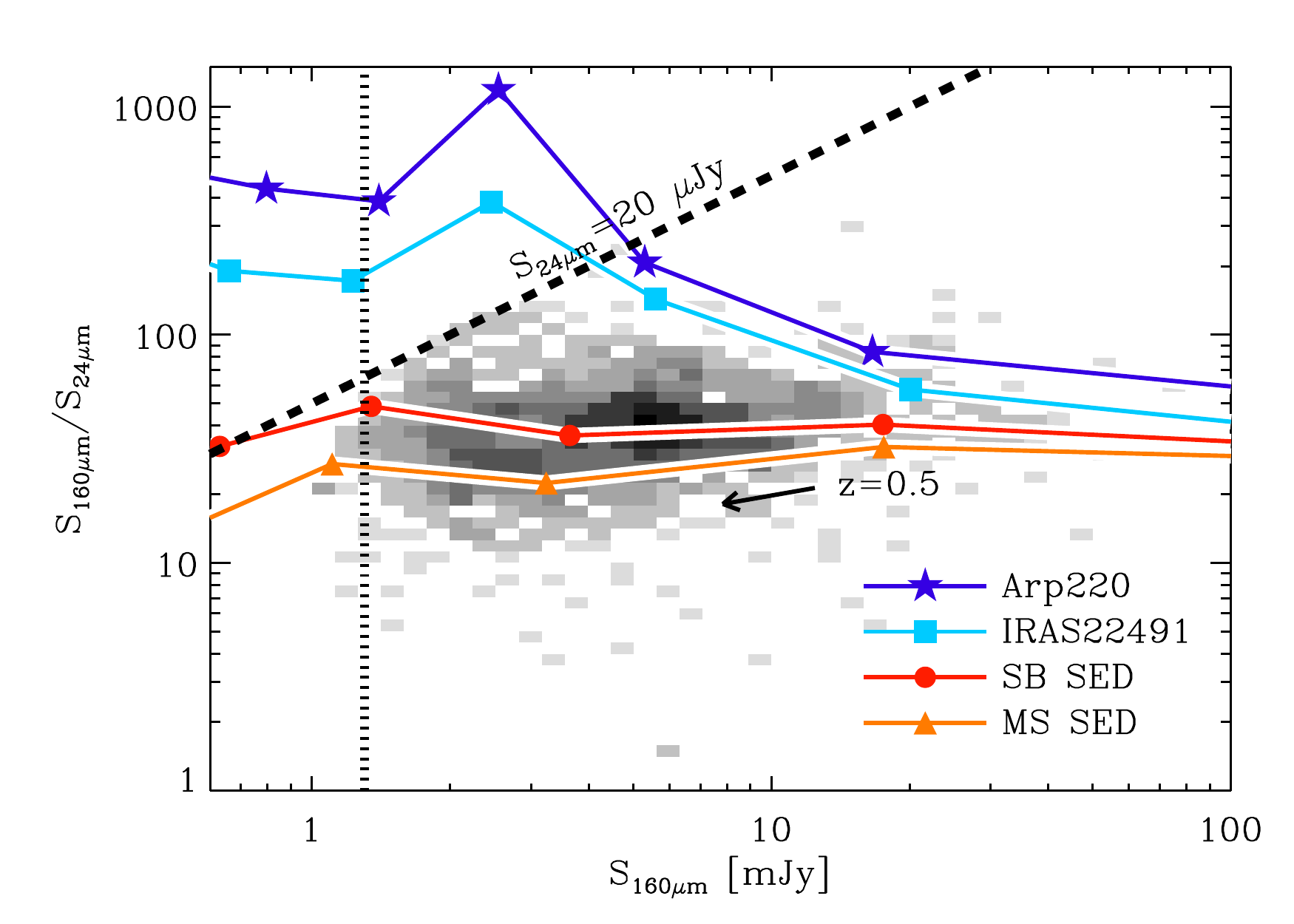}
		\end{center}
	\caption{\label{fig:ratio}
	PACS-to-24$\,\mu$m flux density ratio as a function of the PACS flux density inferred from the ``blind'' source catalogue cross-matched with the MIPS-24$\,\mu$m catalogue.
	Shaded regions show the space density distribution of galaxies at 70$\,\mu$m (\textit{top left}), 100$\,\mu$m (\textit{top right}) and 160$\,\mu$m (\textit{bottom}) in the GOODS-S field.
	In this field, the MIPS 24$\,\mu$m catalogue reaches a 3$\sigma$ limit of 20$\,\mu$Jy, while the PACS-70, 100 and 160$\,\mu$m data reach 3$\sigma$ limits of 0.9, 0.6 and 1.3$\,$mJy (vertical dotted black lines).
	The parameter spaces reachable by our prior source extraction method using the MIPS catalogue are located below the dashed black lines: PACS sources located above these lines will not be detected at 24$\,\mu$m.
	Orange triangles, red circles, light blue squares and dark blue stars show the PACS-to-24$\,\mu$m flux density ratios of galaxies with $L_{\rm IR}=10^{11.5}\,$L$_{\odot}$ as predicted at different redshifts using the spectral energy distributions (SEDs) of normal (MS), starburst (SB) galaxies \citep{elbaz_2011}, IRAS22491 \citep{berta_2005} and Arp220 \citep{silva_1998}, respectively.
	The right point of each track corresponds to $z=0.5$, while other points correspond to increasing redshifts, with intervals of $\Delta z=0.5$ (arrows indicate the path followed for increasing redshift).
	Predictions for galaxies with lower or higher infrared luminosities will correspond to a global vertical shift of these tracks towards lower or higher PACS flux densities, respectively.
	Most of the PACS population lies well within the limits of the parameter space reachable by our MIPS 24$\,\mu$m catalogues, i.e., below the dashed lines.
	}
\end{figure*}
\section{Source Extraction\label{sec:source extraction}}
At the resolution of PACS most of the sources in our fields are point sources (i.e FWHM$\,\thicksim\,$$4.7''$, 6.7$''$ and 11$''$ at 70, 100 and 160$\,\mu$m, respectively).
Therefore, we use PSF fitting to derive their flux densities. 
Two catalogues are derived using complementary approaches. 
First, we construct a catalogue using as priors the source positions expected on the basis of a deep 24$\,\mu$m catalogue. 
This provides good deblending of neighbouring sources, but will miss a few \textit{Herschel} sources that are not detected at 24$\,\mu$m (i.e., $<\,$1$\%$ and $<\,$4$\%$ of the PACS sources in the GOODS-N and GOODS-S fields, respectively).
Second, we provide a ``blind'' catalogue using PSF fitting without positional priors.
\subsection{Prior source extraction\label{subsec:prior}}
Starting from the positions of IRAC-3.6$\,\mu$m sources (Infrared Array Camera; FWHM$\,\thicksim\,$1.6\arcsec) from the GOODS \textit{Spitzer} Legacy Program (PI: M.~Dickinson), we extract sources in the MIPS-24$\,\mu$m maps \citep{magnelli_2011a}\footnote{The MIPS-24$\,\mu$m catalogues used here are slightly different from those released by \citet{magnelli_2011a}. See Section \ref{subsec:package} for more details.} and then use the 24$\,\mu$m-detected sources (i.e., with $S_{24}$$\,>\,$3$\sigma$$\,\thicksim\,$20$\,\mu$Jy) as priors for the source extraction in the PACS maps.
The main advantage of this approach is that it deals with a large part of the blending issues encountered in dense fields and provides a straightforward association between IRAC, MIPS and PACS sources \citep{magnelli_2011a}.
The disadvantage of this method is that we must assume that all sources present in our PACS images have already been detected at 24$\,\mu$m.
Based on the relative depth of the mid- and far-infrared images of the GOODS fields, \citet{magdis_2011} have investigated this assumption.
They found that in the GOODS-H observations of the GOODS-S field, less than 2$\%$ of the PACS sources are missed in the MIPS-24$\,\mu$m catalogue.
In our combined dataset, PACS observations of the GOODS-N field are shallower than those analysed in \citet{magdis_2011}, while MIPS-24$\,\mu$m observations are equivalently deep (i.e., 3$\sigma$$\,\thicksim\,$20$\,\mu$Jy in both GOODS-N and GOODS-S).
Consequently, in the GOODS-N field, the fraction of PACS sources with no MIPS-24$\,\mu$m counterparts should be significantly lower than $2\%$.
This result is confirmed by cross-matching our GOODS-N ``blind'' PACS and MIPS-24$\,\mu$m catalogues using a matching radius of 4\arcsec and 6\arcsec at 100 and 160$\,\mu$m, respectively: less than 1$\%$ of the PACS ``blind'' sources are missed in the MIPS-24$\,\mu$m catalogue.
In GOODS-S, the situation is somehow different as in this field our PACS observations are deeper than those analysed in \citet{magdis_2011}.
By cross-matching our GOODS-S ``blind'' PACS and MIPS-24$\,\mu$m catalogues, we find that 4.1$\%$, $4.2$$\%$ and $3.5$$\%$ of the PACS sources have no MIPS-24$\,\mu$m counterparts within 3\arcsec, 4\arcsec and 6\arcsec at 70, 100 and 160$\,\mu$m, respectively.
These fractions should be treated as upper limits, given that most of these sources have faint PACS flux densities (i.e., 80\% with S/N$\,<\,$5) and thus could be spurious (see Section \ref{subsec:noise}).
To understand the spectral properties of PACS sources with no MIPS-24$\,\mu$m counterparts, we study the typical PACS-to-MIPS flux ratio observed in the GOODS-S field (Fig.~\ref{fig:ratio}).
While the bulk of the PACS-100 (-160)$\,\mu$m population lies well within the parameter space reachable by our MIPS-24$\,\mu$m catalogue (i.e., below the dashed lines of Fig.~\ref{fig:ratio}), there is, at faint flux densities, a slight truncation of the high-end of the dispersion of the PACS-to-MIPS flux ratio.
This truncation will translate into faint PACS-100 and -160$\,\mu$m sources with no MIPS-24$\,\mu$m counterparts.
In addition, galaxies with similar spectral properties to Arp220 (i.e., with high PACS-to-MIPS flux ratio and strong silicate absorptions) will also be missed in MIPS-24$\,\mu$m catalogues, especially at $z$$\,\thicksim\,$$0.4$ and $z$$\,\thicksim\,$$1.3$ where the 18$\,\mu$m and $9.4\,\mu$m silicate absorption features are shifted into the MIPS-24$\,\mu$m passband \citep[see also][]{magdis_2011}.
The existence of PACS sources with no MIPS-24$\,\mu$m counterparts will naturally introduce slight incompleteness (i.e., $<\,$1$\%$ and $<\,$4$\%$ in the GOODS-N and GOODS-S fields, respectively) in our prior catalogues.
However, because this incompleteness mostly appears at faint flux densities (i.e., S/N$\,<\,$5), it should not be much larger than that introduced by prior-free source extraction methods at such low S/N.
Indeed, at faint flux densities, there are only small discrepancies, in terms of number of sources, between our blind and prior catalogues (see Appendix \ref{appendix:blind/prior}).
We note that we cannot use IRAC priors to reduce this incompleteness because the high IRAC source density (i.e., 2$\,-\,$4 priors per PACS beam) would force each far-infrared source to be deblended into several unrealistic counterparts.

From the expected positions of sources, we fit empirical PSFs (see Section \ref{subsec:PSF}) to our PACS maps.
PACS flux densities are then defined as the intensity of the scaled PSFs.
The simultaneous fit of nearby sources optimizes the deblending of their flux densities.
As for a standard aperture measurement, the photometry of each source is corrected to account for the finite size of our empirical PSFs, i.e., they do not include the power contained in the wings of the real PSFs (see Section \ref{subsec:PSF}).
Finally, because of flux losses from the high-pass filtering, additional corrections are applied to our flux density measurements (see Section \ref{sec: PACS images}).

Flux uncertainties are estimated on residual maps.
These uncertainties are defined as the pixel dispersion, around a given source, of the residual map convolved with the appropriate PSF.
This method has the advantage of taking into account, at the same time, the rms of the map (including correlated noise and inhomogeneities in the exposure time) and the quality of our fitting procedure.
Because our far-infrared observations have been designed to reach the confusion limit of \textit{Herschel},  our flux uncertainties also contain a part of the confusion noise left in the residual maps (i.e., confusion noise, due to sources fainter than PACS detection limits).
In this context of complex combination of instrumental and confusion noise, we test the accuracy of our flux uncertainties using extensive Monte Carlo (MC) simulations (see Section \ref{subsec:noise}).
We find that the flux uncertainties inferred from the residual maps and the photometric accuracies inferred from the MC simulations are in good agreement.
\subsection{Blind source extraction\label{subsec:blind}}
PACS flux densities are also extracted using a ``blind'' PSF-fitting analysis, i.e., without taking, as prior information, the expected positions of sources.
This blind source extraction is performed with \texttt{Starfinder} \citep{diolaiti_2000a,diolaiti_2000b}.
This code is appropriate for the extraction of the unrevolved PACS sources, because it was especially designed to obtain high precision astrometry and photometry of point sources in crowded fields.
Basically, \texttt{Starfinder} carries out source extraction using three steps:
(i) detection of candidate point sources; (ii) verification of the likelihood of the candidate point sources; (iii) fits of empirical PSFs to the map, at positions of the candidate point sources.

Input parameters of \texttt{Starfinder} are fine-tuned using MC simulations (See Section \ref{subsec:noise}).
\texttt{Starfinder} performed its PSF-fitting analysis with the same empirical PSFs as those used for the prior source extraction (see Section \ref{subsec:PSF}).
Flux densities inferred by \texttt{Starfinder} are corrected for the finite size of the empirical PSFs and for the effect of high-pass filtering (see Sections \ref{sec: PACS images} and \ref{subsec:PSF}).
From the MC simulations we observe that, despite the fine-tuning, \texttt{Starfinder} still tends to overestimate the flux density of sources at low S/N levels (about $10\%$ at S/N$\,\thicksim\,$$3$).
This behaviour is corrected using factors derived from the MC simulations.

\texttt{Starfinder} being exclusively designed for point source extraction, the presence of a few extended sources ($\thicksim$8 sources per field) in the PACS maps affects its detection process: extended sources are split into sub-components.
To fix this problem, we run \texttt{Sextractor} on these extended sources using Kron elliptical apertures.
Extended sources are identified using an empirical method exploiting the isophotal area vs flux parameter space.
In this parameter space, extended sources have large isophotal areas compared to their flux densities.
PACS flux densities of the sub-components of the extended sources are erased from our blind source catalogue and replaced by flux densities inferred by \texttt{SExtractor}.
We note that the accuracy of the \texttt{Sextractor} flux densities was verified using the \texttt{Sextractor} residual maps.

Flux uncertainties are estimated with \texttt{Starfinder} using our PACS uncertainty maps (see Section \ref{sec: PACS images}) and empirical PSFs.
These flux uncertainties are corrected for the finite size of the PSFs and for the effect of high-pass filtering.
In addition, because our PACS uncertainty maps do not account for correlated noise, these flux uncertainties have to be corrected upwards using mean correlation corrections (Section \ref{sec: PACS images}).
We note that after correction, the flux uncertainties inferred by \texttt{Starfinder} are in good agreement with the photometric accuracies inferred from the MC simulations.

\subsection{PSF reconstruction\label{subsec:PSF}}
Empirical PSFs used for the prior and blind source extraction are derived directly from the PACS maps using \texttt{Starfinder}.
A number of isolated point-like sources present in the maps are stacked and then normalized to unit total flux.
Because of the limited number of isolated point-like sources, the wings of these empirical PSFs have limited S/N ratio.
Therefore, they have to be truncated to smaller radii ($9.6\arcsec$, $7.2\arcsec$ and $12\arcsec$ at 70, 100 and 160$\,\mu$m, respectively) and then renormalized to a unit total flux.

Because of their finite extent, the empirical PSFs do not include all the power contained in the wings of the real PSFs.
Consequently, flux densities inferred from these PSFs have to be corrected, as would be done for any standard aperture flux measurement.
Aperture corrections are derived by comparing the empirical PSFs with the reference in-flight PSF for our observing mode, obtained from observations of the asteroid Vesta.
First, the Vesta observations are manipulated to account for variations of the position angle (P.A.) between our AORs, i.e., the Vesta observations are rotated to the P.A. of each AOR and then stacked.
Second, Vesta observations are smoothed to match the slightly broader FWHM of the empirical PSFs (broader by a factor 1.04, 1.02 and 1.01 at 70, 100 and 160$\,\mu$m, respectively).
These broader FWHM can be explained by residual pointing uncertainties left by our recentring procedure (see Section \ref{sec: PACS images}).
Finally, aperture corrections are defined as the fraction of the total energy contained in these manipulated Vesta PSFs within the radii of the empirical PSFs.
At 70, 100 and 160$\,\mu$m, the empirical PSFs contain 77, 68 and 69$\,\%$ of the total energy of the manipulated Vesta PSFs, respectively.

\subsection{Monte Carlo simulations\label{subsec:noise}}
 \begin{figure*}
 	\begin{center}
	\includegraphics[width=8.5cm]{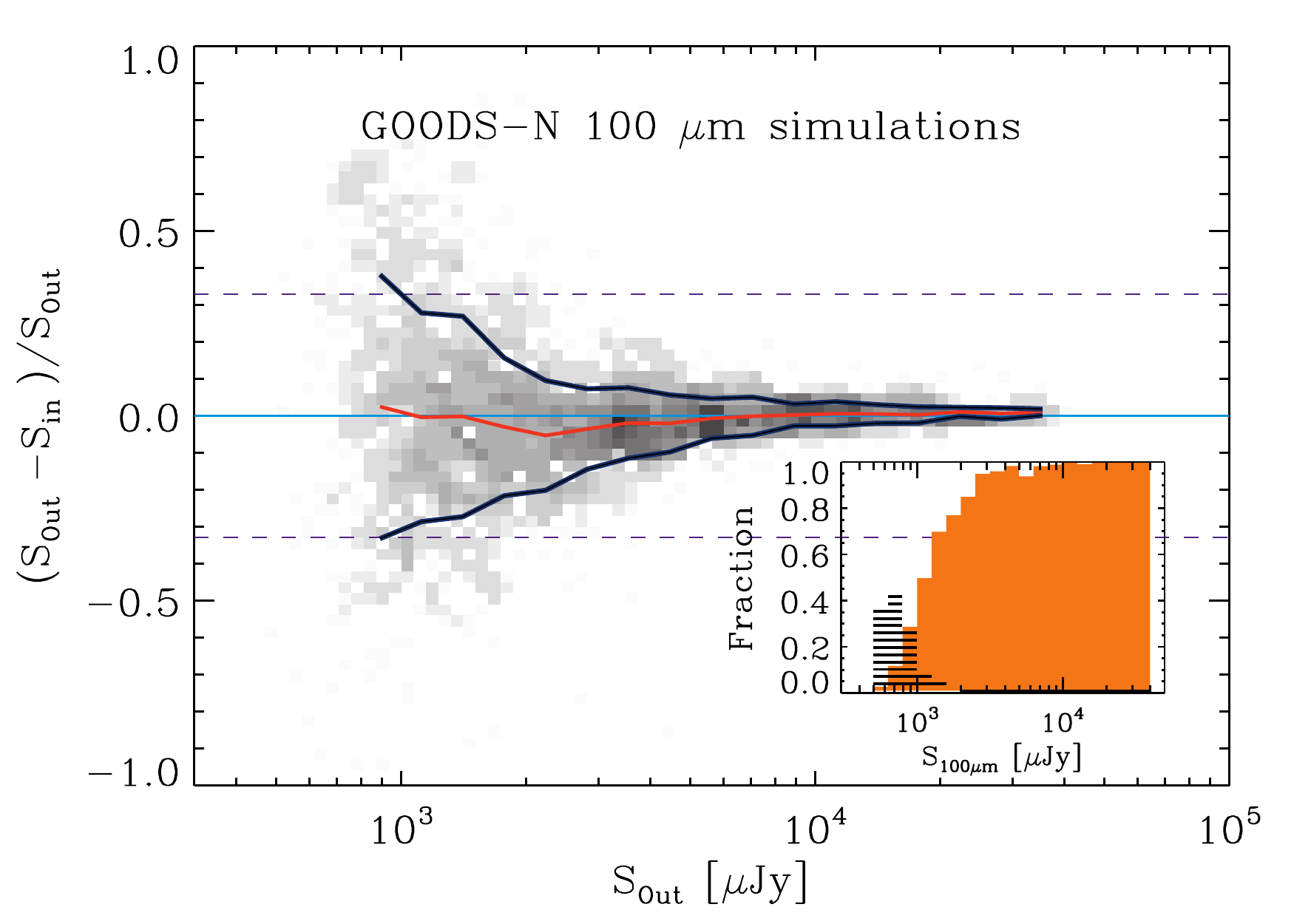}
	\includegraphics[width=8.5cm]{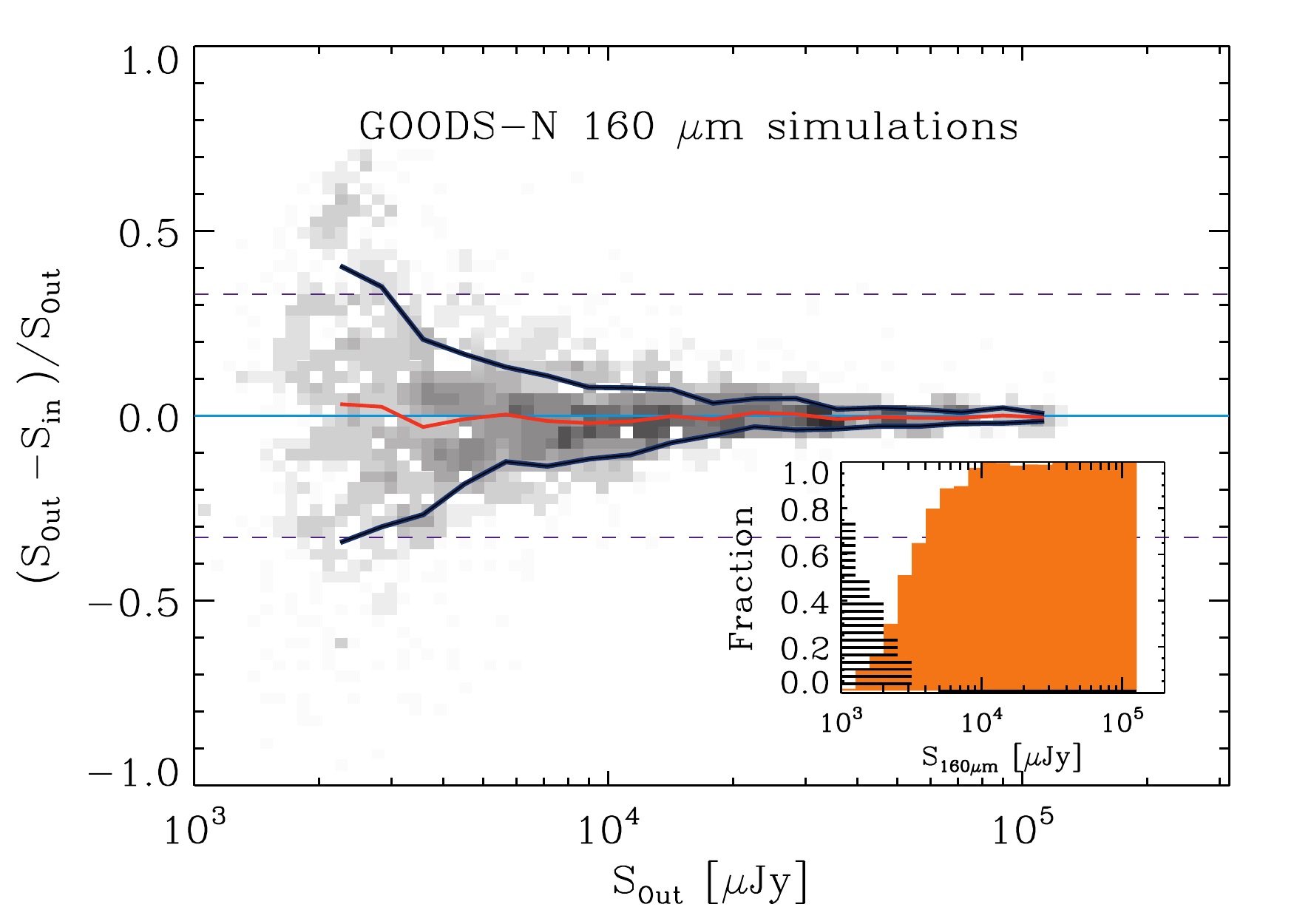}\\
	\includegraphics[width=8.5cm]{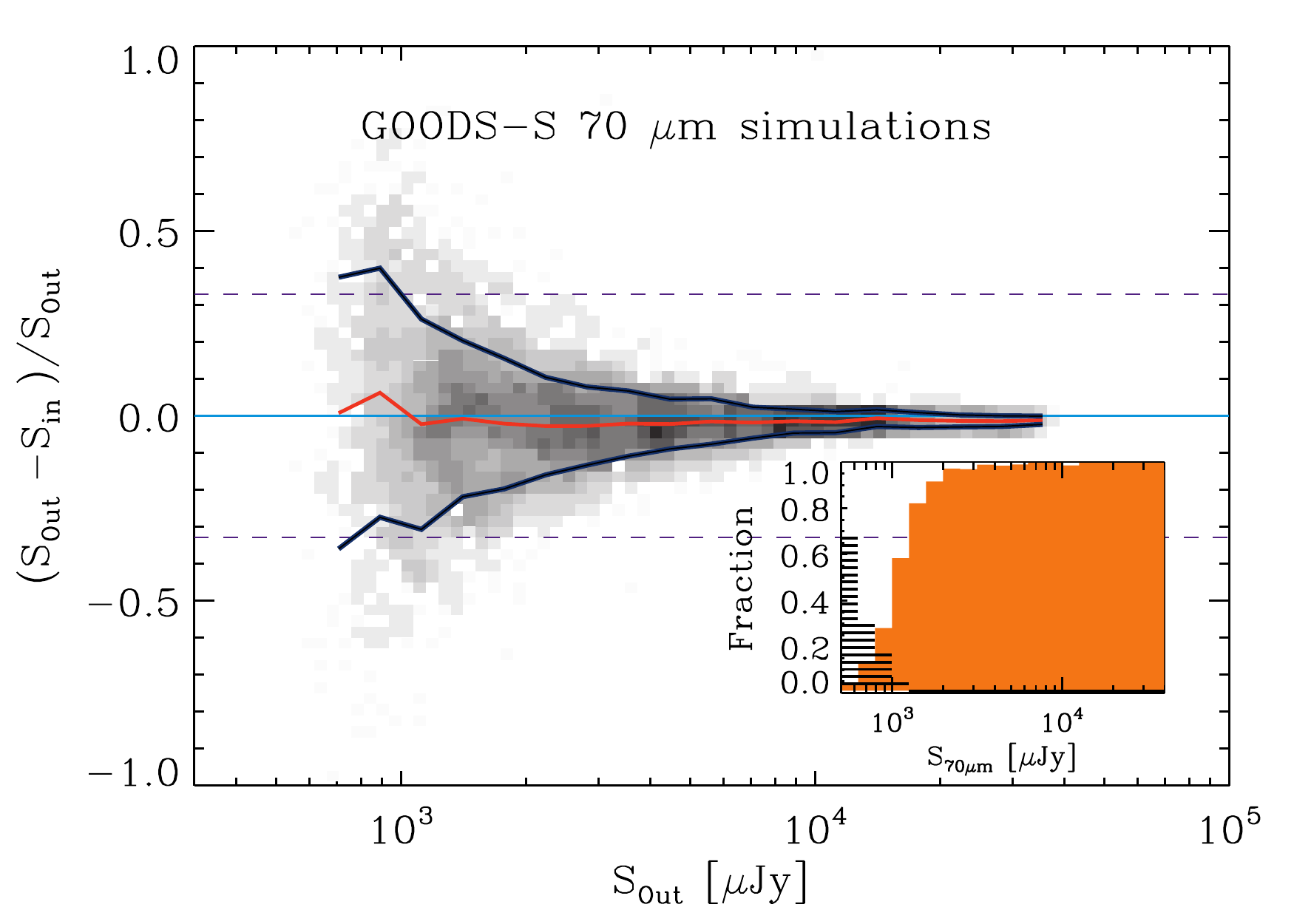}
	\includegraphics[width=8.5cm]{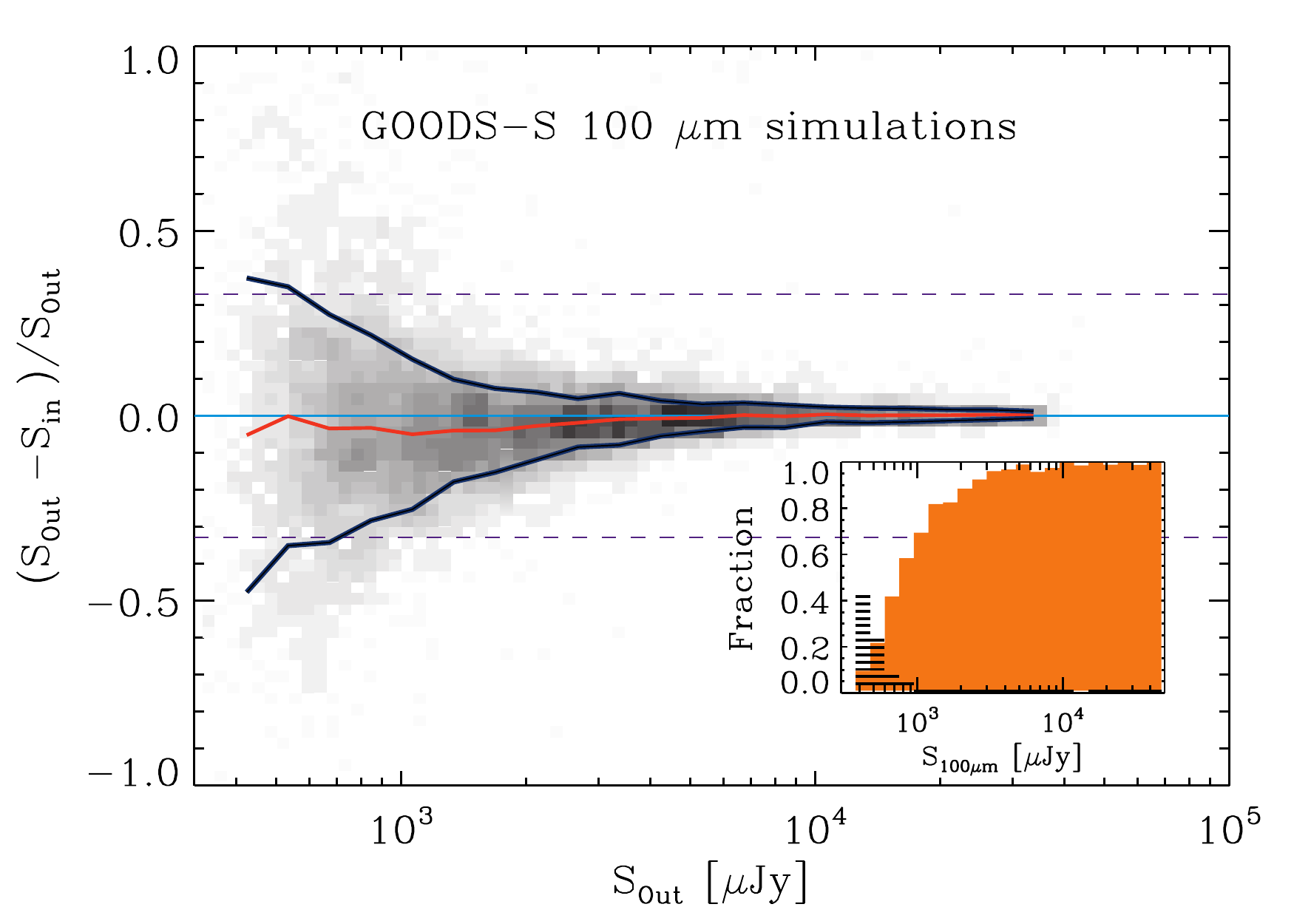}\\
	\includegraphics[width=8.5cm]{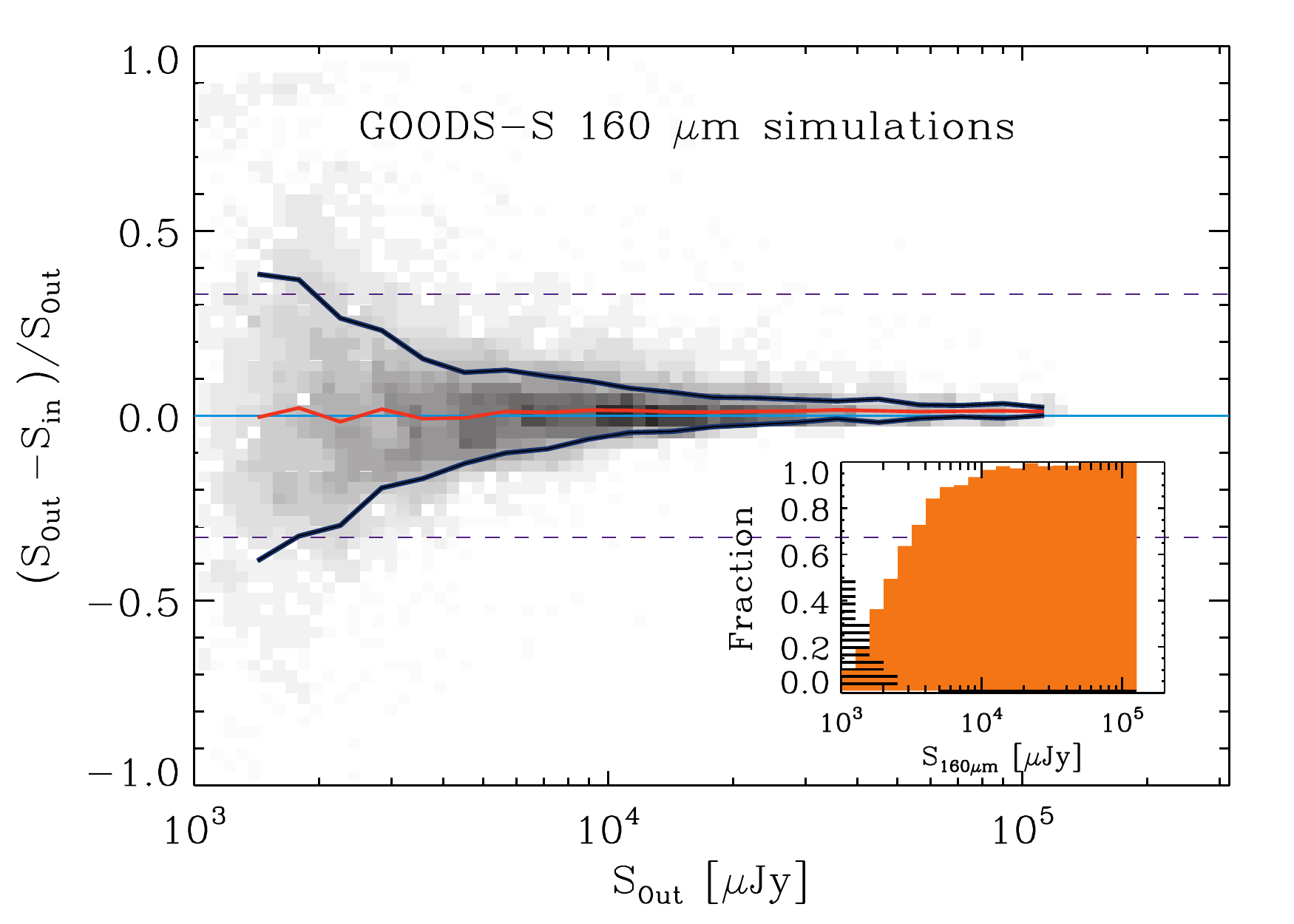}
	\end{center}
	\caption{Results of MC simulations in the GOODS-North and -South fields using the prior source extraction method.
	These MC simulations correspond to the central deepest regions of each field, i.e., avoiding the rather noisy edges of the GOODS-N field and concentrating on the GOODS-S-ultradeep part of the GOODS-S field.
	Blue lines represent the average photometric accuracy defined as the standard deviation of the $(S_{\rm out}-S_{\rm in})/S_{\rm out}$ distribution in each flux bin (after 3$\sigma$ clipping).
	Red lines show the mean value of the $(S_{\rm out}-S_{\rm in})/S_{\rm out}$ distribution in each flux bin.
	 Inset plots show the fraction of artificial sources detected in the image (i.e., completeness) as a function of input flux (\textit{orange plain histogram}) and the fraction of \textit{spurious} sources (i.e., contamination) as a function of flux density (\textit{striped black histogram}).
	\label{fig:MC GOODS}}
\end{figure*}
\begin{table*}
\center
\footnotesize
\caption{ \label{tab:MC}
Statistics of released catalogues, for regions above the specified exposure time.}
\begin{tabular}{lccccccccc}
\hline\hline
Field &  $t_{{\rm exposure}}^{\rm a}>$ & $S(1\sigma)$$^{\rm b}$ & Num.$^{\rm c}$ & Num.$^{\rm c}$ & Compl.$^{\rm d}$ & Contam.$^{\rm e}$ & Compl.$^{\rm d}$ & Contam.$^{\rm e}$ & $S(80\%)^{\rm f}$ \\
\& bands  & {\small hrs/pixel} & {\small mJy} & {\small S/N$\,>\,$$3$} & {\small S/N$\,>\,$$5$} & {\small \% at $3\sigma$} &  {\small \% at $3\sigma$} & {\small \% at $5\sigma$} & {\small \% at $5\sigma$} & {\small mJy} \\
\hline
\multicolumn{10}{c}{ \rule{0pt}{2ex}Prior source extraction} \\
\hline
GOODS-N 100  & $1.3$ & $ 0.32 $ & $ 914  $ & $ 574 $ & $ 36 $ & $ 14 $ & $ 72 $ & $ 5 $ & $ 1.97 $ \\
GOODS-N 160  & $1.3$ & $ 0.70 $ & $ 838  $ & $ 546 $ & $ 24 $ & $ 32 $ & $ 62 $ & $ 5 $ & $ 4.56 $  \\
 \rule{0pt}{3ex}GOODS-S 70    & $1.0$ & $ 0.30 $ & $ 476   $ & $ 245 $ & $ 21 $ & $ 24 $ & $ 77 $ & $ 2 $ & $ 1.39 $ \\
GOODS-S 100 ultradeep  & $6.4$ & $ 0.18 $ & $ 472 $ & $ 346 $ & $ 21 $ & $ 27 $ & $ 65 $ & $ 5 $ & $ 1.22 $ \\
GOODS-S 160 ultradeep & $8.3$ & $ 0.43 $ & $ 445   $ & $ 313 $ & $ 18 $ & $ 38 $ & $ 45 $ & $ 14  $ & $ 3.63 $ \\
\hline
\multicolumn{10}{c}{ \rule{0pt}{2ex}Blind source extraction} \\
\hline
GOODS-N 100  & $1.3$ & $ 0.32 $ & $865 $ & $ 596 $ & $ 13 $ & $ 37 $ & $ 56 $ & $ 7 $ & $ 2.04 $ \\
GOODS-N 160  & $1.3$ & $ 0.70 $ & $ 867 $ & $ 521 $ & $ 16 $ & $ 48 $ & $ 53 $ & $ 17 $ & $ 5.88 $ \\
 \rule{0pt}{3ex}GOODS-S 70      & $1.0$ & $ 0.33 $ & $ 396   $ & $ 205 $ & $ 19 $ & $ 19 $ & $ 72 $ & $ 11 $ & $ 1.51 $\\
GOODS-S 100  ultradeep  & $6.4$ & $ 0.18 $ & $ 513 $ & $ 377  $ & $ 21 $ & $ 42 $ & $ 74 $ & $ 9 $ & $ 1.02 $ \\
GOODS-S 160  ultradeep & $8.3$ & $ 0.43 $ & $ 485 $ & $ 368  $ & $ 15 $ & $ 47 $ & $ 45 $ & $ 8 $ & $ 4.89 $ \\
\hline
\end{tabular}
\begin{list}{}{}
\item[$^{\rm a}$] Numbers listed in this table are only suitable for sources situated in the regions of the field with exposure time higher than that reported in this column.To convert the PACS coverage maps provided in the released package into hrs/pixel, users should multiply them by $1.06$$\,\times\,$$10^{-2}$.
\item[$^{\rm b}$]  The 1$\sigma$ flux density levels have been computed from $10\,000$ random extractions on residual maps.
\item[$^{\rm c}$]  Number of sources above a given S/N threshold.
\item[$^{\rm d}$]  Completeness is defined as the fraction of simulated sources with $S_{\rm in}$$\,\thicksim\,$$3(5)\sigma$ and extracted with a flux accuracy better than 50\%, i.e., $-0.5$$\,<\,$$(S_{\rm out}-S_{\rm in})/S_{\rm out}$$\,<\,$$0.5$.
\item[$^{\rm e}$] Contamination is defined as the fraction of simulated sources introduced with $S_{\rm in}$$\,<\,$$2\sigma$ but extracted with $S_{\rm out}$$\,>\,$$3\sigma$.
\item[$^{\rm f}$] Flux densities above which our catalogues are 80\% complete.
\end{list}
\end{table*}

The PEP/GOODS-H observations have been designed to reach the confusion limit of the \textit{Herschel Space Observatory} at 100 and 160$\,\mu$m.
Flux uncertainties are therefore a complex combination of instrumental and confusion noise.
In order to estimate these complex flux uncertainties and to characterize the quality of our catalogues we perform extensive MC simulations.

We add simulated sources to our PACS 70, 100 and 160$\,\mu$m images with a flux distribution, approximately matching the measured number counts \citep{berta_2010,berta_2011}.
The flux densities of the faintest simulated sources are defined as the PACS flux densities (i.e., $S_{\lambda}$) expected for the faintest MIPS-24$\,\mu$m (i.e., $S_{24}=20\,\mu$Jy) sources, i.e., $S_{\lambda}^{\rm min}$$\,=\,$${\rm min}(S_{24})$$\,\times\,$$ {\rm mean}(S_{\lambda}/S_{24})$, where ${\rm mean}(S_{\lambda}/S_{24})$ is 10, 20 and 30 at $\lambda$$\,=\,$70, 100 and 160$\,\mu$m, respectively (see Fig.~\ref{fig:ratio}).
To preserve the original statistics of the image (especially its source number density), the number of simulated objects added at one time is kept small (i.e., 20 sources). 
To recreate the clustering properties of PACS sources, simulated objects are positioned, with respect to the MIPS-24$\,\mu$m sources, to reproduce the distance to the closest neighbour distribution observed in the MIPS-$24\,\mu$m prior catalogues.
Simulated sources are created using the manipulated Vesta PSFs which contain the wing of the real PSFs (see Section \ref{subsec:PSF}).
On these simulated images, we then perform both our blind and prior source extraction using the empirical PSFs and compare the resulting flux densities to the input values.
To improve the statistics, this process is repeated a large number of times using each time different positions and fluxes for the simulated sources.
For each field and at each wavelength, a total of $20,000$ artificial sources are used.
Figure \ref{fig:MC GOODS} shows, as an example, results from the MC simulations performed in the GOODS fields using the prior source extraction method.
In both GOODS fields, these MC simulations correspond to the central deepest regions (i.e., avoiding the rather noisy edges of the GOODS-N field and concentrating on the GOODS-S-ultradeep part of the GOODS-S field).
MC simulations on the outskirts of the GOODS-S field (i.e., GOODS-S-deep) have been run, but are not shown here.
In any case, results from these MC simulations are very similar to those for the GOODS-N field, as expected from the similarity between the GOODS-S-deep and GOODS-N average exposure times.

From these MC simulations we derive three important quantities: the completeness; the contamination; and the photometric accuracy of our catalogues as a function of flux density.
Completeness is defined as the fraction of simulated sources extracted with a flux accuracy better than 50\%.
The contamination is defined as the fraction of simulated sources introduced with $S$$\,<\,$$2\sigma$ but extracted with $S$$\,>\,$$3\sigma$.
The photometric accuracy is defined as the standard deviation of the $(S_{\rm out}-S_{\rm in})/S_{\rm out}$ distribution as a function of $S_{\rm out}$ (blue lines in Fig.~\ref{fig:MC GOODS}).
These photometric accuracies have the advantage of taking into account simultaneously nearly all sources of noise, i.e., including confusion. 
Table \ref{tab:MC} summarises these quantities for both the blind and prior source extraction approaches.

From these MC simulations we conclude that the prior and blind PSF-fitting methods perform accurate extraction of PACS sources.
The prior and blind source catalogues are characterized by high completeness and low contamination levels, as well as by good photometric accuracy (i.e., $<\,$$33\%$).
We observe in Fig.~\ref{fig:MC GOODS} that, using our flux uncertainties, a S/N$\,>\,$$3$ cut almost always translates into a photometric accuracy better than 33\% (i.e., as expected from sources with S/N$\,>\,$$3$).
We can thus conclude that our flux uncertainties are fairly accurate.
However, we also note that in the GOODS-S field the photometric accuracy is worse than 33\% for flux densities below $0.6\,$mJy and $2.0\,$mJy at 100 and 160$\,\mu$m, respectively.
At such faint flux densities our PACS 100 and 160$\,\mu$m maps are likely to be affected by confusion noise (i.e., $\sigma_{\rm c}(100\,\mu{\rm m})$$\,\thicksim\,$$0.15\,$mJy and $\sigma_{\rm c}(160\,\mu{\rm m})$$\,\thicksim\,$$0.68\,$mJy; see Sect. \ref{sec:confusion}) that is not fully accounted in flux uncertainty estimated by our source extraction methods.
This conclusion is further confirmed by the analysis of the $(S_{\rm in}-S_{\rm out})/\sigma_{\rm s}$ distribution ($\sigma_{\rm s}$ being the flux uncertainty inferred by our source extraction methods).
Indeed, in all but the GOODS-S 100 and 160$\,\mu$m fields, the $(S_{\rm in}-S_{\rm out})/\sigma_{\rm s}$ distribution follows, as expected, a Gaussian distribution with a dispersion almost equal to 1.
Instead, in the GOODS-S 100 and 160$\,\mu$m fields, we find that the $(S_{\rm in}-S_{\rm out})/\sigma_{\rm s}$ distribution follow a Gaussian distribution with a dispersion of $\thicksim\,$$1.2$, indicating that our flux uncertainties are underestimated and do not fully account for confusion noise.
\textit{Therefore, we recommend caution in using flux densities lower than $0.6\,$mJy and $2.0\,$mJy at 100 and 160$\,\mu$m, respectively.}

From this analysis we conclude that in the GOODS-N field the combined PEP/GOODS-H data are 3.0 (1.16) and 2.8 (1.16) times deeper than the PEP (GOODS-H) data at 100 and 160$\,\mu$m, respectively. 
Similarly, in the GOODS-S-ultradeep subarea covered by both projects, the combined PEP/GOODS-H data are 2.3 (1.5) and 2.0 (1.5) times deeper than the PEP (GOODS-H) data at 100 and 160$\,\mu$m, respectively.

\subsection{Content of the PEP/GOODS-H released package\label{subsec:package}}
The PEP/GOODS-H released package\footnote{\texttt{http://www.mpe.mpg.de/ir/Research/PEP/public\_data\_releases.php}} contains the scientific, uncertainty and coverage PACS maps.
We reiterate that the GOODS-S 100/160$\,\mu$m coverages and noise levels are inhomogeneous due to the combination of observations with different layouts (see Section \ref{sec:observations}). 
In contrast, the coverages and noise levels for GOODS-N 100/160$\,\mu$m and GOODS-S 70$\,\mu$m are fairly homogeneous, except for degradation near the edges.

The released package contains PACS blind and prior source catalogues, down to $3\sigma$ significance.
Because our source extraction methods might be inaccurate on the noisy edge of the PACS maps, we crop these regions from our catalogues.
In order to track coverage variations across the fields, we provide for each source its exposure time (or equivalent) in each passband.
The completeness and contamination levels inferred from the MC simulations are part of the released package.
To use them, users should restrict their sample to sources with exposure time greater than that quoted in the second column of Table \ref{tab:MC}.
Finally, we remind users that because our flux uncertainties do not fully account for confusion noise, \textit{sources with flux densities below $0.6\,$mJy at 100$\,\mu$m and $2.0\,$mJy at 160$\,\mu$m have to be treated with caution}.

Calibration factors used to generate the final PACS maps are derived assuming an in-band SED of $\nu S_{\nu}=\,$constant (see the \textit{Herschel} data handbook for more details). 
Thus, some moderated colour corrections (i.e., $<7\%$) might need to be applied to our flux density measurements, as in-band SEDs of distant galaxies could be different from those assumed here.
However, because these colour corrections depend on the redshift of the source, we decided \textit{not} to apply any colour correction to the released catalogues.

PACS blind catalogues have been cross-matched to our MIPS-24$\,\mu$m catalogues using a maximum likelihood analysis \citep{ciliegi_2001,sutherland_1992}.
This method simultaneously accounts for fluxes and positions of MIPS-24$\,\mu$m counterparts as well as positional errors in both the PACS and MIPS samples.
The cross-identification of PACS and MIPS sources is included in the released package.
In appendix \ref{appendix:blind/prior}, we compare the PACS blind and prior source catalogues using this cross-identification.

Following the results of \citet{elbaz_2011} and \citet{hwang_2010}, we provide for each source of the blind and prior catalogues, its ``clean index''.
Because this ``clean index'' is a measure of the number of bright neighbours for a given source in all passbands, it supplies information on the potential contamination of its flux densities.
Here, ``bright'' neighbours are defined as sources brighter than half of the flux density of the source of interest, and closer than 20\arcsec, 6.7\arcsec\ and 11\arcsec\ at 24, 100 and 160, respectively.
Having counted the number of neighbours of a given source (i.e., \texttt{Neib24}, \texttt{Neib100} and \texttt{Neib160}), its ``clean index'' is given by
\begin{equation}
{\rm \texttt{clean\_index}=\texttt{Neib24}+\texttt{Neib100}\times10 +\texttt{Neib160}\times100},
\end{equation}
We note that the radius of 20\arcsec\ at 24$\,\mu$m was defined in order to provide a ``clean region'' to the SPIRE-250$\,\mu$m galaxy population \citep[FWHM$\,\thicksim\,$$18\arcsec$; see][]{elbaz_2011}.
Therefore a ``clean index'' corresponding to \texttt{Neib24}$\,\leq\,$$1$ would provide a very conservative selection for accurate PACS flux density estimates.
Thus, we recommend the use of this conservative criterion (i.e., \texttt{Neib24}$\,\leq\,$$1$; 36$\%$ and 33$\%$ of the PACS sources in the GOODS-N and GOODS-S fields, respectively) not to select PACS sources with accurate flux densities but only to test whether any scientific results obtained with the full catalogue do not change when this criterion is applied.

We stress that the MIPS-24$\,\mu$m catalogues released here are slightly different from those released by \citet{magnelli_2011a}.
These new MIPS-24$\,\mu$m catalogues are generated using, as prior information, newer versions of the IRAC catalogues which are publicly available as part of the GOODS-H data release.
These MIPS-24$\,\mu$m catalogues are identical to those used for the GOODS-H data release\footnote{\texttt{http://hedam.oamp.fr/GOODS-Herschel}} \citep{elbaz_2011}.

Finally, we note that the PEP/GOODS-H released package also contains ancillary products, i.e., the uncertainty and coverage maps, the empirical PSFs used by our source extraction methods, the Vesta PSFs used to measure aperture corrections, the results of the MC simulations and the PACS residuals maps.

\section{Confusion noise\label{sec:confusion}}
\begin{figure}
	\includegraphics[width=9.2cm]{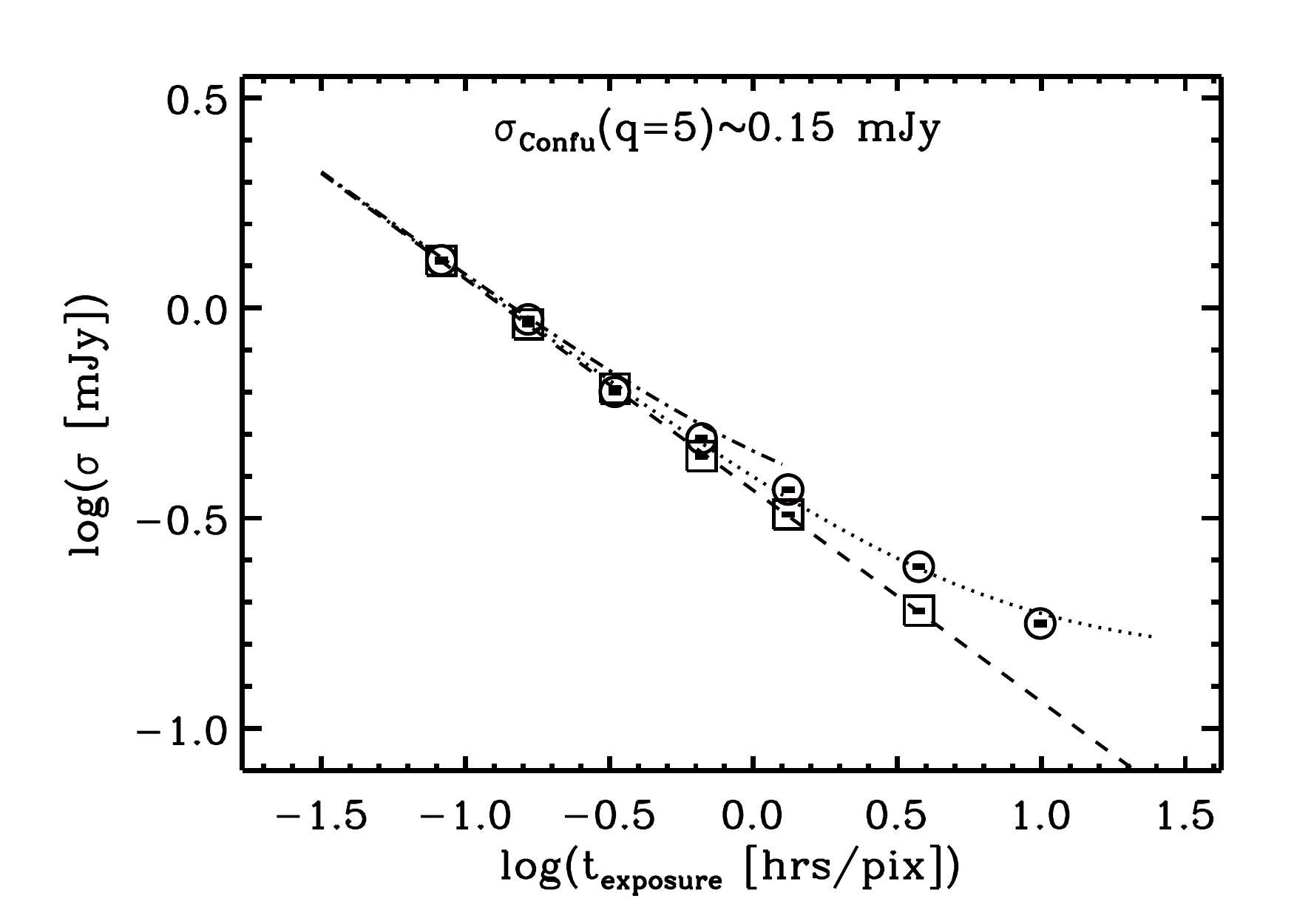}
		\includegraphics[width=9.2cm]{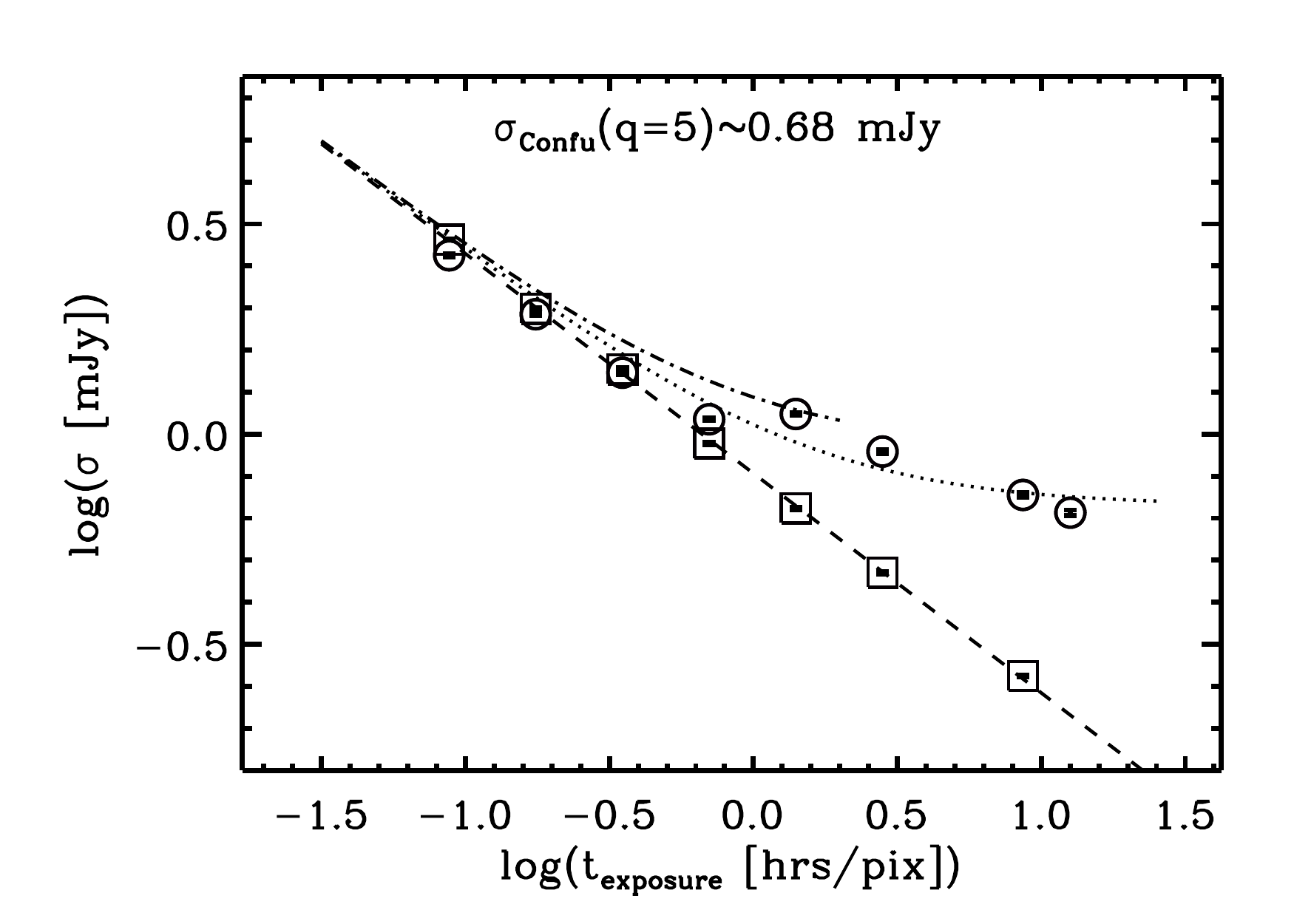}
		\caption{\label{fig:confusion}
		Noise in the PACS-100$\,\mu$m (\textit{top panel}) and PACS-160$\,\mu$m (\textit{bottom panel}) maps as a function of exposure time.
		Empty circles represent the total noise $\sigma_{\rm T}$ of the maps.
		The total noise is fitted with two noise components added in quadrature:  an instrumental noise $\sigma_{\rm I}$ component following a $t^{-0.5}$ trend (dashed line and empty squares) and an constant confusion noise $\sigma_{\rm c}$ component.
		The dotted lines present the two components fitted to $\sigma_{\rm T}$.
		The dot-dashed lines present the two components fitted to $\sigma_{\rm T}$ made in \citet{berta_2011} and illustrate the smaller exposure time range probed in their study.
		}
\end{figure}

Following the formalism of \citet{dole_2003}, deep PACS observations might be affected by two different types of confusion due to extragalactic sources.
First, the photometric confusion noise due to sources below the detection limit  $S_{\rm lim}$ which produce signal fluctuations (i.e., $\sigma_{\rm c}$) within the beam of the PACS sources, i.e., $S_{\rm lim}/\sigma_{\rm c}$$\,=\,$$q$ with $q=3$ or $5$.
Second, the density confusion due to the high density of sources above $S_{\rm lim}$ which increases the probability to miss objects that are blended with bright neighbours.
Because our GOODS-S 100 and 160$\,\mu$m maps are the deepest blank field observations obtained by the \textit{Herschel Space Observatory}, they are the most suitable datasets to estimate the PACS-100 and -160$\,\mu$m photometric and density confusion noise.
These estimates are presented in this section while the PACS-70$\,\mu$m confusion noise is derived and discussed in \citet[][they find that the GOODS-S PACS-70$\,\mu$m observations are too shallow to be affected by either type of confusion]{berta_2011}.

The photometric confusion noise is estimated empirically following the procedure described in \citet{frayer_2006} and \citet{berta_2011}.
First, we built several GOODS-S maps using only a fraction of the \textit{Herschel} observations available.
These partial-depth maps regularly probe the logarithmic exposure time parameter space from $\thicksim\,$$0.1\,$hrs/pixel to $\thicksim\,$$10\,$hrs/pixel.
Secondly, we performed prior source extractions on these partial-depth maps and produced the corresponding residual maps removing sources with $S$$\,>\,$$S_{\rm lim}$, with $S_{\rm lim}/\sigma_{\rm c}$$\,=\,$$q$ using $q=5$.
Thirdly, we measured the total noise $\sigma_{\rm T}$ in these residual maps using the procedure described in Sect. \ref{subsec:prior}, i.e., $\sigma_{\rm T}$ is defined as the pixel dispersion of the residual map convolved with the Vesta PSF.
Finally, we estimated $\sigma_{\rm c}$ by analyzing the variation of $\sigma_{\rm T}$ as a function of exposure time (see Fig.~\ref{fig:confusion}).
This procedure was iterated until convergence at $S_{\rm lim}/\sigma_{\rm c}$$\,=\,$$5$ was reached.

Noise in partial-depth maps with short exposure time is dominated by the instrumental noise $\sigma_I$ and therefore decreases as $t^{-0.5}$.
In contrast, noise in partial-depth maps with long exposure time departs from the $t^{-0.5}$ trend and is a combination of instrumental noise and confusion noise.
Assuming that both components follow a Gaussian distribution, $\sigma_{\rm T}$ can then be approximated by $\sigma_{\rm T}=\sqrt{\sigma_{I}^{2}+\sigma_{\rm c}^{2}}$.
Because $\sigma_{\rm c}$ does not depend on the exposure time, one can thus estimate $\sigma_{\rm c}$ by fitting $\sigma_{\rm T}$ with a two components function, i.e., $\sigma_{\rm c}$ and $\sigma_{\rm I}$$\,\propto\,$$t^{-0.5}$.

We find a photometric confusion noise $\sigma_{\rm c}$ of $0.15$ and $0.68\,$mJy in the PACS-100 and -160$\,\mu$m passbands, respectively.
The total noise of the PEP/GOODS-H observations is thus nearly dominated by the photometric confusion noise at 100$\,\mu$m ($\sigma_{\rm I}^{100}=0.11\,$mJy) and it is fully dominated by the photometric confusion noise at 160$\,\mu$m ($\sigma_{\rm I}^{160}=0.21\,$mJy). 
In both bands, this significant contribution of the confusion noise to the total noise affects our flux uncertainty estimates: our flux uncertainties (1$\sigma$$\,\thicksim\,$0.18 and 0.43 mJy at 100 and 160$\,\mu$m, respectively) do not fully account for confusion noise and sources with flux densities below $0.6\,$mJy at 100$\,\mu$m and $2.0\,$mJy at 160$\,\mu$m have to be treated with caution (see Sect. \ref{subsec:noise}).
\begin{figure*}
	\includegraphics[width=9.2cm]{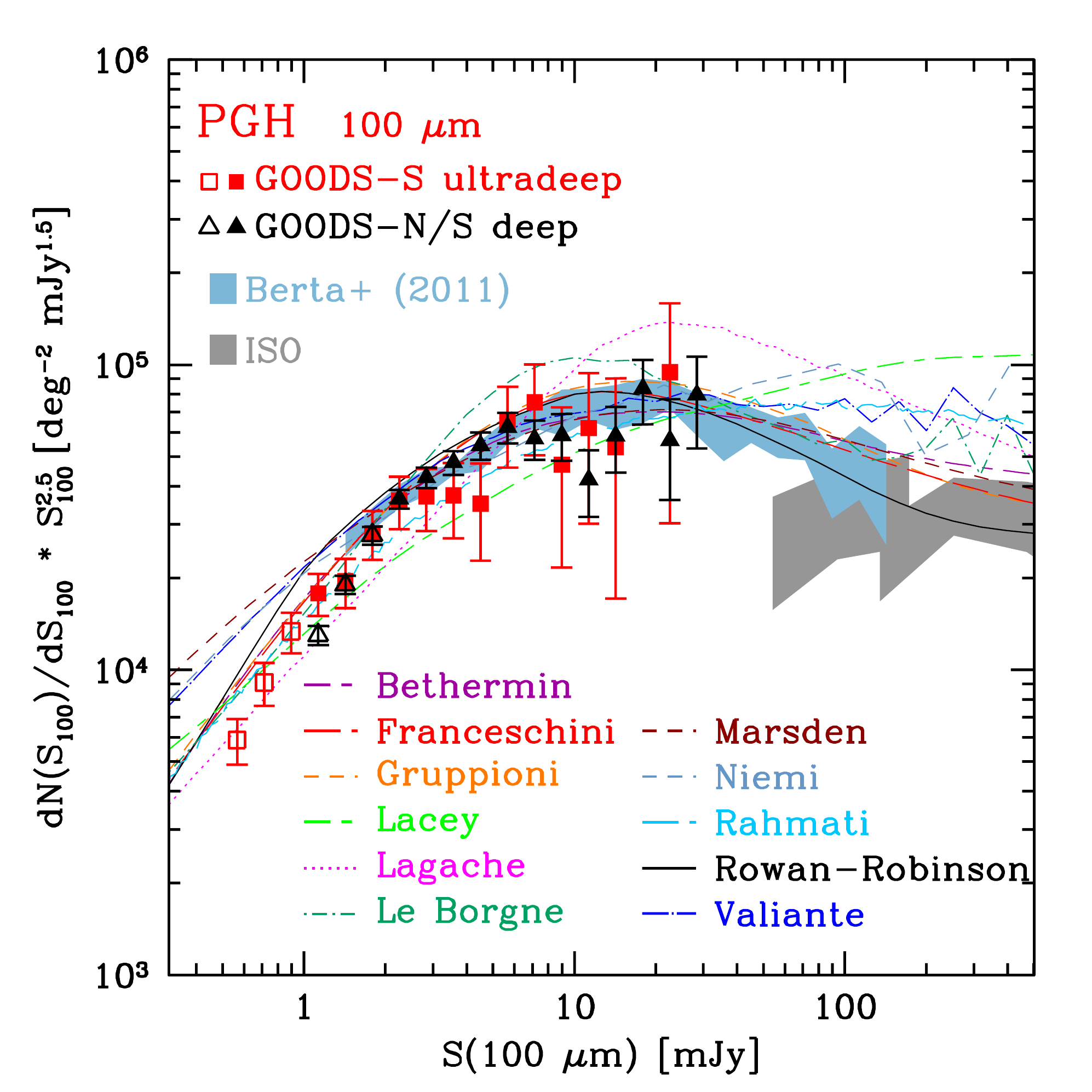}
	\includegraphics[width=9.2cm]{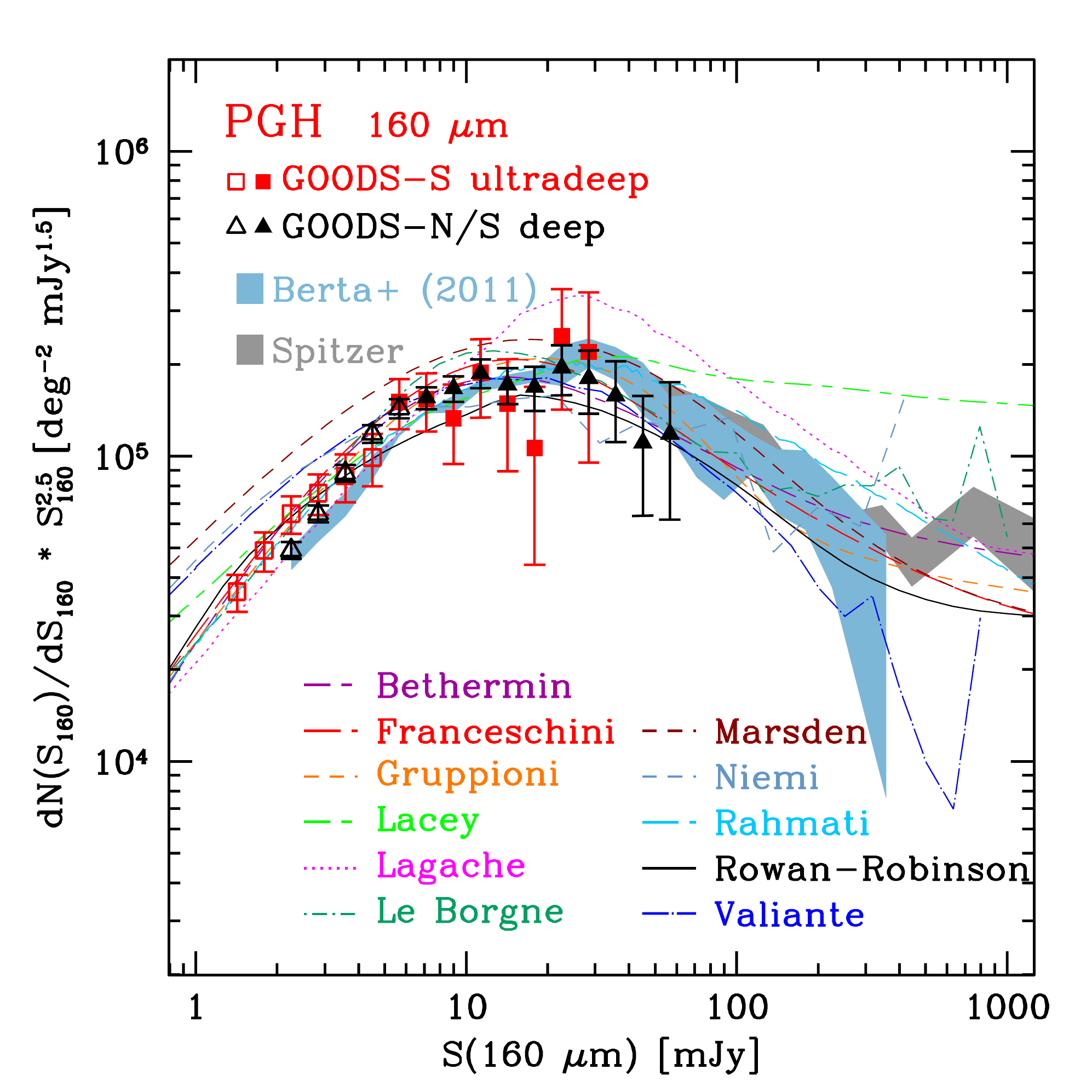}
	\caption{\label{fig:counts}
	PACS 100 and 160$\,\mu$m differential number counts, normalized to the Euclidean slope ($dN/dS\propto S^{-2.5}$). 
	Filled and open symbols show flux density bins above and below the 80\% completeness limit, respectively. 
	Grey shaded areas present estimates obtained using pre-\textit{Herschel} observations.
	Blue shaded areas present estimates obtained using PACS observations \citep{berta_2011}.
	Lines represent predictions from backwards or forwards evolutionary models \citep{lagache_2004,rowan_robinson_2009, valiante_2009, leborgne_2009, franceschini_2010, gruppioni_2010,lacey_2010, marsden_2011b,rahmati_2011,niemi_2012,bethermin_2012b}.
		}
\end{figure*}

We note that our photometric confusion noise estimates are lower than those of \citet[][; $\sigma_{\rm c}$$\,\thicksim\,$$0.27$ and $0.92\,$mJy at 100 and 160$\,\mu$m, respectively]{berta_2011}.
These discrepancies likely come from the fact that the observations used in \citet{berta_2011} were not deep enough to be dominated by the photometric confusion noise and therefore led to more uncertain estimates.
This limitation is illustrated by the dot-dashed lines in Fig.~\ref{fig:confusion} which represent the range of exposure time probed in \citet{berta_2011}.\\

The confusion due to the high density of bright sources is usually defined as the flux limit $S_{\rm lim}$ at which the source density corresponds to 16.7 beams/source, i.e., the density at which 10\% of the sources are separated by less than $0.8\,\times\,$FWHM and thus ``blended'' \citep[][; where $\Omega=1.14\,\times\,$FWHM$^2$ is the area of a beam]{dole_2003}.
Using the PACS number counts, \citet{berta_2011} found that this density criterion corresponds to $S_{\lim}$$\,\thicksim\,$$2.0$ and $\,\thicksim\,$$8\,$mJy at 100 and 160$\,\mu$m, respectively.
These limits are thus greater than the detection limits derived from our MC simulations.
This discrepancy can be explained by the very conservative approach adopted by \citet{dole_2003}.
The density limit of 16.7 beams/source of \citet{dole_2003} corresponds to 10\% of blended sources (i.e., separated by less than $0.8\,\times\,$FWHM).
This requirement translates into a blending-completeness of 90\% while sources can be accurately extracted at lower completeness: at the 3$\sigma$ level a catalogue has a typical completeness value of 40-60\% (i.e., 60-40\% of blended sources). 
Using this more realistic assumption (i.e., measuring the density for which 40\% of the sources are separated by less than $0.8\,\times\,$FWHM), we infer a confusion density limit of $\thicksim\,$$3.5$ beams/source.
In our GOODS-S-ultradeep catalogues, the density of sources corresponds to 8 and 4 beams/source at 100 and 160$\,\mu$m, respectively.
In these catalogues, the density of sources is thus much higher than the density confusion limit defined by \citet[][i.e., 16.7 beams/source]{dole_2003} but lower than those defined using more realistic assumptions (i.e., 3.5 beams/source).
We stress that while an analytical estimate of the confusion density limit is useful, it should be used with caution as it does not account for the fact that our ability to separate pairs of sources depends on their S/N.
In contrast, empirical estimates through MC simulations take this effect into account.
\section{Number counts\label{sec:counts}}
Using the blind catalogues (i.e., without applying any ``clean index'' selection) we inferred the PACS 100 and 160$\,\mu$m differential number counts down to an unprecedented depths \citep[PACS-70$\,\mu$m differential number counts are derived in ][]{berta_2011}.
For that purpose, we applied the method described in \citet{berta_2010,berta_2011}, which accounts for the incompleteness and contamination of our catalogues using results from the MC simulations \citep[see also][]{chary_2004,smail_1995}.
In this method, observations with different depths cannot be treated simultaneously.
Therefore we divided our PACS observations into two sub-samples:
(i) a ultradeep sub-sample containing sources in the GOODS-S-ultradeep field; and (ii) a deep sub-sample containing sources in the GOODS-S-deep and GOODS-N fields (these two fields having the same depth).
The ultradeep sub-sample covers an effective area of $47\,$arcmin$^{2}$, while the deep sub-sample covers an effective area of $327\,$arcmin$^{2}$.

The PACS 100 and 160$\,\mu$m differential number counts inferred from these two sub-samples are presented in Fig.~\ref{fig:counts} and provided in Table \ref{tab:counts}.
These differential number counts are normalized to the Euclidean slope (i.e., $dN/dS$$\,\propto\,$$S^{-2.5}$), expected for a uniform distribution of galaxies in Euclidean space.
Error bars include Poisson statistics, flux calibration uncertainties, and photometric uncertainties, as described in \citet{berta_2011}.
Filled and open symbols show flux density bins above and below the 80\% completeness limit, respectively.
Our differential number counts are compared with various estimates from the literature based on \textit{Spitzer} or \textit{Herschel} observations \citep[for more details see][]{berta_2011}.
There is good agreement between all these estimates over the flux density range in common.
However, thanks to the use of deeper PACS observations, our differential number counts extend to fainter flux densities than any previous estimates.
This extension of $\thicksim\,$$0.5\,$dex and $\thicksim\,$$0.2\,$dex at 100 and 160$\,\mu$m, respectively, allows us to resolve into individual galaxies an even larger fraction of the cosmic infrared background (CIB).
Using Figure 10 and CIB estimates of \citet[][ i.e., $12.61^{+8.31}_{-1.74}$ and $13.63^{+3.53}_{-0.85}$ nW$\,$m$^{-2}$$\,$sr$^{-1}$ at 100 and 160$\,\mu$m, respectively, by power-law fitting of PEP data]{berta_2011}, we find that in the GOODS-S-ultradeep field our PACS observations resolve $\thicksim\,$$75^{+12}_{-30}\%$ and $\thicksim\,$$75^{+7}_{-15}\%$ of the CIB at 100 and 160$\,\mu$m, respectively.
Furthermore, from the wealth of ancillary data available for the GOODS-S and -N fields (see Section \ref{sec:lf}), we can also study the redshift distribution of the PACS sources.
Faint PACS 100$\,\mu$m sources (i.e., $S_{100}\,[{\rm mJy}]$$\,<\,$$1.5$) have a median redshift of $z$$\,=\,$$1.37_{-0.52}^{+0.58}$ (errors give the interquartile range), while brighter sources (i.e., $S_{100}\,[{\rm mJy}]$$\,>\,$$1.5$) have a median redshift of $z$$\,=\,$$0.85_{-0.33}^{+0.41}$.
Similarly, faint PACS 160$\,\mu$m sources  (i.e., $S_{160}\,[{\rm mJy}]$$\,<\,$$2.5$) have a median redshift of $z$$\,=\,$$1.22_{-0.41}^{0.68}$ while brighter sources (i.e., $S_{160}\,[{\rm mJy}]$$\,>\,$$2.5$) have a median redshift of $z$$\,=\,$$0.94_{-0.38}^{+0.52}$.
At 100$\,\mu$m (160$\,\mu$m), sources at $0$$\,<\,$$z$$\,<\,$$0.5$, $0.5$$\,<\,$$z$$\,<\,$$1.0$, $1.0$$\,<\,$$z$$\,<\,$$2.0$ and $z$$\,>\,$$2.0$ contribute 24~(17)\%, 36 (33)\%, 26 (28)\% and 14 (22)\% of the CIB resolved by our PACS observations.

The PACS 100 and 160$\,\mu$m counts are finally compared to predictions from backwards or forwards evolutionary models \citep{lagache_2004,rowan_robinson_2009, valiante_2009, leborgne_2009, franceschini_2010, gruppioni_2010,lacey_2010, marsden_2011b,rahmati_2011,niemi_2012,bethermin_2012b}.
Although most of the models reproduce the observed PACS 100 and 160$\,\mu$m number counts fairly well, some of them can be ruled out because of significant discrepancies with our estimates.
In particular, we note that the models of \citet{marsden_2011b}, \citet{valiante_2009} and \citet{niemi_2012} cannot reproduce the steep faint-end slope of the observed 100 and 160$\,\mu$m counts.
From this cursory comparison, it is clear that ultradeep PACS number counts allow for a better refinement of the models and thus for better constraints on the evolution of star-forming galaxies.
\section{The infrared luminosity function\label{sec:lf}}
\begin{figure}
	\includegraphics[width=9.2cm]{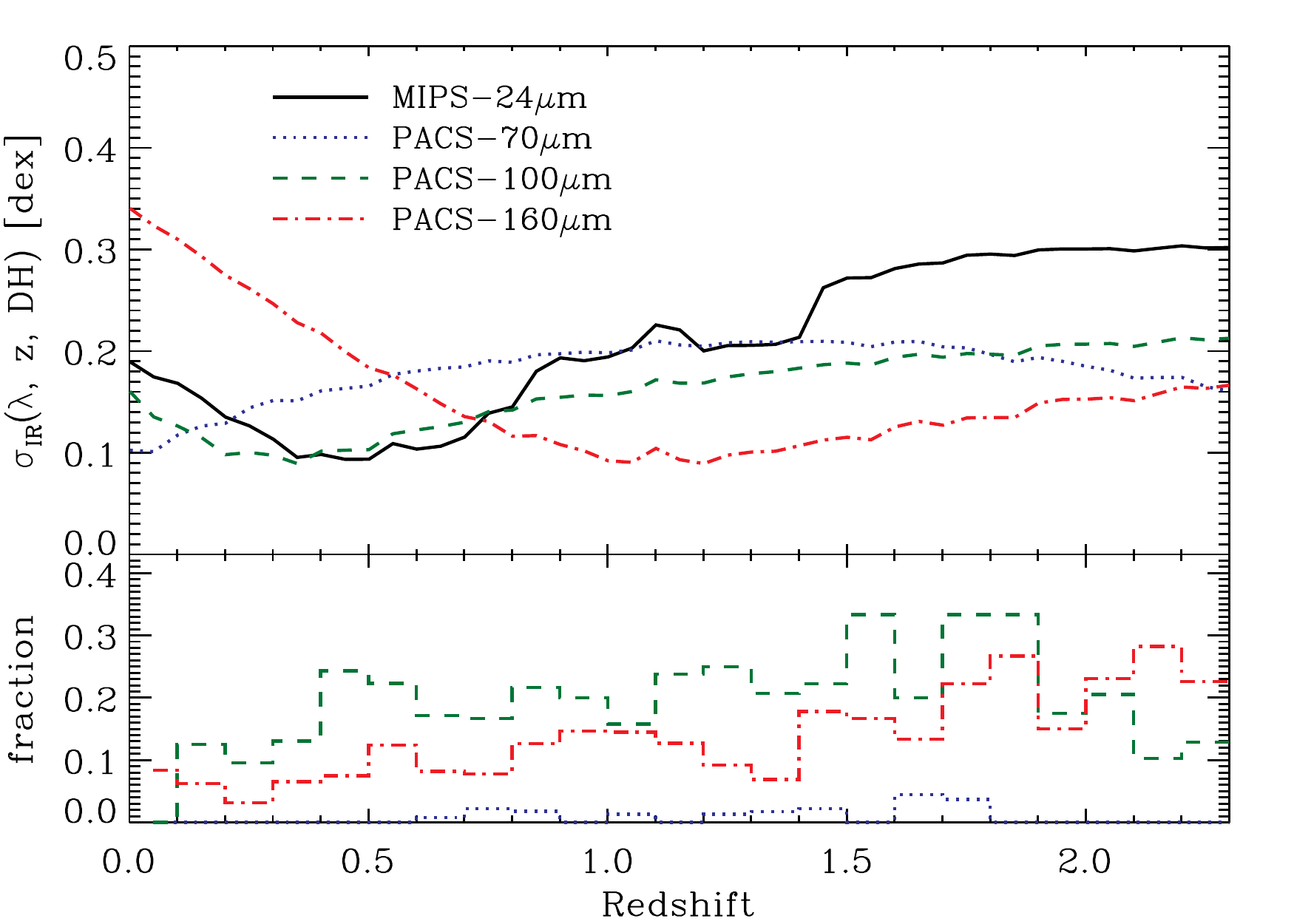}
		\caption{\label{fig:uncertainties}
		(\textit{Top panel}) Uncertainties in determining $L_{\rm IR}$ from monochromatic observations (i.e., MIPS-24$\,\mu$m, PACS-70$\,\mu$m, PACS-100$\,\mu$m or PACS-160$\,\mu$m bands) and the \citet{dale_2002} SED library.
		These uncertainties are derived by taking the standard deviation of the log($L_{\rm IR}$) distribution provided when normalizing all \citet{dale_2002} SED templates to the same monochromatic flux density ($\pm20\%$ to account for typical photometric uncertainties, i.e., S/N$\,\thicksim\,$5) at a given observed wavelength (i.e., this observed wavelength depends on the band and redshift of interest).
		(\textit{Bottom panel}) Fraction of PACS sources detected (i.e., S/N$\,>\,$3) in only one of our PACS passbands (i.e., not with a 70+100$\,\mu$m, 70+160$\,\mu$m, 100+160$\,\mu$m or 70+100+160$\,\mu$m detections) as a function of redshift.
				}
\end{figure}

Deep \textit{Herschel} observations give us the opportunity to determine the infrared luminosity function (LF) of galaxies with an unprecedented accuracy.
Indeed, far-infrared observations provide more accurate infrared luminosity estimates than mid-infrared observations from \textit{Spitzer}: for sources with multiple far-infrared detections (i.e., $\thicksim\,$70\% of sources in the PEP/GOODS-H fields; see Fig.~\ref{fig:uncertainties}), the SED-shape-$L_{\rm IR}$ degeneracy is broken and infrared luminosity estimates are only limited by photometric uncertainties; for sources with only one far-infrared detection, uncertainties on the monochromatic-to-$L_{\rm IR}$ conversion are significantly reduced compared to those provided by single mid-infrared detection (see Fig.~\ref{fig:uncertainties}).
The importance of far-infrared data increases further in cases where the mid-infrared may be contaminated by the emission from active galaxy nuclei (AGN).

Taking advantage of PACS and SPIRE far-infrared observations for several multi-wavelength fields, \citet{gruppioni_2013} derived the infrared LF of galaxies.
Thanks to the large area covered by their observations ($\thicksim\,$2.5$\,$deg$^{2}$), they were able to robustly constrain the intermediate and bright-end part of the infrared LF up to $z$$\,\thicksim\,$$2$ and the bright-end of the infrared LF up to $z$$\,\thicksim\,$$4$.
Here, we extend such study down to unprecedented depths using deeper PACS observations ($\thicksim\,$2 deeper in term of flux density than those of Gruppioni et al.).
From these deeper observations, we are able to better constrain the faint-end and intermediate part of the infrared LF up to $z$$\,\thicksim\,$$2$ and therefore to obtain better constraints on its redshift evolution.

The GOODS-N and -S fields benefit from an extensive multi-wavelength coverage necessary to obtain redshift information for the PACS sources.
In the GOODS-N field, we use a $z+K$ bands selected PSF-matched catalogue created for the PEP survey\footnote{publicly available at \texttt{http://www.mpe.mpg.de/ir/Research/PEP/}} \citep{berta_2010,berta_2011}, with photometry in 16 bands and a collection of spectroscopic redshifts \citep[mainly from][]{barger_2008}.
In the GOODS-S field, we use the GOODS-MUSIC $z+K$ bands selected PSF-matched catalogue \citep{grazian_2006,santini_2009}, with photometry in 15 bands and a collection of spectroscopic redshifts.
These multi-wavelength catalogues also include photometric redshift estimates computed using all optical and near-infrared data available \citep[see][]{berta_2011,grazian_2006,santini_2009}.
The quality of these photometric redshifts was tested by comparing them with the redshifts of spectroscopically confirmed galaxies.
The high quality of these photometric redshifts is characterized by a relatively small scatter in $\Delta z/(1+z)$ of 0.04 and 0.06 in the GOODS-N and -S fields, respectively \citep{berta_2011,santini_2009}.

Multi-wavelength catalogues were cross-matched with our MIPS-PACS catalogues using their IRAC positions and a matching radius of 0.8$\arcsec$ (i.e., approximately the HWHM of the IRAC-3.6$\,\mu$m observations). 
In case of multiple associations ($\thicksim\,$$10\%$), we select the closest optical counterparts.
In the GOODS-N and GOODS-S fields, the common area covered by these catalogues is $164\,$arcmin$^2$ and $184\,$arcmin$^2$, respectively.
In these regions, 97\% and 96\% of the PACS sources have a multi-wavelength counterpart.
Among those sources, 64\% and 61\% have a spectroscopic redshift in the GOODS-N and -S fields, respectively.
The rest of the sources has photometric redshift estimates.

\begin{figure}
	\includegraphics[width=9.4cm]{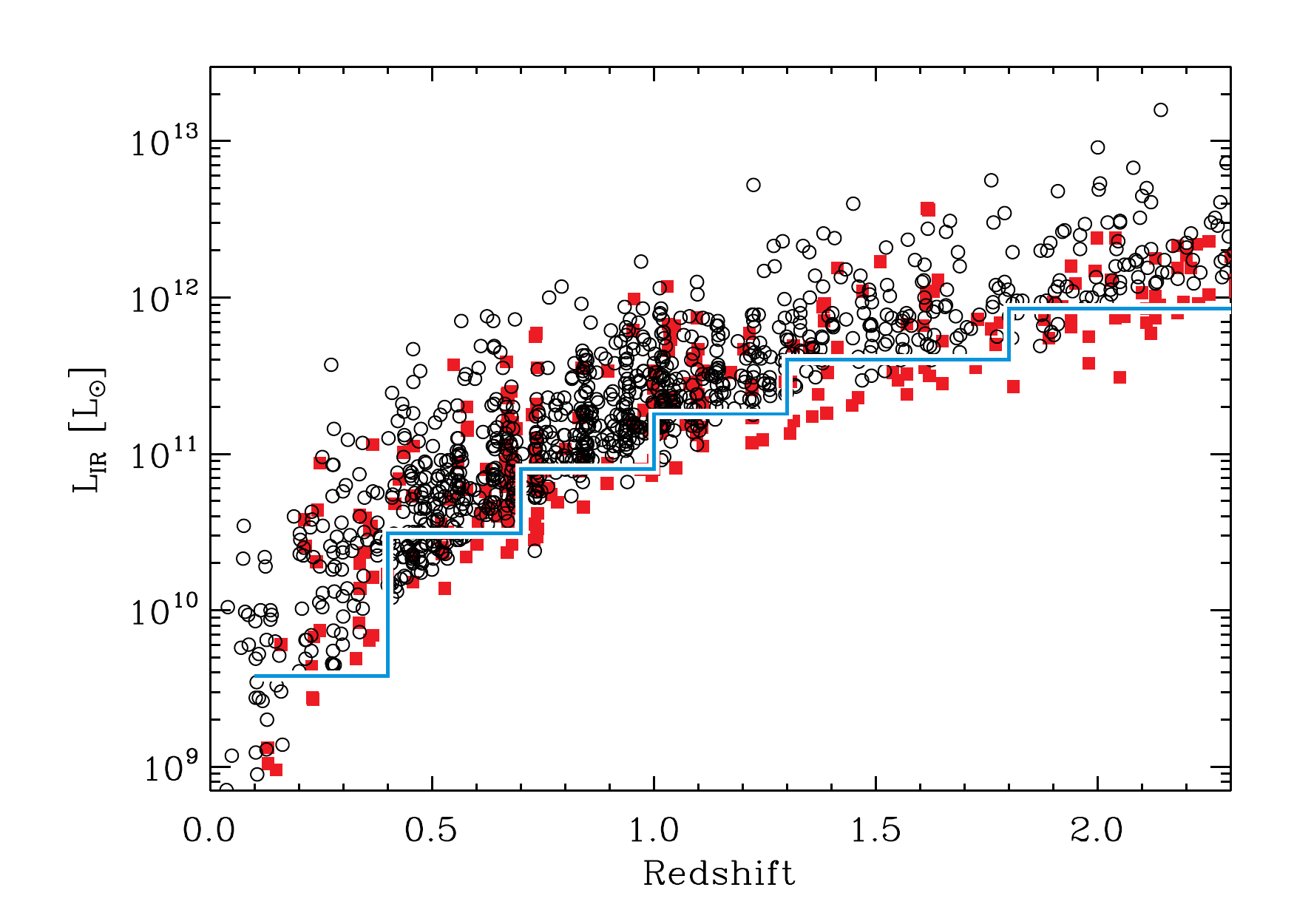}
		\caption{\label{fig:lir_z}
		Infrared luminosities as a function of redshift for \textit{Herschel} sources situated in the GOODS-S-ultradeep field (red squares) and situated in the GOODS-S-deep and GOODS-N fields (open circles).
		The blue line on a white background shows the $L_{\rm IR}$ limits above which the GOODS-S-ultradeep sample could be considered as a unbiased sample of the star-forming galaxy population at this redshift.
		These limits are inferred in Fig.~\ref{fig:lf biais}, and ``steps'' correspond to the redshift bins used for the LF analysis.
		For the GOODS-S-deep and GOODS-N fields, these limits are shifted by $\thicksim\,$$0.2\,$dex towards higher $L_{\rm IR}$.
				}
\end{figure}

The total infrared luminosities (8-1000$\,\mu$m) of PACS sources with redshift estimates were inferred by fitting their far-infrared flux densities (i.e., 70, 100 and 160$\,\mu$m) with the SED template library of \citet[][]{dale_2002}, i.e., leaving the normalization of each SED template as a free parameter.
For sources with only one far-infrared detection, infrared luminosities were defined as the geometric mean across the range of infrared luminosities given by all SED templates.
As shown on Fig.~\ref{fig:uncertainties}, even in this case of only one far-infrared detection, the uncertainties in the inferred infrared luminosities are small, i.e., better than $\thicksim\,$0.2$\,$dex.
To use the MIPS-24$\,\mu$m flux density of the PACS sources during our fitting procedure does not change our results.
Indeed, the $L_{\rm IR}^{\rm PACS+MIPS}/L_{\rm IR}^{\rm PACS} $ distribution has a mean value of 1 and a dispersion of 3\%.
Naturally,  for sources with only one far-infrared detection, the addition of the MIPS-24$\,\mu$m flux densities allow us to break the SED-shape-$L_{\rm IR}$ degeneracy and thus to reduce uncertainties on our infrared luminosity estimates.
However, because the fraction of PACS sources with single far-infrared detection is low (i.e., $\thicksim\,$30\%; see Fig.~\ref{fig:uncertainties}) and because the MIPS-24$\,\mu$m flux density might be affected by emission from an AGN, we decided not to use the MIPS-24$\,\mu$m flux densities to derive $L_{\rm IR}$.
We note that using the SED template library of \citet{chary_2001} instead of that of \citet{dale_2002} to derive $L_{\rm IR}$ (again leaving the normalisation as a free parameter), has no impact on our results.
Indeed, the $L_{\rm IR}^{\rm DH}/L_{\rm IR}^{\rm CE01} $ distribution has a mean value of 1 and a dispersion of 13\%.
Figure \ref{fig:lir_z} illustrates the detection limits of our PACS samples in term of $L_{\rm IR}$ as a function of redshift.

Uncertainties in determining $L_{\rm IR}$ from monochromatic observations (i.e., Fig.~\ref{fig:uncertainties}) are inferred assuming that the \citet{dale_2002} SED library reproduces both the full range of models appropriated for star-forming galaxies and the correct distribution of SEDs within this population.
Because neither of these assumptions are necessarily true, the \textit{absolute} values of these uncertainties should be taken with caution.
However, even with these limitations, uncertainties derived here are fully consistent with those inferred in \citet{elbaz_2011} by analyzing the mid-to-far-infrared SEDs (i.e., based on \textit{Spitzer}, PACS and SPIRE observations) of a large sample of star-forming galaxies at $0$$\,<\,$$z$$\,<\,$$2$.

The infrared (IR) LFs were derived using the standard $1/V_{\rm max}$ method \citep{schmidt_1968}.
The comoving volume of a given source is defined as $V_{\rm max}$$\,=\,$$V_{\rm z_{\rm max}}-V_{\rm z_{\rm min}}$ where $z_{\rm min}$ is the lower limit of the redshift bin being used, and $z_{\rm max}$ is the maximum redshift at which the source could be seen given the flux density limits of our observations, with a maximum value corresponding to the upper limit of the redshift bin.
Here $z_{\rm max}$ was defined by redshifting the \citet{dale_2002} template fitted to the far-infrared flux densities of our sources until it fell below the detection limits of our PACS observations, or until $z_{\rm max}$ is greater than the upper limit of our redshift bin.
For each luminosity bin, the LF is then given by
\begin{equation}
\phi=\frac{1}{\Delta L}\sum{\frac{1}{V_{{\rm max},i}\times w_{i}}},
\end{equation}
where $V_{{\rm max},i}$ is the comoving volume over which the $i$th galaxy of the luminosity bin could be observed, $\Delta L$ is the size of the luminosity bin, and $w_{i}$ is the completeness correction factor of the $i$th galaxy.
The value of $w_{i}$ is given by the MC simulations (Sect.~\ref{subsec:noise}) and depends on the flux densities of each source: $w_{i}$ equals 1 for bright PACS sources and decreases at faint flux densities due to the incompleteness of our PACS catalogues.
For sources with multiple PACS detections it is defined as the maximum of the three passbands, i.e., $w_{i}={\rm max}(w_{i}^{70},w_{i}^{100},w_{i}^{160})$.
Because we limit our LFs to infrared luminosities where $w_{i}>0.5$, none of our results strongly depend on these corrections.
\begin{figure}
	\includegraphics[width=9.4cm]{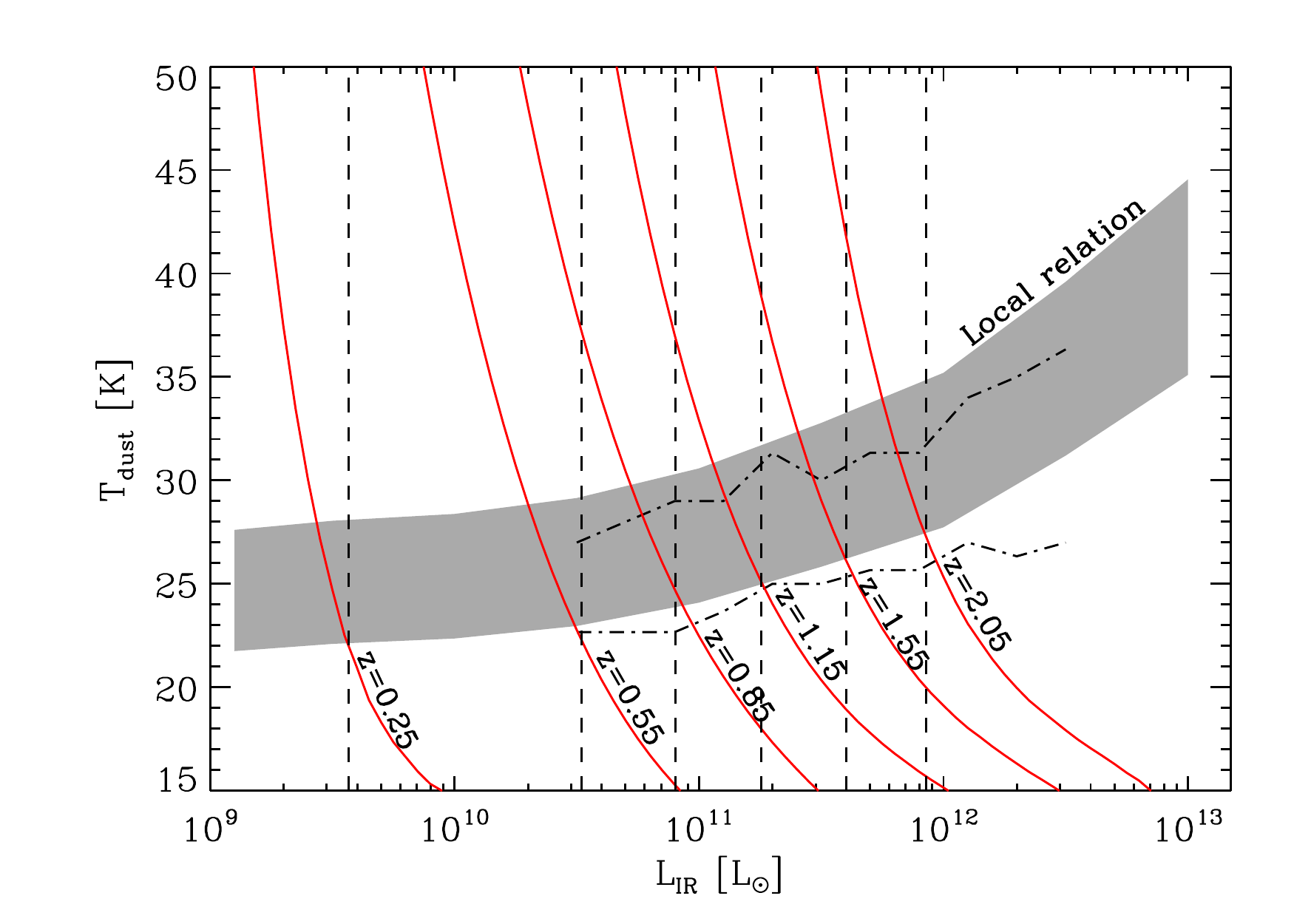}
		\caption{\label{fig:lf biais}
		Selection limits introduced in the $T_{\rm dust}$$-$$L_{\rm IR}$ parameter space by our deepest PACS observations, i.e., in GOODS-S-ultradeep.
		Red continuous lines are these selection limits at different redshifts using our PACS 5$\sigma$ detection limits.
		Each line corresponds to the central redshift of our redshift bins, i.e., $z$$\,=\,$$0.25$ (further left), $z$$\,=\,$$0.55$, $z$$\,=\,$$0.85$, $z$$\,=\,$$1.15$, $z$$\,=\,$$1.55$ and $z$$\,=\,$$2.05$ (further right).
		The shaded area shows the local $T_{\rm dust}$$-$$L_{\rm IR}$ relation found by \citet{chapman_2003}, linearly extrapolated to $10^{13}\,$L$_{\odot}$.
		Dot-dashed lines show the $T_{\rm dust}$$-$$L_{\rm IR}$ relation inferred by \citet{symeonidis_2013} using a sample of high-redshift (i.e., $0.2$$\,<\,$$z$$\,<\,$$1.2$) \textit{Herschel}-detected galaxies.
		Dashed black lines show, for each redshift, the lowest infrared luminosities probed by our ultradeep PACS observations without any dust temperature biases and yet populated by star-forming galaxies.
		}
\end{figure}

The minimum infrared luminosities that can be probed by the IR LFs in each of our redshift bins (i.e., $0.1$$\,<\,$$z$$\,<\,$$0.4$, $0.4$$\,<\,$$z$$\,<\,$$0.7$, $0.7$$\,<\,$$z$$\,<\,$$1.0$, $1.0$$\,<\,$$z$$\,<\,$$1.3$, $1.3$$\,<\,$$z$$\,<\,$$1.8$ and $1.8$$\,<\,$$z$$\,<\,$$2.3$) depend on the depth of our observations: at a given infrared luminosity, a large fraction of the galaxies has to be observable (i.e., small completeness correction; $w_i$$\,>\,$$0.5$) over at least half of our redshift bin (i.e., small $V_{\rm max}$ correction; $z_{\rm max}>z_{\rm bin}^{\rm lower}+(z_{\rm bin}^{\rm upper}-z_{\rm bin}^{\rm lower})/2$).
From this definition, it is clear that observations with different depths cannot be treated simultaneously.
Therefore, as for the number counts, we first divided our PACS sources into two sub-samples: (i) a ultradeep sub-sample from the GOODS-S-ultradeep field; and (ii) a deep sub-sample from the GOODS-S-deep and GOODS-N fields.
Then, for each of our redshift bins, we estimated the lowest infrared luminosities probed by these sub-samples.

At a given redshift, the minimum infrared luminosity observable by PACS depends on the dust colour temperature of galaxies: at a given $L_{\rm IR}$, galaxies with warmer dust have brighter PACS flux densities.
Therefore, we estimated, for each point of the $T_{\rm dust}$$-$$L_{\rm IR}$ parameter space its detectability by our PACS observations using the SED templates of \citet{dale_2002}, i.e, a dust temperature was assigned to each Dale \& Helou template in a manner that is consistent with procedures used in \citet{chapman_2003} to derive the $L_{\rm IR}-T_{\rm dust}$ relation.
Then, assuming that the local $T_{\rm dust}$$-$$L_{\rm IR}$ correlation of \citet{chapman_2003} remains the same at high-redshift \citep[see also][]{chapin_2009b}, we defined the minimum $L_{\rm IR}$ of our IR LFs as the minimum $L_{\rm IR}$ observable by PACS without any $T_{\rm dust}$ biases and populated by star-forming galaxies (i.e., within the $T_{\rm dust}$$-$$L_{\rm IR}$ correlation).
In this analysis we used the 5$\sigma$ limits of our PACS observations (i.e., where $w_i$$\,>\,$$0.5$) and the central redshift of our redshift bins (i.e, $z_{\rm max}>z_{\rm bin}^{\rm lower}+(z_{\rm bin}^{\rm upper}-z_{\rm bin}^{\rm lower})/2$).
Figure \ref{fig:lf biais} presents the results of this analysis for our ultradeep sub-sample.
The minimum $L_{\rm IR}$ observable by PACS strongly increases with increasing redshift, and sources with hotter dust temperatures can be detected down to fainter infrared luminosities (red lines of Fig.~\ref{fig:lf biais}).
Combined with the expected positions of galaxies in the $T_{\rm dust}$$-$$L_{\rm IR}$ parameter space, we can define the minimum $L_{\rm IR}$ of our IR LFs (dashed lines of Fig.~\ref{fig:lf biais}).
The same analysis was performed for our deep sub-sample. 
Results are very similar but systematically shifted by $\thicksim\,$$0.2\,$dex towards higher $L_{\rm IR}$.
We note that the modest evolution of the $T_{\rm dust}$$-$$L_{\rm IR}$ relation with redshift (see dot-dashed lines of Fig.~\ref{fig:lf biais}\footnote{To be consistent with \citet{chapman_2003}, the $T_{\rm dust}$$-$$L_{\rm IR}$ relation of \citet{symeonidis_2013} is obtained from their $f60/f100$$-$$L_{\rm IR}$ relation converted using the $f60/f100$ and assigned dust temperature of each Dale \& Helou template.}; Symeonidis et al. \citeyear{symeonidis_2013}; Hwang et al. \citeyear{hwang_2010}; Magnelli et al. in prep.) only has a minor effect on these limits, i.e., they are shifted by $0.05$$\,-\,$$0.1\,$dex towards higher luminosities.
We also note that using the dust temperature of galaxies situated on the main sequence (MS) of the SFR-$M_{\ast}$ plane, we would derive similar limits to those obtained using the $T_{\rm dust}$$-$$L_{\rm IR}$ relation.
Indeed, MS galaxies have $T_{\rm dust}$ of $27\pm3$, $28\pm3$, $29\pm3$, $30\pm4$, $32\pm4$ and $34\pm5\,$K at $z$$\,=\,$$0.25$, $z$$\,=\,$$0.55$, $z$$\,=\,$$0.85$, $z$$\,=\,$$1.15$, $z$$\,=\,$$1.55$ and $z$$\,=\,$$2.05$ \citep[Magnelli et al. in prep; see also][]{magdis_2012}.

Uncertainties in the IR LFs depend on the number of sources per luminosity bin, on the photometric redshift errors and on the infrared luminosity errors.
For the plot, they were defined as the quadratic sum of the Poissonian errors ($\propto\,$$1/\sqrt{N}$) and errors computed from MC simulations which account for both photometric redshift and infrared luminosity uncertainties.
The methodology of these MC simulations is described in \citet{magnelli_2009,magnelli_2011a}.\\
 \begin{figure*}
	 \begin{center}
	\includegraphics[width=18.5cm]{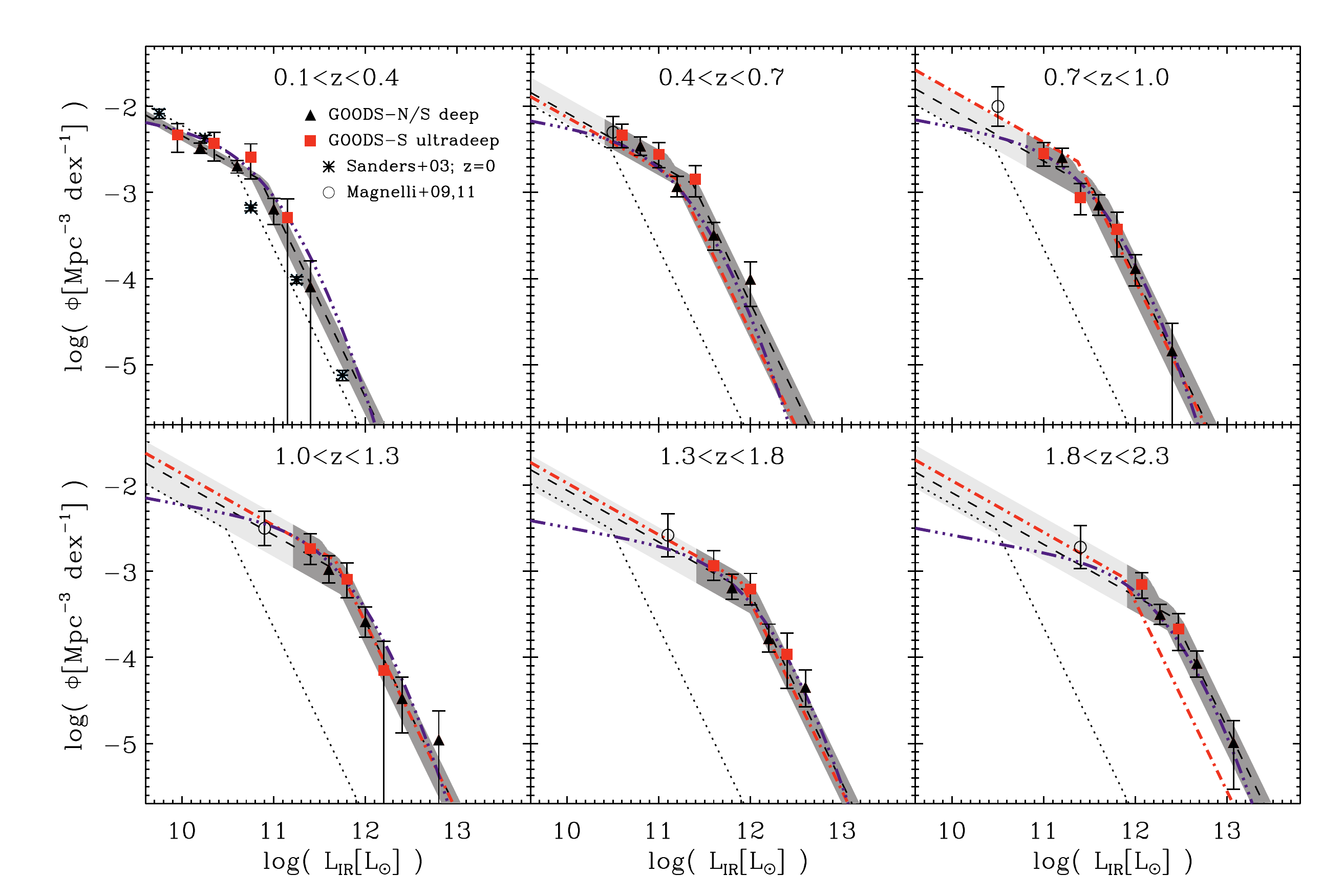}
	\end{center}
	\caption{\label{fig:lf}
	Infrared luminosity functions estimated in six redshift bins with the $1/V_{\rm max}$ method.
	Red squares and black triangles show results from our ultradeep (i.e., GOODS-S-ultradeep) and deep (i.e., GOODS-S-deep and GOODS-N) samples, respectively.
	Dashed lines represent the best fits to these data points with a double power-law function with fixed slopes. 
	The shaded areas span all the solutions of these fits which are compatible, within 1$\sigma$, with our data points: the dark shaded parts of these areas highlight the luminosity ranges directly constrained by our PACS observations, while the light shaded parts highlight the luminosity ranges where our constraints rely upon extrapolations based on $\phi_{\rm knee}$ and $L_{\rm knee}$ and a faint-end slope fixed at its $z$$\,\thicksim\,$$0$ value.
	Asterisks show the local reference, taken from \citet{sanders_2003}, and the dotted line is the best fitted to these data points with our double power-law function with fixed slopes.
	Red dot-dashed lines are results from \citet{magnelli_2009,magnelli_2011a} using deep MIPS-24$\,\mu$m observations.
	\citet{magnelli_2009,magnelli_2011a} used the same double power-law function that is used here.
	To illustrate the infrared luminosity range constrained by \citet{magnelli_2009,magnelli_2011a}, we show, as open circles, their lowest infrared luminosity bins. 
	Blue triple-dot-dashed lines present results from \citet{gruppioni_2013} who used a different analytical function to fit their data points, in particular, a much shallower faint-end slope.
		}
\end{figure*}
 \begin{figure*}
	 \begin{center}
	\includegraphics[width=18.5cm]{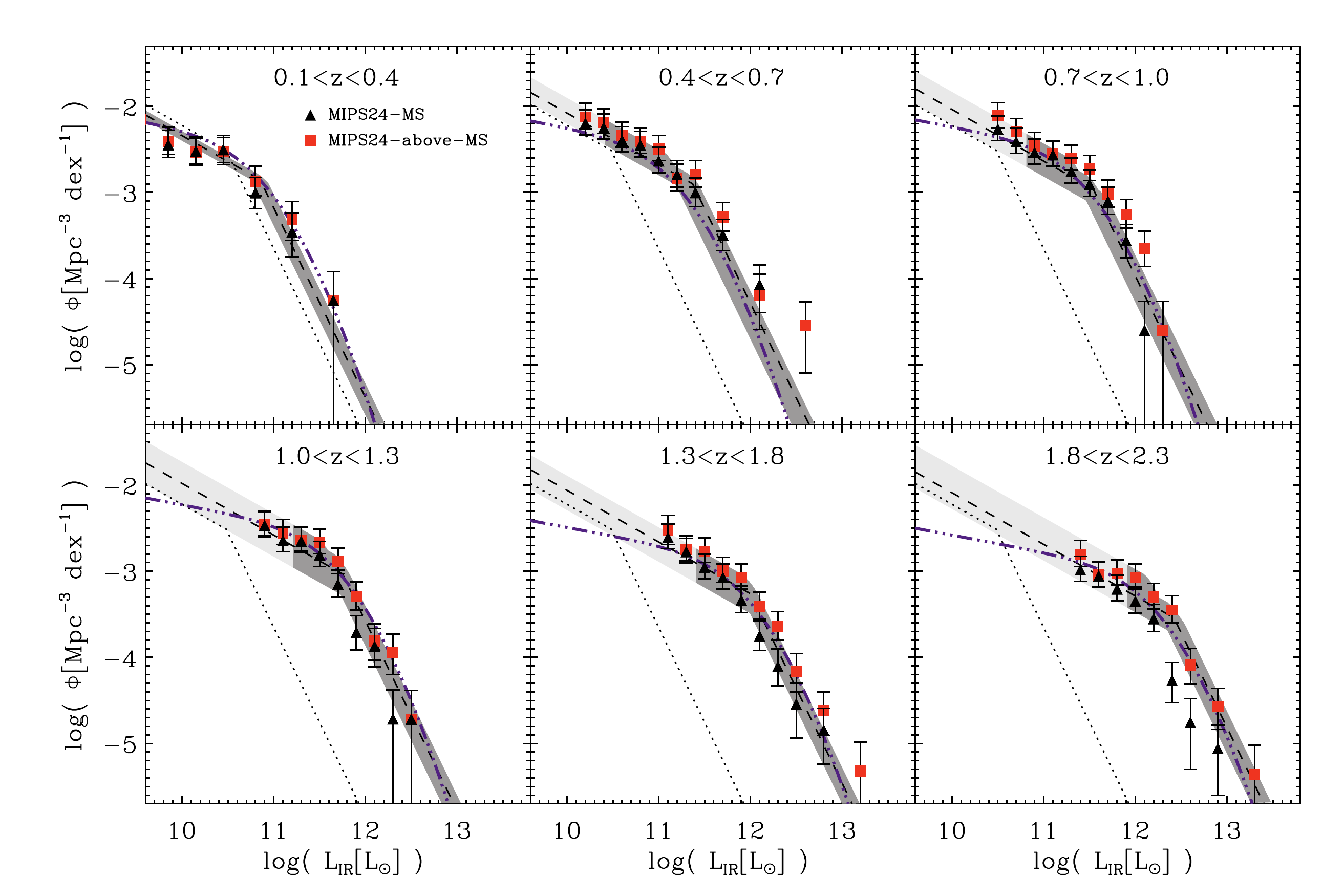}
	\end{center}
	\caption{\label{fig:lf 24}
	Comparisons between the infrared luminosity functions derived using PACS observations and those derived using MIPS observations.
	PACS-based IR LFs are shown as dark/light shaded areas and dashed lines (see also Fig.~\ref{fig:lf}).
	MIPS-based IR LFs are derived using MS-based 24$\,\mu$m-to-$L_{\rm IR}$ conversion factors (black triangles) and above-MS-based 24$\,\mu$m-to-$L_{\rm IR}$ conversion factors (red squares; see text for more details).
	The dotted lines show the local reference taken from \citet{sanders_2003}.
	Blue triple-dot-dashed lines present results from \citet{gruppioni_2013}.
	At low luminosities, the agreement between PACS-based and MIPS-based IR LFs confirms that PACS extrapolations towards lower infrared luminosities (i.e., light shaded areas) are reliable, at least down to the infrared luminosities probed by \textit{Spitzer}.
		}
\end{figure*}

\begin{figure*}
 	\includegraphics[width=9.0cm]{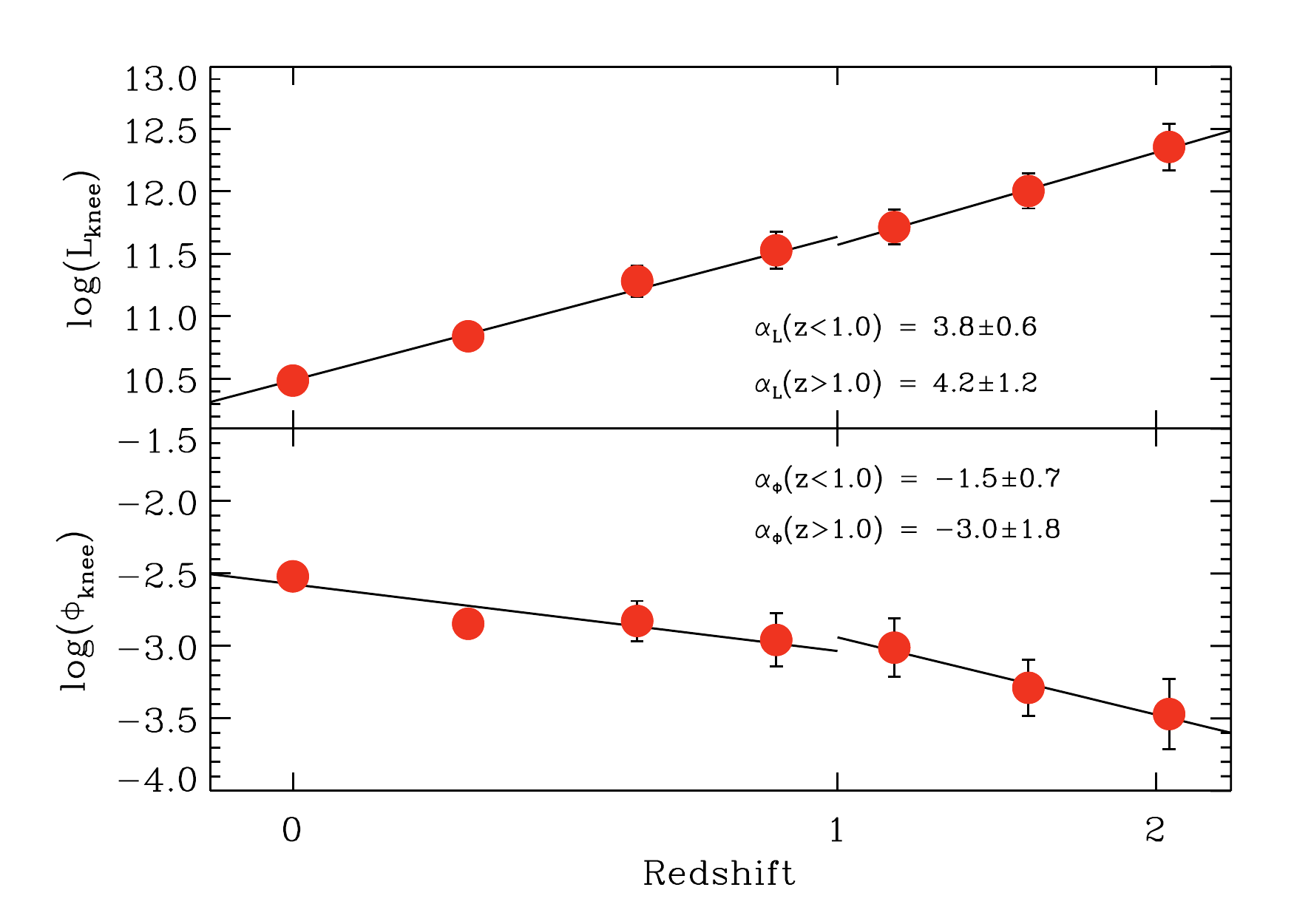}
		\includegraphics[width=9.0cm]{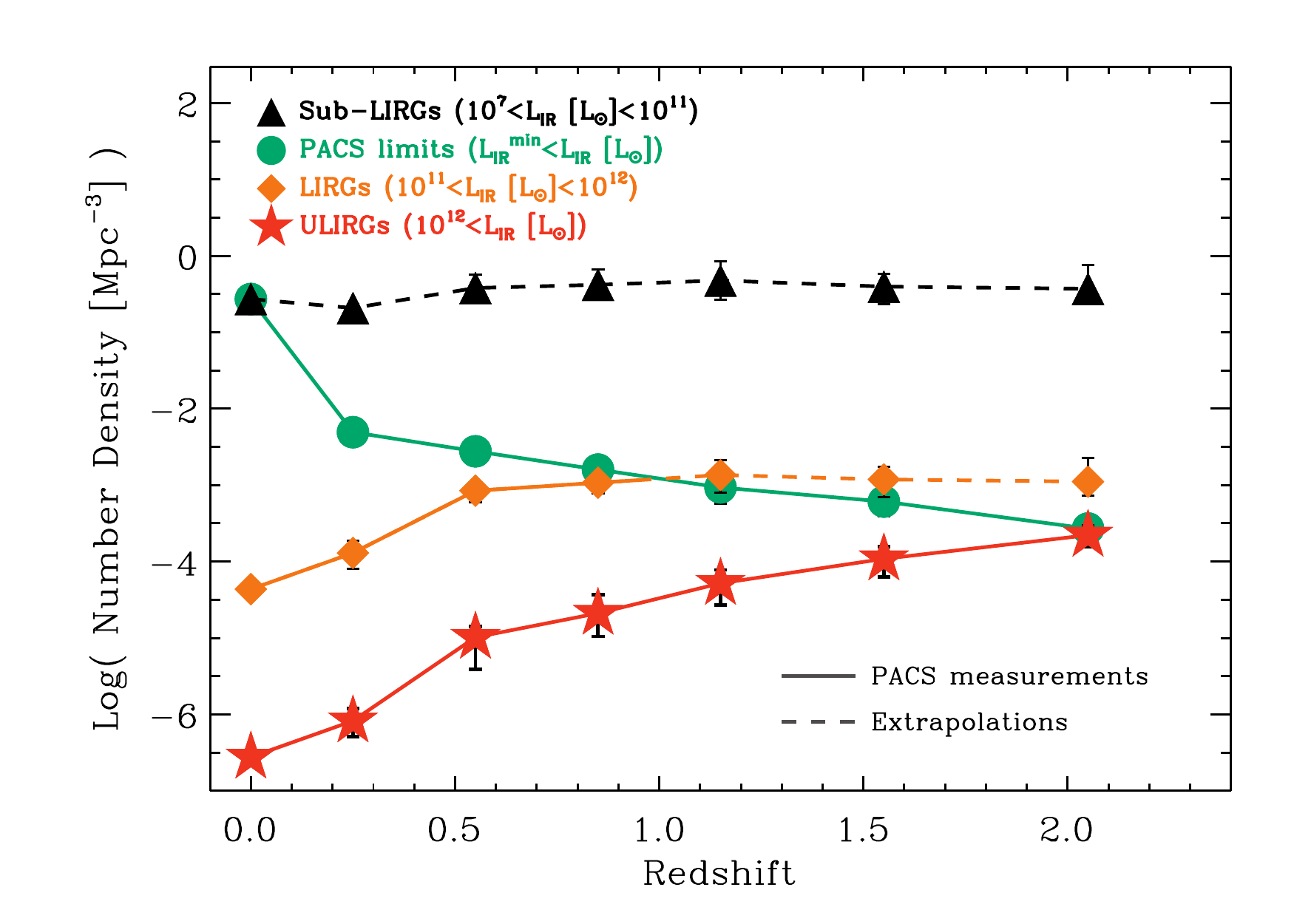} \\
	\includegraphics[width=9.5cm]{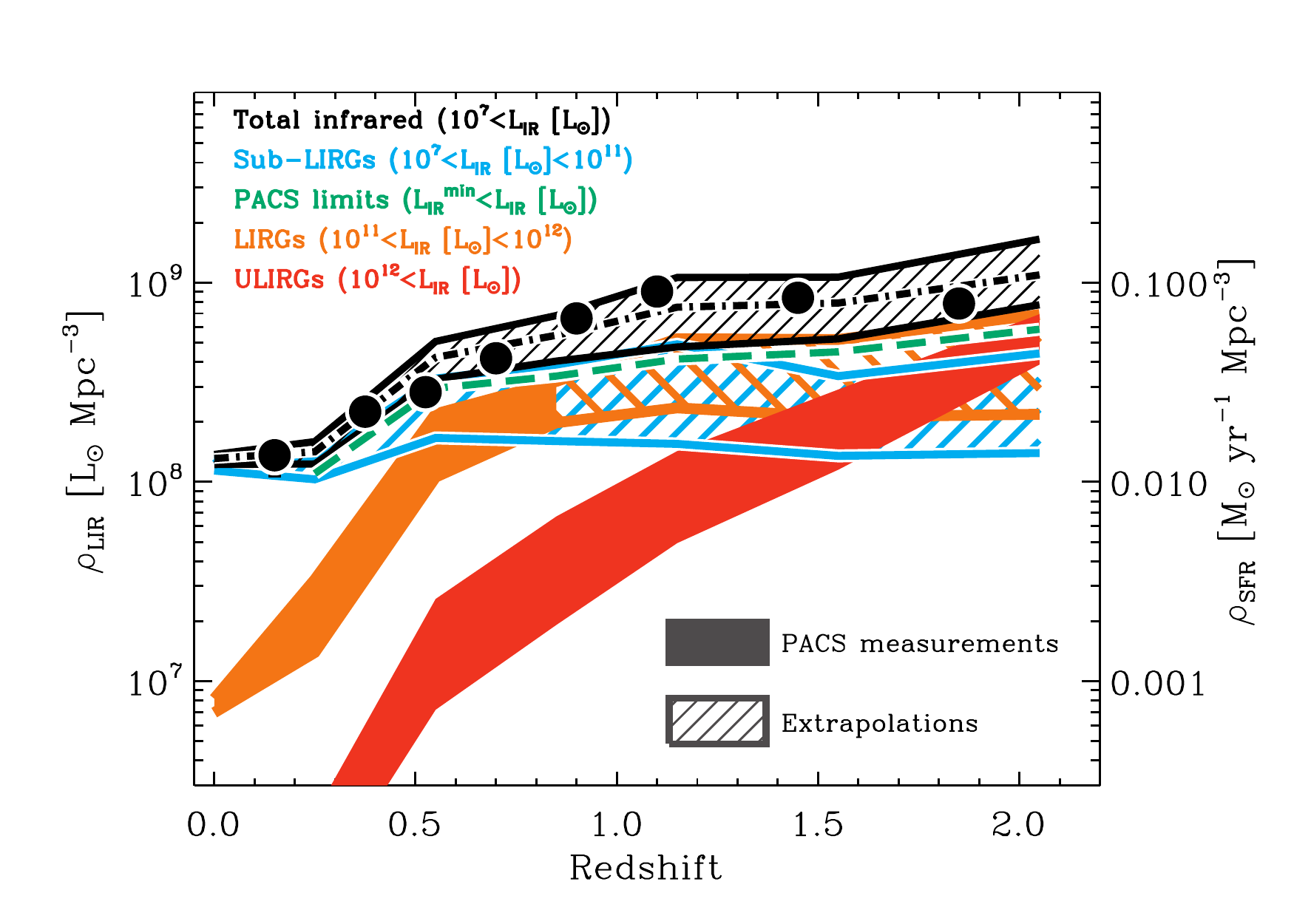}
	\includegraphics[width=9.5cm]{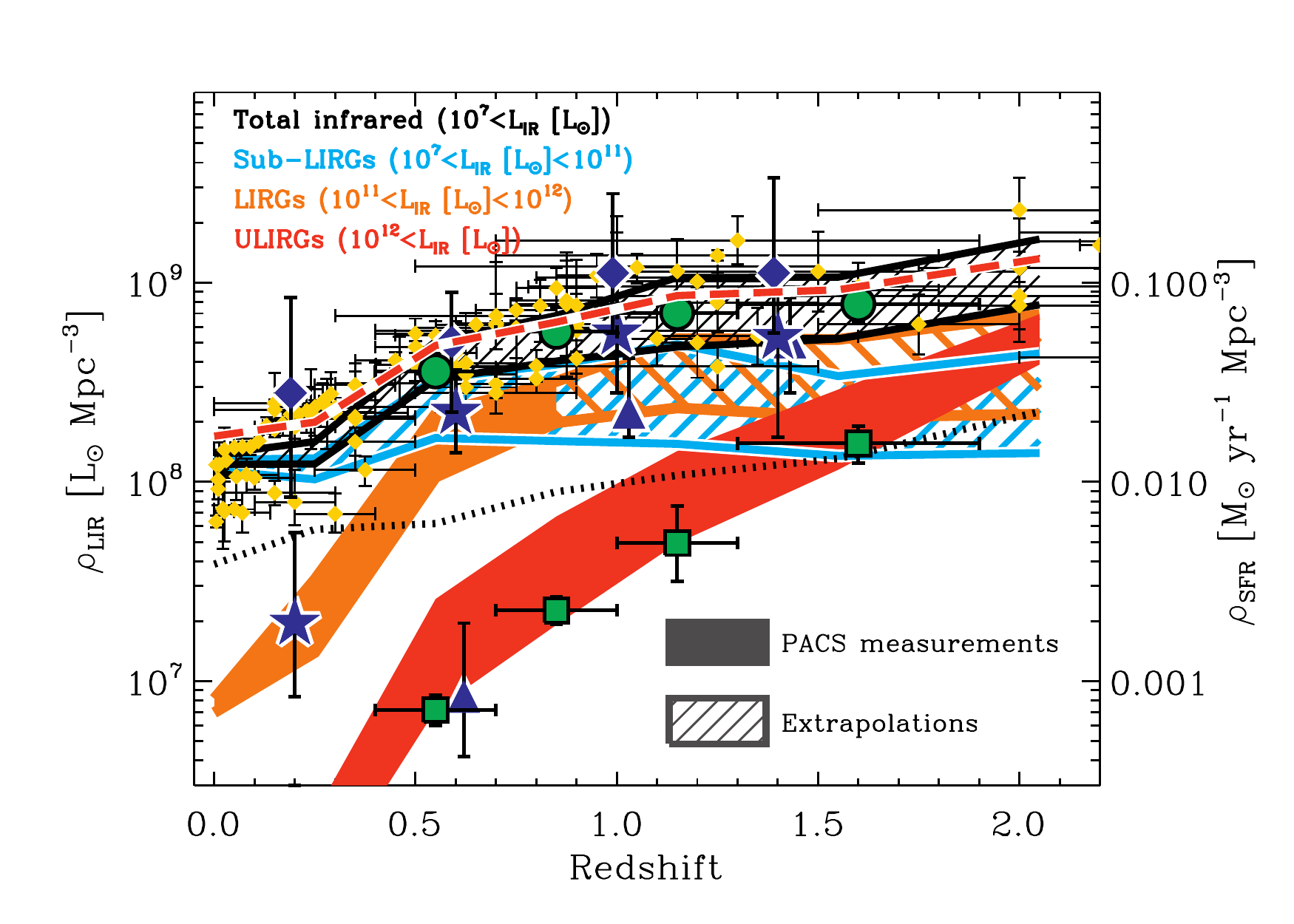}
	\caption{\label{fig:lf side}
	(\textit{Top left panel}) Evolution of $\phi_{\rm knee}$ and $L_{\rm knee}$ as a function of redshift.
	(\textit{Top right panel}) Evolution of the comoving number density of ``faint'' galaxies (i.e., $10^7\,$L$_{\odot}$$\,<\,$$L_{\rm IR}$$\,<\,$$10^{11}\,$L$_{\odot}$; black triangles), LIRGs ($10^{11}\,$L$_{\odot}$$\,<\,$$L_{\rm IR}$$\,<\,$$10^{12}\,$L$_{\odot}$; orange diamonds) and ULIRGs ($L_{\rm IR}$$\,>\,$$10^{12}\,$L$_{\odot}$; red stars).
	Green circles show the number density of galaxies which are above the PACS detection limit of our ultradeep sub-sample (i.e., $>\,$$L_{\rm IR}^{\rm min}$, defined in Fig.~\ref{fig:lf biais}).
	The local reference is taken from \citet{sanders_2003}.
	(\textit{Bottom left panel}) Evolution of the total comoving IR energy density (black area) and the relative contribution of the ``faint'' galaxies (light blue area), LIRGs (orange area) and ULIRGs (red area).
	The black dot-dashed line shows the best fit of the total comoving IR energy density, i.e., defined by integrating the IR LFs best-fitting our data points (see dashed lines in Fig.~\ref{fig:lf}).
	The green dashed line shows the best fit of the comoving IR energy density of galaxies which are above the PACS detection limit of our ultradeep sub-sample (i.e., $>\,$$L_{\rm IR}^{\rm min}$).
	Black circles represent the total comoving IR energy density inferred by \citet{gruppioni_2013}.
	The right axis of the panel shows the evolution of the obscured SFR density assuming that SFR$\,$[M$_{\odot}\,$yr$^{-1}$]$=10^{-10}$$\,\times\,$$ L_{\rm IR}\,$[L$_{\odot}$] for a \citet{chabrier_2003} IMF.
	(\textit{Bottom right panel}) Evolution of the comoving IR energy density. 
	Areas are the same as in the bottom left panel.
	The dotted lines represents the unobscured SFR density of the Universe \citep[i.e., not corrected for extinction;][]{cucciati_2012}.
	The red dashed line on a white background shows the total SFR density of the Universe, defined as the sum of the obscured and unobscured SFR densities.
	Yellow diamonds are a compilation of SFR density estimates from \citet{hopkins_2006}.
	Green circles and squares are the total SFR density of the Universe and the contribution of ULIRGs estimated by \citet{murphy_2011} using deep MIPS-24 and -70$\,\mu$m observations.
	Dark blue diamonds, stars and triangles are estimates from \citet{casey_2012} for all galaxies, LIRGs and ULIRGs, respectively, using SPIRE (i.e., submillimetre) observations.
	One has to keep in mind that even with our ultradeep PACS data we only constrain the IR LFs of galaxies down to $L_{\rm IR}$$\,=\,$$10^{11}\,$L$_{\odot}$ at $z$$\,\thicksim\,$$1$ and $L_{\rm IR}$$\,=\,$$10^{12}\,$L$_{\odot}$ at $z$$\,\thicksim\,$$2$, respectively.
Therefore, number densities and IR LDs of galaxies below these limits rely upon extrapolations based on $\phi_{\rm knee}$ and $L_{\rm knee}$ and a faint-end slope fixed at its $z$$\,\thicksim\,$$0$ value.
The redshift and luminosity ranges for which the inferred number densities and IR LDs rely upon these extrapolations are highlighted by dashed lines and striped regions, respectively.
	We note that these extrapolations are nevertheless corroborated by direct constraints based on deep MIPS-24$\,\mu$m observations (see Fig.~\ref{fig:lf 24}).
		}
\end{figure*}

Figure \ref{fig:lf} represents the IR LFs derived in six redshift bins using our ultradeep and deep PACS observations (Tables \ref{tab:LF 1} and \ref{tab:LF 2}).
We fit the IR LFs with a double power-law function similar to that used to fit the local IR LF which we also plot for reference \citep[][$\phi\propto L^{-0.6}$ for log($L/$L$_{\odot})$$\,<\,$$L_{\rm knee}$ and $\phi\propto L^{-2.2}$ for log($L/$L$_{\odot})$$\,>\,$$L_{\rm knee}$]{sanders_2003}.
In this fitting procedure, the normalization (i.e., $\phi_{\rm knee}$) and the transition luminosity (i.e., $L_{\rm knee}$) of the double power-law function are left as free parameters.
The shaded areas of Fig.~\ref{fig:lf}  present the solutions compatible with the data within 1$\sigma$.
The evolution of $\phi_{\rm knee}$ and $L_{\rm knee}$ with redshift is presented in the upper left panel of Fig.~\ref{fig:lf side} and given in Table \ref{tab:fit parameter}.

We compare our IR LFs with estimates made by \citet{magnelli_2009,magnelli_2011a} using deep MIPS-24$\,\mu$m observations and \textit{Herschel}-based estimates from \citet{gruppioni_2013} that use shallower PACS observations covering a wider effective area.
At z$\,<\,$1.8, we find good agreement with the \textit{Spitzer} analysis of \citet{magnelli_2009,magnelli_2011a}.
In particular, we observe good agreement, within the uncertainties, between our low luminosity extrapolations (based on $\phi_{\rm knee}$ and $L_{\rm knee}$ and a faint-end slope fixed at its $z$$\,\thicksim\,$$0$ value) and direct constraints obtained by \citet[][empty circles of Fig.~\ref{fig:lf}]{magnelli_2009,magnelli_2011a}.
This agreement shows that even though our deepest PACS data do not allow us to probe luminosities far below the ``knee'' of the IR LFs, we obtain accurate low luminosity extrapolations, at least down to the infrared luminosities probed by \textit{Spitzer}.
The most significant difference between the present results and those of \citet{magnelli_2009,magnelli_2011a} is observed at $z\thicksim2$ and at the highest infrared luminosities ($L_{\rm IR}$$\,>\,$$10^{12}\,$L$_{\odot}$).
There, our new IR LF has a higher normalization.
Because, the MIPS-24$\,\mu$m band can be affected by a significant contribution from an AGN, \citet{magnelli_2009,magnelli_2011a} excluded from their sample all X-ray AGNs, i.e., sources detected in the GOODS-S/N \textit{Chandra} catalogues \citep{alexander_2003,lehmer_2005} with either $L_{\rm X}$[0.5$-$8.0$\,$keV]$\,>\,$3$\,\times\,$10$^{42}\,$erg$\,$s$^{-1}$ or a hardness ratio greater than 0.8 \citep{bauer_2004}.
We excluded in the same manner all X-ray AGNs from our sample and found that these exclusions could not reconcile our IR LFs at $z$$\,\thicksim\,$$2$.
The observed difference is thus likely due to the fact that at $z$$\,\thicksim\,$$2$, estimates from \citet{magnelli_2009,magnelli_2011a} are affected by large pre-\textit{Herschel} uncertainties on the 24$\,\mu$m-to-$L_{\rm IR}$ conversion factors \citep[][ and reference therein]{nordon_2010,elbaz_2010,elbaz_2011,nordon_2012}.
At this redshift, PACS estimates are more accurate.

At all redshifts we observe very good agreement between our estimates and those of \citet{gruppioni_2013}.
At high luminosities this agreement is very encouraging, because the data analysed by \citet{gruppioni_2013} sample a larger volume at shallower flux limits, and thus can more accurately measure the number densities for rare, bright sources.
The only disagreement between our results and those of \citet{gruppioni_2013} appear at very low infrared luminosities, i.e., at luminosities not probed by both studies and which depend on the analytic function used to fit the data.
\citet{gruppioni_2013} used a faint end slope shallower than that used here, i.e., $-0.2$ compared with $-0.6$.

We find an evolution of $\phi_{\rm knee}$ ($\phi_{\rm knee}$$\,=\,$$10^{-2.57\pm0.12}\times(1+z)^{-1.5\pm0.7}$ for $z$$\,<\,$$1.0$ and $\phi_{\rm knee}$$\,=\,$$10^{-2.03\pm0.72}\times(1+z)^{-3.0\pm1.8}$ for $z$$\,>\,$$1.0$) and $L_{\rm knee}$ ($L_{\rm knee}$$\,=\,$$10^{10.48\pm0.10}\times(1+z)^{3.8\pm0.6}$ for $z$$\,<\,$$1.0$ and $L_{\rm knee}$$\,=\,$$10^{10.31\pm0.47}\times(1+z)^{4.2\pm1.2}$ for $z$$\,>\,$$1.0$) in broad agreement with what was observed in \citet{magnelli_2011a} and \citet{gruppioni_2013}.
However, we note that in our study we find a stronger evolution of $L_{\rm knee}$ at $z$$\,>\,$$1.0$ than in \citet{magnelli_2011a}.\\

With our ultradeep PACS data we can only constrain the IR LFs of galaxies down to $L_{\rm IR}$$\,=\,$$10^{11}\,$L$_{\odot}$ at $z$$\,\thicksim\,$$1$ and $L_{\rm IR}$$\,=\,$$10^{12}\,$L$_{\odot}$ at $z$$\,\thicksim\,$$2$. 
Consequently, at lower infrared luminosities, our constraints only rely on extrapolations based on $\phi_{\rm knee}$ and $L_{\rm knee}$ and a faint-end slope fixed at its $z$$\,\thicksim\,$$0$ value.
To test these extrapolations, we take advantage of deep MIPS-24$\,\mu$m observations using appropriate 24$\,\mu$m-to-$L_{\rm IR}$ conversion factors.
Thanks to recent \textit{Herschel} studies, we know that the 24$\,\mu$m-to-$L_{\rm IR}$ conversion factors depend not only on the MIPS-24$\,\mu$m luminosity (as, e.g., in the Chary \& Elbaz SED library) but also on the localization of galaxies with respect to the ``main sequence'' of star formation \citep[MS; log(SFR)$=\alpha\times {\rm log}(M_{\ast})+C(z)$, where $0.5<\alpha<1.0$;][]{elbaz_2011,nordon_2012}: galaxies situated on the MS have different 24$\,\mu$m-to-$L_{\rm IR}$ conversion factors than galaxies situated above the MS \citep{elbaz_2011,nordon_2012}.
Consequently, we derive MIPS-based IR LFs using, on one hand, MS-based 24$\,\mu$m-to-$L_{\rm IR}$ conversion factors and on the other hand, above-MS-based 24$\,\mu$m-to-$L_{\rm IR}$ conversion factors \citep[][Fig.~\ref{fig:lf 24}]{elbaz_2011}.
The real IR LFs should be situated between these two estimates, but with a faint-end closer to the MS-based estimates (i.e., in a luminosity range dominated by MS galaxies) and a bright-end closer to the above-MS-based estimates (i.e., in a luminosity range dominated by above-MS galaxies).
Up to $z$$\,\thicksim\,$$1.3$, MS-based and above-MS-based estimates are in agreement: in this redshift range, MS-based and above-MS-based 24$\,\mu$m-to-$L_{\rm IR}$ conversion factors are not significantly different.
In contrast, at $z>1.3$, above-MS-based IR LFs have higher normalization than MS-based estimates: at $z>1.3$, above-MS-based 24$\,\mu$m-to-$L_{\rm IR}$ conversion factors are much larger than those for MS galaxies \citep{elbaz_2011,nordon_2012}.
At $z$$\,\thicksim\,$$2$ and at high infrared luminosities, the agreement found between PACS-based and above-MS-based estimates confirms the assumption that the bright-end of the IR LF is dominated by above-MS galaxies.
At all redshifts and at low infrared luminosities (i.e., where MS galaxies should dominate the IR LFs), MS-based estimates are in perfect agreement with PACS extrapolations towards lower infrared luminosities (i.e., dashed lines and light shaded areas).
We note that these PACS extrapolations also agree, within the error bars, with the above-MS-based estimates.
All these agreements confirm that our PACS extrapolations are reliable, at least down to the infrared luminosities probed by \textit{Spitzer}.\\

By integrating our PACS-based IR LFs we derive the evolution of the comoving number density (top right panel of Fig.~\ref{fig:lf side}) and comoving infrared luminosity density (IR LD, bottom panels of Fig.~\ref{fig:lf side} and Table \ref{tab:IR LD}) of ``faint'' galaxies (i.e., $10^7\,$L$_{\odot}$$\,<\,$$L_{\rm IR}$$\,<\,$$10^{11}\,$L$_{\odot}$), LIRGs (i.e., $10^{11}\,$L$_{\odot}$$\,<\,$$L_{\rm IR}$$\,<\,$$10^{12}\,$L$_{\odot}$) and ULIRGs (i.e., $L_{\rm IR}$$\,>\,$$10^{12}\,$L$_{\odot}$).
Here, one has to keep in mind that even with our ultradeep PACS data we can only constrain the IR LFs of galaxies down to $L_{\rm IR}$$\,=\,$$10^{11}\,$L$_{\odot}$ at $z$$\,\thicksim\,$$1$ and $L_{\rm IR}$$\,=\,$$10^{12}\,$L$_{\odot}$ at $z$$\,\thicksim\,$$2$, respectively.
Therefore, number densities and IR LDs of galaxies below these limits rely upon \textit{extrapolations} based on $\phi_{\rm knee}$ and $L_{\rm knee}$ and a faint-end slope fixed at its $z$$\,\thicksim\,$$0$ value.
\textit{The redshift and luminosity ranges for which the inferred number densities and IR LDs rely upon these extrapolations are highlighted by dashed lines and striped regions, respectively.}
Although these extrapolations seem to be corroborated by direct constraints from \textit{Spitzer} (see Fig.~\ref{fig:lf 24}), we recommend caution when interpreting values not directly constrained by our PACS observations.
We also emphasize that the LIRG and ULIRG designations are used here strictly to segregate the luminosity bins, but not to imply physical properties.
Indeed, \textit{Herschel} studies have unambiguously revealed that high-redshift (U)LIRGs do not have the same properties as their local counterparts \citep[e.g.,][]{elbaz_2011,wuyts_2011b,nordon_2012,magnelli_2012}.

We find that the number density of ULIRGs strongly evolves from $z$$\,=\,$$0$ to $z$$\,\thicksim\,$$2$, as it is multiplied by a factor $\thicksim\,$$800$.
Similarly, the number density of LIRGs significantly evolves with redshift: using direct constraints from PACS, we find that the LIRGs number density is multiplied by a factor $\thicksim\,$$25$ between $z$$\,=\,$$0$ and $z$$\,\thicksim\,$$1$, while relying on extrapolations we find that it is multiplied by a factor $\thicksim\,$$30$ between $z$$\,=\,$$0$ and $z$$\,\thicksim\,$$2$.
Naturally, the redshift evolution of the LIRGs and ULIRGs number densities translate into the redshift evolution of their IR LDs.
The IR LDs of LIRGs and ULIRGs are multiplied by a factor $\thicksim\,$130 ($\thicksim\,$1000) and $\thicksim\,$40 ($\thicksim\,$45; based on extrapolations) between $z$$\,=\,$$0$ and $z$$\,\thicksim\,$$1$ (2), respectively.
The redshift evolution of the IR LD of ULIRGs is consistent with that found by \citet{murphy_2011} using deep MIPS-24/70$\,\mu$m observations (dark blue squares in the bottom right panel of Fig.~\ref{fig:lf side}).
In contrast, at $z$$\,\thicksim\,$$2$, \citet{magnelli_2011a} found an IR LD of ULIRGs lower by a factor $\,\thicksim\,$3. 
This inconsistency reflects the discrepancies observed at $z$$\,\thicksim\,$$2$ between their and our IR LF (see Fig.~\ref{fig:lf}).

Relying upon extrapolations to low infrared luminosities, we find that the total IR LD strongly increases from $z$$\,=\,$$0$ and $z$$\,\thicksim\,$$2$, with most of the evolution happening before $z$$\,\thicksim\,$$1.2$ driven by the strong increase of the IR LD of LIRGs.
At $z$$\,\thicksim\,$$1$, LIRGs account for $50\pm26$\% of the total IR LD while ULIRGs contribute only $10\pm6$\% of it.
In contrast, at $z$$\,\thicksim\,$$2$ ULIRGs contribute $50\pm24$\% while LIRGs contribute only $30\pm20$\% of the total IR LD.

Although the total IR LDs derived here are consistent, within the uncertainties, with those from \citet{magnelli_2011a}, there are some differences.
At $z$$\,\thicksim\,$$1$, the total IR LD derived here is lower by a factor $\thicksim\,$1.3 than that of \citet{magnelli_2011a} while at $z$$\,\thicksim\,$$2$ it is higher by a factor $\thicksim\,$1.3.
More importantly, at $z$$\,\thicksim\,$$2$, the ULIRGs contribution to the total IR LD found here (i.e., $50\%\pm24$) is much higher than that found in \citet[][$\thicksim17\%$]{magnelli_2011a}.
Naturally, this finding reflects the disagreement at $z$$\,\thicksim\,$$2$ between our two IR LFs.

The total IR LDs inferred in \citet[][black circles in the bottom left panel of Fig.~\ref{fig:lf side}]{gruppioni_2013} agree with our findings.
Noticeably, their and our estimates are obtained by integrating IR LFs which significantly differ at infrared luminosities lower than those probed by \textit{Spitzer}, i.e., \citet{gruppioni_2013} used a much shallower faint-end slope (see Fig. ~\ref{fig:lf 24}).
This demonstrates that extrapolations towards very low infrared luminosities do not affect much the estimates of the total IR LDs.
Indeed, unless the faint-end slope of the IR LFs is significantly steeper than that used here, most of the total comoving IR LD is emitted by galaxies with luminosities reachable with PACS (i.e., $\thicksim\,$50\%; see black dot-dashed and green dashed lines in the bottom left panel of Fig.~\ref{fig:lf side}) and \textit{Spitzer} (i.e., $\thicksim\,$75\%).

While the IR LDs derived here for all galaxies and for LIRGs are consistent with SPIRE-based results from \citet{casey_2012}, the IR LDs of ULIRGs at $z$$\,>\,$$0.8$ derived here are lower than those inferred by  \citet{casey_2012}.
This disagreement might be due to the large spectroscopic redshift incompleteness corrections applied to the IR LFs of \citet{casey_2012} at $z$$\,>\,$$0.8$ (i.e., corrections $>\,$50\%).
This seems to be confirmed by the good agreement, at $z$$\,>\,$$0.8$, between the IR LFs of \citet[][i.e., also in agreement with our IR LFs]{gruppioni_2013} and the SPIRE-based IR LFs of Vaccari et al. (in prep).
Finally we note that the total IR LD derived here, as well as the significant contribution of ULIRGs at $z$$\,\thicksim\,$$2$, are consistent with results from \citet{murphy_2011}.\\

Assuming that the IR LD is totally produced by star formation (i.e., without any contribution from AGNs), it can be converted into the obscured SFR density of the Universe using the relation of \citet{kennicutt_1998}, scaled to a \citet{chabrier_2003} initial mass function\footnote{\citet{kennicutt_1998} adopts a Salpeter initial mass function so that we divide his normalization by $1.72$.} :
\begin{equation}
{\rm SFR\,[M_{\odot}\,yr^{-1}]=10^{-10}\,\times}\,L_{\rm IR}\,[{\rm L_{\odot}]},
\end{equation}
About 10-15\% of our PACS sources are associated with X-ray AGNs that might significantly contribute to the IR LD derived here.
However, \textit{Herschel} studies of X-ray AGNs \citep{rosario_2013,shao_2010,mullaney_2012,rosario_2012} have demonstrated that in the vast majority of cases (i.e., $>\,$94\%) the PACS flux densities are dominated by emission from the host galaxy and thus provide an uncontaminated view of their star-formation activities.
Consequently, we assume that the IR LD derived here has no significant contribution from AGNs and can be converted into the obscured SFR density of the Universe (right axis of the bottom panels of Fig.~\ref{fig:lf side}).
Combined with the unobscured SFR density of the Universe derived by \citet{cucciati_2012} using rest-frame UV observations, we can then infer the total SFR density and its evolution up to $z$$\,\thicksim\,$$2$ (dashed red line on a white background in the bottom right panel of Fig.~\ref{fig:lf side}).
The cosmic star-formation history strongly evolves from $z$$\,=\,$$0$ to $z$$\,\thicksim\,$$1$ and flattens at $z$$\,>\,$$1$.
The unobscured SFR density accounts for about $\thicksim\,$25\%, $\thicksim\,$12\% and $\thicksim\,$17\% of the total SFR density of the Universe at $z$$\,\thicksim\,$$0$, $z$$\,\thicksim\,$$1$ and $z$$\,\thicksim\,$$2$, respectively.
The contribution of the unobscured SFR density and its evolution with redshift is consistent with the redshift evolution of the mean rest-frame UV dust attenuation \citep[e.g.,][]{cucciati_2012,tresse_2007}.
Finally, we note that the cosmic star-formation history derived here is fully consistent with the combination of indicators, either obscured or corrected for dust extinction, as compiled by \citet{hopkins_2006}, \citet{seymour_2008} and \citet[][not shown in Fig.~\ref{fig:lf side}]{karim_2011}.

\section{Summary\label{sec:conclusion}}
By combining observations of the GOODS fields from the PEP and GOODS-\textit{Herschel} key programmes, we obtain the deepest PACS far-infrared blank field extragalactic survey carried out by the \textit{Herschel Space Observatory}.
In particular, in the GOODS-S field the combination of these observations is not limited by the exposure time but by confusion.
PACS flux densities are extracted from the maps using two complementary PSF-fitting approaches.
Firstly, we extract PACS flux densities using, as prior information, the expected positions of the sources on the basis of deep 24$\,\mu$m catalogues. 
Secondly, PACS flux densities are extracted ``blindly'', i.e., without positional priors.
The accuracy of both approaches is tested through MC simulations.
In the deepest parts of the GOODS-S field, these catalogues reach 3$\sigma$ depths of 0.9, 0.6, 1.3 mJy at 70, 100, 160$\,\mu$m, respectively. 
From these catalogues we derive number counts down to these unprecedented depths, and determine the infrared luminosity function down to $L_{\rm IR}$$\,=\,$$10^{11}\,$L$_{\odot}$ at $z$$\,\thicksim\,$$1$ and $L_{\rm IR}$$\,=\,$$10^{12}\,$L$_{\odot}$ at $z$$\,\thicksim\,$$2$, respectively.
By integrating these infrared luminosity functions, we estimate the evolution of the SFR density of the Universe up to $z$$\,\thicksim\,$$2$.
We find that the obscured SFR density of the Universe strongly increases from $z$$\,=\,$$0$ to $z$$\,\thicksim\,$$1$ and then increases more moderately up to $z$$\,\thicksim\,$$2.3$.
The obscured SFR density of the Universe is dominated by the contribution of the LIRGs at $z$$\,\thicksim\,$$1$ (i.e., $50\pm26$\%) while it is dominated by the contribution of ULIRGs ($50\pm24$\%) and LIRGs ($30\pm20$\%) at $z$$\,\thicksim\,$$2$.

Maps and source catalogues ($>\,$$3\sigma$) are now publicly released.
Combined with the large wealth of multi-wavelength data available for the GOODS fields, they provide a powerful new tool for studying galaxy evolution over a broad range of redshifts.
\begin{acknowledgements}
We thank the anonymous referee for suggestions which greatly enhanced this work. 
PACS has been developed by a consortium of institutes led by MPE (Germany) and including: UVIE (Austria); KU Leuven, CSL, IMEC (Belgium); CEA, LAM (France); MPIA (Germany); INAF-IFSI/OAA/OAP/OAT, LENS, SISSA (Italy); and IAC (Spain). This development has been supported by the funding agencies BMVIT (Austria), ESA-PRODEX (Belgium), CEA/CNES (France), DLR (Germany), ASI/INAF (Italy), and CICYT/MCYT (Spain).
Support for this work was provided by the NASA through an award issued by JPL/Caltech
\end{acknowledgements}
\bibliographystyle{aa}

\begin{appendix}
\section{Comparison between blind and prior catalogues\label{appendix:blind/prior}}
\begin{figure*}
\center
	\includegraphics[width=15.cm]{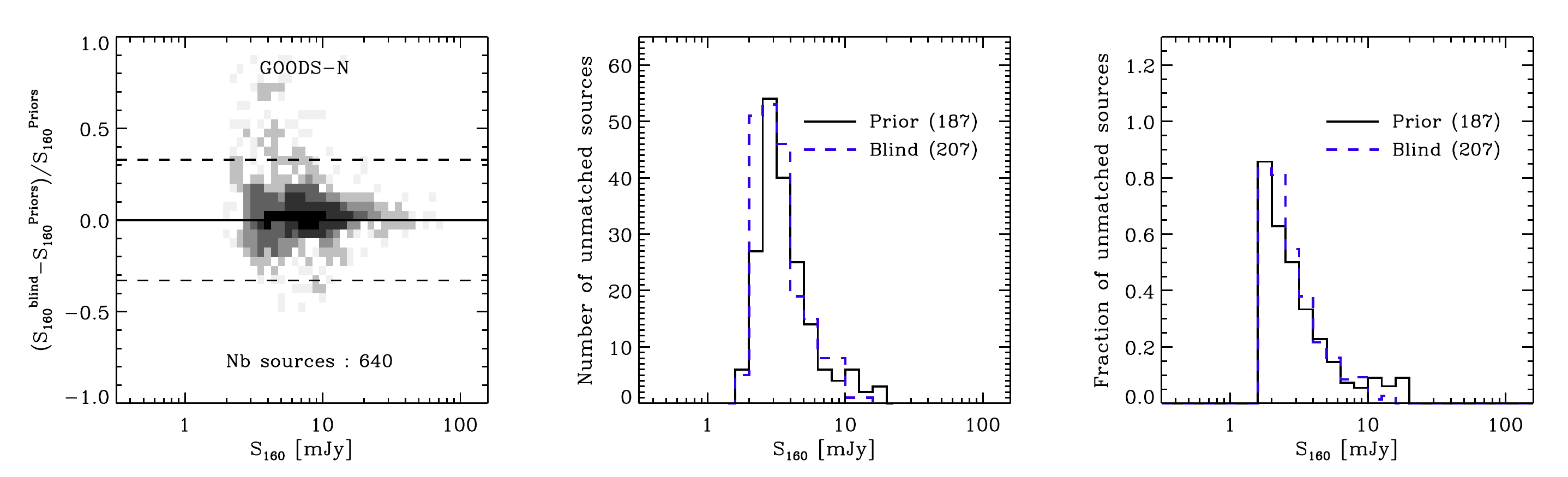}
	\includegraphics[width=15.cm]{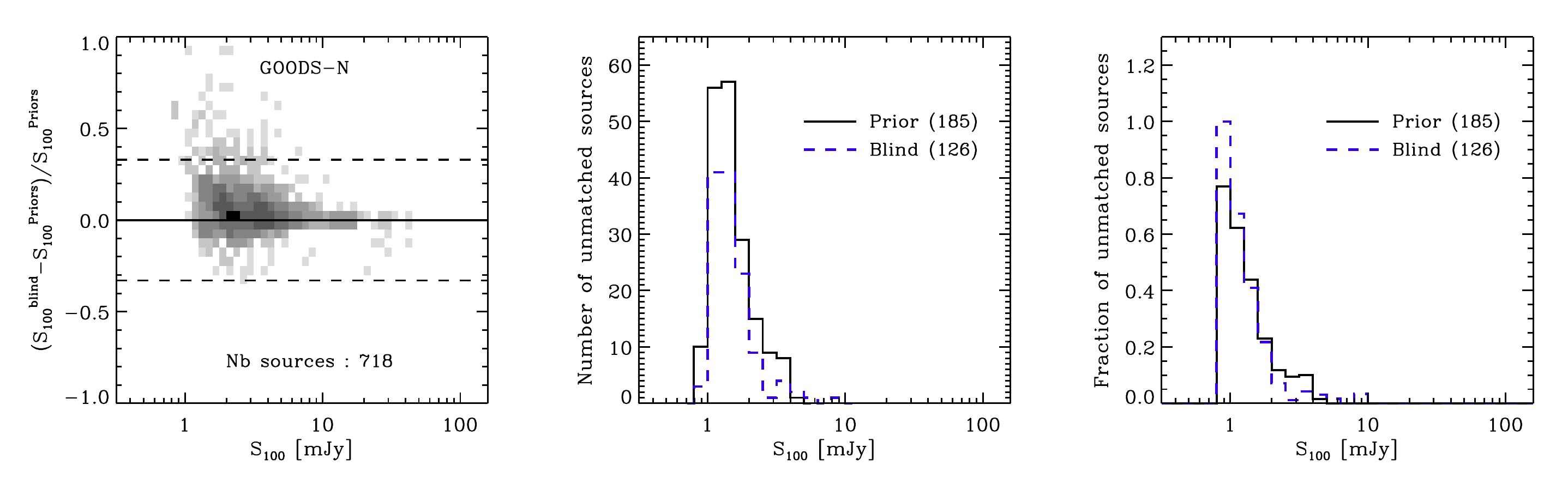}
	\includegraphics[width=15.cm]{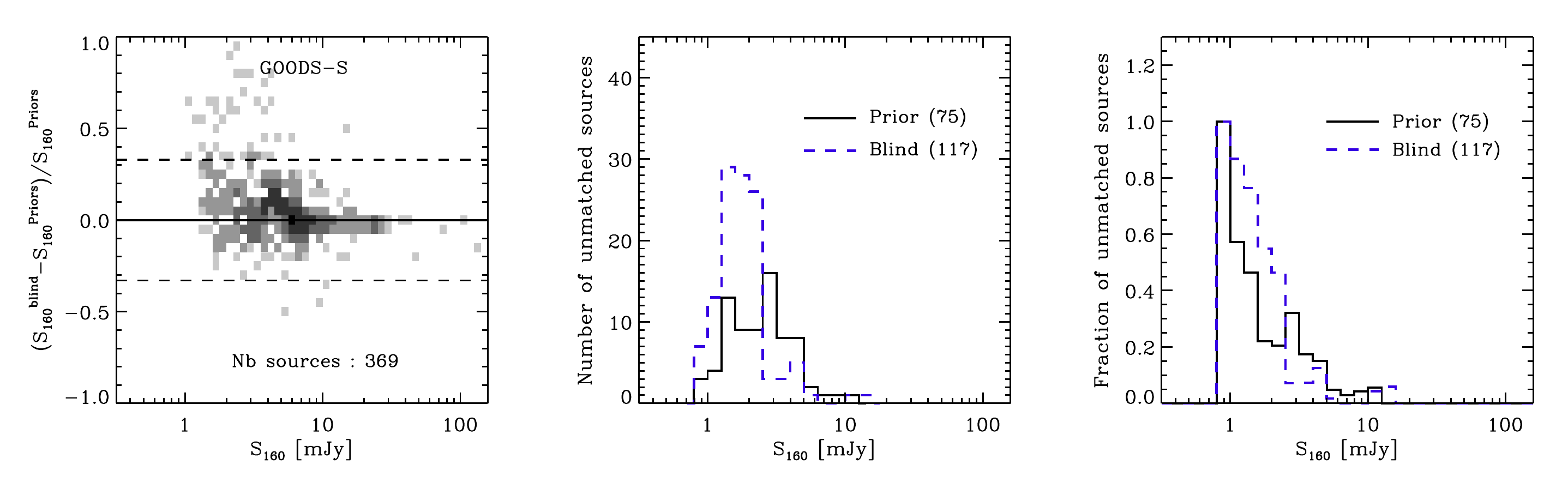}
	\includegraphics[width=15.cm]{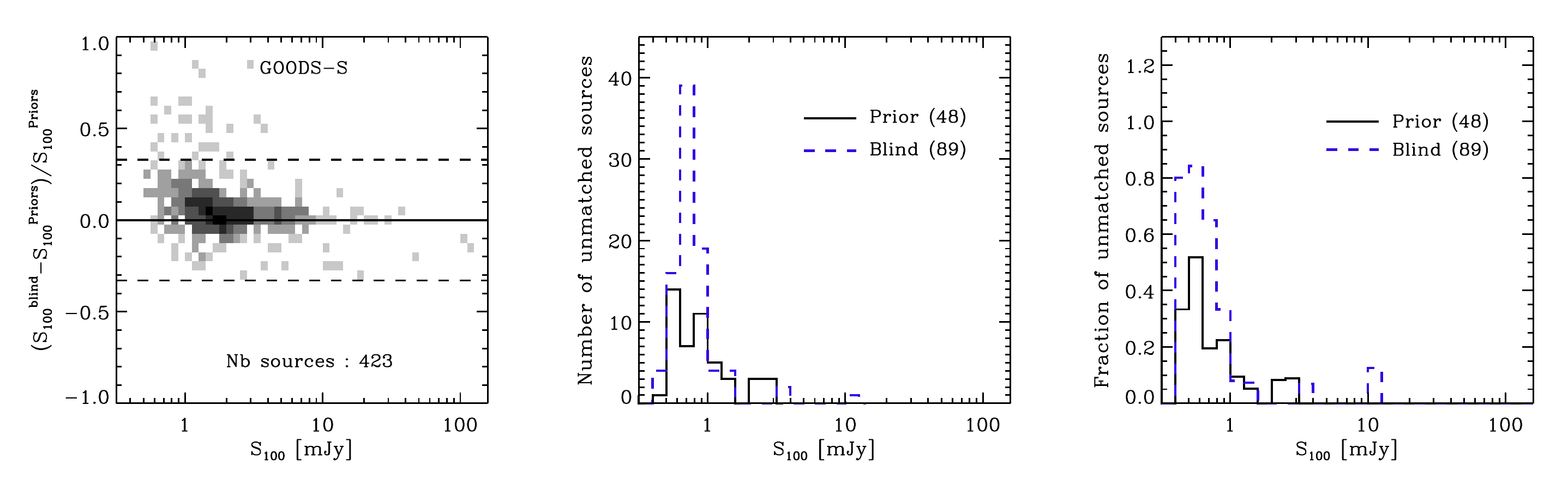}
	\includegraphics[width=15.cm]{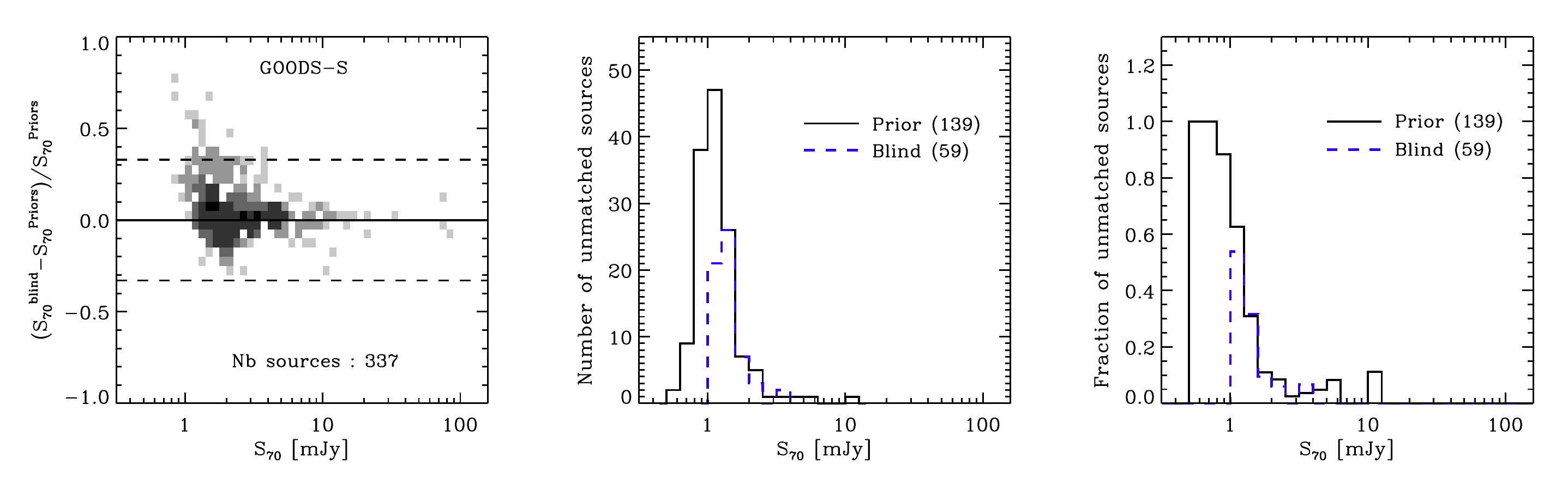}
	\caption{\label{fig:blind/prior}
	Comparison between blind and prior source catalogues.
	(\textit{Left panel}) Direct comparison of flux densities for sources in common to both catalogues.
	(\textit{Centre panel}) Distribution of unmatched sources in absolute number.
	(\textit{Right panel}) Distribution of unmatched sources in fraction relative to total in the given flux density bin.
			}
\end{figure*}

Figure \ref{fig:blind/prior} presents the comparison between the blind and prior catalogues in the GOODS-N and GOODS-S fields.
These comparisons are restricted to the region of the field with exposure time higher than those mentioned in the second column of Table \ref{tab:MC}.
For the GOODS-S 100 and 160$\,\mu$m maps, this restriction corresponds to the central deepest region of the field, i.e., GOODS-S-ultradeep.
The cross-identification of PACS blind and prior sources is based on their MIPS-24$\,\mu$m positions, i.e., for the blind catalogues we use the PACS/MIPS-24$\,\mu$m cross-identification performed using a maximum likelihood analysis (see Section \ref{subsec:package}).
For each field and passband we present the direct comparison of flux densities for sources in common to both catalogues; the distribution of unmatched sources in absolute number; and the distribution of unmatched sources in fraction relative to total in the given flux density bin.

Flux densities extracted with these two independent source extraction methods are consistent with each other.
At faint flux densities, there are however few outliers, with higher flux densities in the blind catalogues than in the prior catalogues.
Such outliers are expected since blind source extraction is more severely affected by blending issues leading to the overestimation of PACS flux densities.

In the GOODS-N field the absolute number of unmatched sources in the blind and prior source catalogues are very similar.
At faint flux densities, the increase of the fraction of unmatched sources is likely explained by the increase of the incompleteness of our catalogues.
Indeed, because our two source extraction methods are independent, at faint flux densities their incomplete samples might not fully overlap.

The blind GOODS-S 100 and 160 $\,\mu$m source catalogues contain a larger number of unmatched sources than the prior source catalogues.
Some sources in ``excess'' in the blind catalogues should correspond to sources effectively missed by our prior extraction due to the lack of MIPS-24$\,\mu$m counterparts (see Section \ref{subsec:prior}).
However, a significant fraction of the sources in ``excess'' in the blind catalogue should correspond to spurious detections, as in this field the contamination of the blind catalogues is supposed to be higher than that of the prior catalogues (see Table \ref{tab:MC}).
The GOODS-N 100$\,\mu$m, GOODS-S 100$\,\mu$m and GOODS-S 160$\,\mu$m blind catalogues also contain each one a bright source (i.e., $>\,$$8\,$mJy) missed by the prior extraction.
Examining these sources, it turns out that they are not missed because of a lack of MIPS-24$\,\mu$m priors, but likely correspond to "spurious" bright sources created by the blind extraction by over-deblending very bright and crowded regions of the field.
Finally, we observe that the prior GOODS-S 70$\,\mu$m source catalogue contains a larger number of unmatched sources as it reaches a slightly lower $3\sigma$ detection limit than the blind catalogue.
\begin{table*}
\center
\footnotesize
\caption{ \label{tab:counts}
PACS 100 and 160$\,\mu$m number counts, normalized to the Euclidean slope.}
\begin{tabular}{ccccccccccccc}
\hline\hline
 \multicolumn{6}{c}{\rule{0pt}{3ex} PACS 100$\,\mu$m} &  & \multicolumn{6}{c}{\rule{0pt}{3ex} PACS 160$\,\mu$m} \\
\cline{1-6} \cline{8-13}
 &  \multicolumn{2}{c}{\rule{0pt}{3ex} GOODS-S ultradeep} & & \multicolumn{2}{c}{\rule{0pt}{3ex} GOODS-N/S deep} & & &  \multicolumn{2}{c}{\rule{0pt}{3ex} GOODS-S ultradeep} & &\multicolumn{2}{c}{\rule{0pt}{3ex} GOODS-N/S deep}\\
 \cline{2-3} \cline{5-6} \cline{9-10} \cline{12-13}
 \rule{0pt}{3ex}$S_{\rm centre}$ & Counts & Err.  & & Counts & Err.  & & $S_{\rm centre}$ & Counts & Err.  & & Counts & Err.  \\
 {\small [mJy]} & \multicolumn{2}{c}{\rule{0pt}{3ex}{\small [$10^4\,$deg$^{-2}\,$mJy$^{1.5}$]} }&   & \multicolumn{2}{c}{\rule{0pt}{3ex}{\small [$10^4\,$deg$^{-2}\,$mJy$^{1.5}$]} } &  &  {\small [mJy]} & \multicolumn{2}{c}{\rule{0pt}{3ex}{\small [$10^4\,$deg$^{-2}\,$mJy$^{1.5}$]} }&   & \multicolumn{2}{c}{\rule{0pt}{3ex}{\small [$10^4\,$deg$^{-2}\,$mJy$^{1.5}$]} }\\
 \hline
 0.56  & 0.59  & 0.10     & & $-$     & $-$  & & 1.42   & 3.59   & 0.49         & & $-$     & $-$  \\
 0.71  & 0.91  & 0.14    & & $-$     & $-$   & & 1.79   & 4.91   & 0.72         & & $-$     & $-$   \\
 0.89  & 1.34  & 0.20     & & $-$     & $-$  & & 2.25   & 6.49   & 0.91         & & 4.92  & 0.31  \\
 1.13  & 1.78  & 0.28     & & 1.29  & 0.10 & & 2.84   & 7.56   & 1.15        & & 6.48  & 0.41   \\
 1.42  & 1.95  & 0.36     & & 1.90  & 0.13 & & 3.57   & 8.62   & 1.54        & & 8.80  & 0.56   \\
 1.79  & 2.80  & 0.51     & & 2.76  & 0.19 & & 4.49   & 9.89   & 1.93        & & 11.85  & 0.78   \\
 2.25  & 3.59  & 0.70     & & 3.63  & 0.26 & & 5.66   & 15.08  & 2.80       & & 14.36  & 1.03   \\
 2.83  & 3.70  & 0.85     & & 4.27  & 0.33 & & 7.13   & 15.37  & 3.32       & & 15.55  & 1.28   \\
 3.57  & 3.73  & 1.03     & & 4.79  & 0.42 & & 8.97   & 13.30  & 3.86       & & 16.68  & 1.59   \\
 4.49  & 3.51  & 1.23     & & 5.44  & 0.55 & & 11.29 & 18.79  & 5.41      & & 18.72  & 2.02    \\
 5.66  & 6.54  & 1.93     & & 6.25  & 0.72 & & 14.22 & 14.87  & 5.93      & & 17.15  & 2.32    \\
 7.13  & 7.53  & 2.48     & & 5.72  & 0.84 & & 17.91 & 10.63  & 6.23      & & 16.86  & 2.79    \\
 8.97  & 4.71  & 2.54     & & 5.87  & 1.02 & & 22.54 & 24.76  & 10.55    & & 19.48  & 3.63    \\
 11.29  & 6.19  & 3.18   & & 4.21  & 1.04 & & 28.38 & 21.99  & 12.47    & & 17.99  & 4.21    \\
14.22  & 5.37  & 3.65    & & 5.85  & 1.42 & & 35.72 &$-$       & $-$         & & 15.81  & 4.69    \\
17.90  & $-$     & $-$    & & 8.36  & 1.99  & &  44.98 &$-$      & $-$         & & 11.07  & 4.68    \\
22.54  & 9.46  & 6.44   & & 5.66  & 2.05  & & 56.62 &$-$       & $-$          & & 11.82  & 5.63   \\ 
28.38  & $-$     & $-$    & & 7.97  & 2.65  & &    $-$   & $-$      & $-$         & & $-$     & $-$       \\
\hline
\end{tabular}
\end{table*}
\begin{table*}
\caption{\label{tab:LF 1} The infrared LF derived from the $1/V_{\rm max}$ analysis in the GOODS-S ultradeep field}
\centering
\begin{tabular}{ccccc}
\hline \hline
{${\rm log}(L_{\rm IR}^{\rm low})-{\rm log}(L_{\rm IR}^{\rm high})$} & {\rm{log($\phi$)}} &  & {${\rm log}(L_{\rm IR}^{\rm low})-{\rm log}(L_{\rm IR}^{\rm high})$} & {\rm{log($\phi$)}} \\
{\tiny{\rm{[log(L$_{\odot}$)]}}} & {\tiny{\rm{[log(${\rm Mpc^{-3}dex^{-1}}$)]}}} & & {\tiny{\rm{[log(L$_{\odot}$)]}}} & {\tiny{\rm{[log(${\rm Mpc^{-3}dex^{-1}}$)]}}}\\
\hline
\multicolumn{2}{c}{\rule{0pt}{3ex}$0.1$$\,<\,$$z$$\,<\,$$0.4$} &  & \multicolumn{2}{c}{\rule{0pt}{3ex}$0.4$$\,<\,$$z$$\,<\,$$0.7$} \\
\cline{1-2} \cline{4-5}
\rule{0pt}{3ex}  9.7 - 10.1 & $-2.33^{+0.13}_{-0.20}$ & & 10.4 - 10.8 & $-2.33^{+0.13}_{-0.14}$\\ 
\rule{0pt}{3ex}10.1 - 10.5 & $-2.43^{+0.13}_{-0.20}$ & & 10.8 - 11.2 & $-2.56^{+0.14}_{-0.15}$\\ 
\rule{0pt}{3ex}10.5 - 10.9 & $-2.59^{+0.15}_{-0.25}$ & & 11.2 - 11.6 & $-2.86^{+0.16}_{-0.20}$\\ 
\rule{0pt}{3ex}10.9 - 11.3 & $-3.29^{+0.21}_{-3.29}$ & & $-$ & $-$ \\ 
\hline
\multicolumn{2}{c}{\rule{0pt}{3ex}$0.7$$\,<\,$$z$$\,<\,$$1.0$} &  & \multicolumn{2}{c}{\rule{0pt}{3ex}$1.0$$\,<\,$$z$$\,<\,$$1.3$} \\
\cline{1-2} \cline{4-5}
\rule{0pt}{3ex}10.8 - 11.2 & $-2.55^{+0.13}_{-0.15}$ & & 11.2 - 11.6 & $-2.74^{+0.17}_{-0.18}$\\ 
\rule{0pt}{3ex}11.2 - 11.6 & $-3.05^{+0.16}_{-0.20}$ & & 11.6 - 12.0 & $-3.09^{+0.19}_{-0.22}$\\ 
\rule{0pt}{3ex}11.6 - 12.0 & $-3.43^{+0.20}_{-0.32}$ & & 12.0 - 12.4 & $-4.15^{+0.34}_{-4.15}$\\ 
\hline
\multicolumn{2}{c}{\rule{0pt}{3ex}$1.3$$\,<\,$$z$$\,<\,$$1.8$} &  & \multicolumn{2}{c}{\rule{0pt}{3ex}$1.8$$\,<\,$$z$$\,<\,$$2.3$} \\
\cline{1-2} \cline{4-5}
\rule{0pt}{3ex}11.4 - 11.8 & $-2.93^{+0.17}_{-0.18}$ & & 11.9 - 12.3 & $-3.15^{+0.14}_{-0.16}$\\ 
\rule{0pt}{3ex}11.8 - 12.2 & $-3.20^{+0.18}_{-0.18}$ & & 12.3 - 12.7 & $-3.67^{+0.18}_{-0.25}$\\ 
\rule{0pt}{3ex}12.2 - 12.6 & $-3.96^{+0.25}_{-0.39}$ & & $-$ & $-$ \\ 
\hline
\end{tabular}
\end{table*}
\begin{table*}
\caption{\label{tab:LF 2} The infrared LF derived from the $1/V_{\rm max}$ analysis in the GOODS-N/S deep fields}
\centering
\begin{tabular}{ccccc}
\hline \hline
{${\rm log}(L_{\rm IR}^{\rm low})-{\rm log}(L_{\rm IR}^{\rm high})$} & {\rm{log($\phi$)}} &  & {${\rm log}(L_{\rm IR}^{\rm low})-{\rm log}(L_{\rm IR}^{\rm high})$} & {\rm{log($\phi$)}} \\
{\tiny{\rm{[log(L$_{\odot}$)]}}} & {\tiny{\rm{[log(${\rm Mpc^{-3}dex^{-1}}$)]}}} & & {\tiny{\rm{[log(L$_{\odot}$)]}}} & {\tiny{\rm{[log(${\rm Mpc^{-3}dex^{-1}}$)]}}}\\
\hline
\multicolumn{2}{c}{\rule{0pt}{3ex}$0.1$$\,<\,$$z$$\,<\,$$0.4$} &  & \multicolumn{2}{c}{\rule{0pt}{3ex}$0.4$$\,<\,$$z$$\,<\,$$0.7$} \\
\cline{1-2} \cline{4-5}
\rule{0pt}{3ex}10.0 - 10.4 & $-2.47^{+0.05}_{-0.07}$ & & 10.6 - 11.0 & $-2.46^{+0.11}_{-0.11}$\\ 
\rule{0pt}{3ex}10.4 - 10.8 & $-2.69^{+0.06}_{-0.08}$ & & 11.0 - 11.4 & $-2.93^{+0.12}_{-0.12}$\\ 
\rule{0pt}{3ex}10.8 - 11.2 & $-3.19^{+0.12}_{-0.18}$ & & 11.4 - 11.8 & $-3.49^{+0.14}_{-0.17}$\\ 
\rule{0pt}{3ex}11.2 - 11.6 & $-4.09^{+0.29}_{-4.09}$ & & 11.8 - 12.2 & $-4.00^{+0.20}_{-0.31}$  \\ 
\hline
\multicolumn{2}{c}{\rule{0pt}{3ex}$0.7$$\,<\,$$z$$\,<\,$$1.0$} &  & \multicolumn{2}{c}{\rule{0pt}{3ex}$1.0$$\,<\,$$z$$\,<\,$$1.3$} \\
\cline{1-2} \cline{4-5}
\rule{0pt}{3ex}11.0 - 11.4 & $-2.59^{+0.10}_{-0.10}$ & & 11.4 - 11.8 & $-2.98^{+0.15}_{-0.16}$\\ 
\rule{0pt}{3ex}11.4 - 11.8 & $-3.14^{+0.11}_{-0.12}$ & & 11.8 - 12.2 & $-3.58^{+0.17}_{-0.18}$\\ 
\rule{0pt}{3ex}11.8 - 12.2 & $-3.88^{+0.16}_{-0.20}$ & & 12.2 - 12.6 & $-4.47^{+0.25}_{-0.40}$\\ 
\rule{0pt}{3ex}12.2 - 12.6 & $-4.83^{+0.31}_{-4.83}$ & & 12.6 - 13.0 & $-4.95^{+0.34}_{-4.95}$\\ 
\hline
\multicolumn{2}{c}{\rule{0pt}{3ex}$1.3$$\,<\,$$z$$\,<\,$$1.8$} &  & \multicolumn{2}{c}{\rule{0pt}{3ex}$1.8$$\,<\,$$z$$\,<\,$$2.3$} \\
\cline{1-2} \cline{4-5}
\rule{0pt}{3ex}11.6 - 12.0 & $-3.19^{+0.16}_{-0.15}$ & & 12.1 - 12.5 & $-3.50^{+0.11}_{-0.12}$\\ 
\rule{0pt}{3ex}12.0 - 12.4 & $-3.78^{+0.16}_{-0.16}$ & & 12.5 - 12.9 & $-4.06^{+0.14}_{-0.16}$\\ 
\rule{0pt}{3ex}12.4 - 12.8 & $-4.34^{+0.20}_{-0.23}$ & & 12.9 - 13.3 & $-4.99^{+0.25}_{-0.54}$\\ 
\hline
\end{tabular}
\end{table*}
\begin{table*}
\caption{\label{tab:fit parameter}Parameter values of the infrared LF}
\centering
\begin{tabular}{ccccc}
\hline \hline
{Redshift} &{$\alpha_{1}\,^\mathrm{a}$} &{$\alpha_{2}\,^\mathrm{a}$} &{\rm{log($L_{{\rm knee}}$)}} & {\rm{log($\phi_{{\rm knee}}$)}} \\
& & &  {\tiny{\rm{[log(${\rm L_{\odot}}$)]}}} & {\tiny{\rm{[log(${\rm Mpc^{-3}dex^{-1}}$)]}}} \\
\hline
$z\thicksim0$ & $-0.60$ & $-2.20$ & $10.48\pm0.02 $ & $-2.52\pm0.03$\\
$0.1<z<0.4$ & $-0.60$ & $-2.20$ & $10.84\pm0.06$ & $-2.85\pm0.04$\\
$0.4<z<0.7$ & $-0.60$ & $-2.20$ & $11.28\pm0.12$ & $-2.82\pm0.14$\\
$0.7<z<1.0$ & $-0.60$ & $-2.20$ & $11.53\pm0.15$ & $-2.96\pm0.18$\\
$1.0<z<1.3$ & $-0.60$ & $-2.20$ & $11.71\pm0.14 $ & $-3.01\pm0.20$\\
$1.3<z<1.8$ & $-0.60$ & $-2.20$ & $12.00\pm0.15$ & $-3.29\pm0.19$\\
$1.8<z<2.3$ & $-0.60$ & $-2.20$ & $12.35\pm0.19 $ & $-3.47\pm0.23$\\
\hline
\end{tabular}
\begin{list}{}{}
\item[$^{\mathrm{a}}$] Fixed slopes of the infrared LF.
\end{list}
\end{table*}
\begin{table*}
\caption{\label{tab:IR LD} Evolution of the comoving IR energy density and the relative contribution of ``faint'' galaxies ($10^7\,$L$_{\odot}$$\,<\,$$L_{\rm IR}$$\,<\,$$10^{11}\,$L$_{\odot}$), LIRGs ($10^{11}\,$L$_{\odot}$$\,<\,$$L_{\rm IR}$$\,<\,$$10^{12}\,$L$_{\odot}$) and ULIRGs ($L_{\rm IR}$$\,>\,$$10^{12}\,$L$_{\odot}$). The local reference is taken from \citet{sanders_2003}. Comoving infrared energy densities constrained using our PACS observations are highlighted in bold. Others values rely upon extrapolations based on $\phi_{\rm knee}$ and $L_{\rm knee}$ and a faint-end slope of the IR LF fixed at its $z$$\,\thicksim\,$$0$ value.}
\centering
\begin{tabular}{ccccc}
\hline \hline
Redshift & Total IR LD & ``faint'' galaxies & LIRGs & ULIRGs \\
 & {\tiny{\rm{[10$^6\,\times\,$L$_{\odot}\,$Mpc$^{-3}$]}}}  & {\tiny{\rm{[10$^6\,\times\,$L$_{\odot}\,$Mpc$^{-3}$]}}}  & {\tiny{\rm{[10$^6\,\times\,$L$_{\odot}\,$Mpc$^{-3}$]}}}  & {\tiny{\rm{[10$^6\,\times\,$L$_{\odot}\,$Mpc$^{-3}$]}}}\\
\hline
\rule{0pt}{3ex}0.00$^{\rm a}$  & $ 131^{+6}_{-9}             $  & $  122^{+6}_{-8}            $  & $    7^{+1}_{-1}                $  & $   0.50^{+0.05}_{-0.05} $ \\
\rule{0pt}{3ex}0.25  & $ 141^{+18}_{-18}        $  & $   117^{+14}_{-14}       $  & $   \mathbf{22^{+10}_{-8}   }        $  & $   \mathbf{1.5^{+0.6}_{-0.6}} $ \\
\rule{0pt}{3ex}0.55  & $ 421^{+85}_{-94}        $  & $   220^{+110}_{-54}     $  & $  \mathbf{ 180^{+40}_{-79}}      $  & $   \mathbf{18^{+7}_{-10} }$ \\
\rule{0pt}{3ex}0.85  & $ 544^{+139}_{-141}    $  & $   242^{+145}_{-82}    $  & $   \mathbf{260^{+59}_{-62}  }     $  & $   \mathbf{38^{+29}_{-19}} $ \\
\rule{0pt}{3ex}1.15  & $ 752^{+254}_{-273}    $  & $   275^{+215}_{-120}  $  & $   379^{+143}_{-144}  $  & $   \mathbf{95^{+47}_{-46} }$ \\
\rule{0pt}{3ex}1.55  & $ 789^{+276}_{-268}    $  & $   229^{+110}_{-94}    $  & $   359^{+159}_{-148}  $  & $   \mathbf{198^{+92}_{-83} } $ \\
\rule{0pt}{3ex}2.05  & $ 1093^{+558}_{-320}  $  & $   215^{+226}_{-76}    $  & $   336^{+354}_{-118}  $  & $   \mathbf{538^{+156}_{-151} } $ \\
\rule{0pt}{0.5ex}\\
\hline
\end{tabular}
\end{table*}
\end{appendix}
\end{document}